\documentclass[twocolumn,superscriptaddress,floatfix,aps,prd,showpacs]{revtex4}

\usepackage{graphicx}
\usepackage{bm}
\usepackage{epsf}

\begin{document}

\def\agt{\mathrel{\raise.3ex\hbox{$>$}\mkern-14mu\lower0.6ex\hbox{$\sim$}}}
\def\alt{\mathrel{\raise.3ex\hbox{$<$}\mkern-14mu\lower0.6ex\hbox{$\sim$}}}

\newcommand{\beq}{\begin{equation}}
\newcommand{\eeq}{\end{equation}}
\newcommand{\beqn}{\begin{eqnarray}}
\newcommand{\eeqn}{\end{eqnarray}}
\newcommand{\pa}{\partial}
\newcommand{\vp}{\varphi}
\newcommand{\varep}{\varepsilon}
\newcommand{\ep}{\epsilon}
\newcommand{\comp}{(M/R)_\infty} 
\def\bI{\hbox{$\,I\!\!\!$--}}

\title{Merger of binary neutron stars to a black hole: 
Disk mass, short gamma-ray bursts, and quasinormal mode ringing}

\author{Masaru Shibata}
\affiliation{Graduate School of Arts and Sciences, 
University of Tokyo, Komaba, Meguro, Tokyo 153-8902, Japan}

\author{Keisuke Taniguchi}
\affiliation{Department of Physics,
University of Illinois at Urbana-Champaign,
Urbana, IL 61801-3080, USA}

\begin{abstract}

Three-dimensional simulations for the merger of binary neutron stars
are performed in the framework of full general relativity. We pay
particular attention to the black hole formation case and to the
resulting mass of the surrounding disk for exploring possibility for
formation of the central engine of short-duration gamma-ray bursts
(SGRBs). Hybrid equations of state are adopted mimicking realistic,
stiff nuclear equations of state (EOSs), for which the maximum allowed
gravitational mass of cold and spherical neutron stars, $M_{\rm sph}$,
is larger than $2M_{\odot}$. Such stiff EOSs are adopted motivated by
the recent possible discovery of a heavy neutron star of mass $\sim
2.1 \pm 0.2 M_{\odot}$. For the simulations, we focus on binary
neutron stars of the ADM mass $M \agt 2.6M_{\odot}$. For an ADM mass
larger than the threshold mass $M_{\rm thr}$, the merger results in
prompt formation of a black hole irrespective of the mass ratio $Q_M$
with $0.65 \alt Q_M \leq 1$. The value of $M_{\rm thr}$ depends on the
EOSs and is approximately written as $1.3$--$1.35M_{\rm sph}$ for the
chosen EOSs.  For the black hole formation case, we evolve the
spacetime using a black hole excision technique and determine the mass
of a quasistationary disk surrounding the black hole.  The disk mass
steeply increases with decreasing the value of $Q_M$ for given ADM
mass and EOS. This suggests that a merger with small value of $Q_M$ is a
candidate for producing central engine of SGRBs. For $M < M_{\rm
thr}$, the outcome is a hypermassive neutron star of a large
ellipticity. Because of the nonaxisymmetry, angular momentum is
transported outward. If the hypermassive neutron star collapses to a
black hole after the longterm angular momentum transport, the disk
mass may be $\agt 0.01M_{\odot}$ irrespective of $Q_M$.  Gravitational
waves are computed in terms of a gauge-invariant wave extraction
technique. In the formation of the hypermassive neutron star,
quasiperiodic gravitational waves of frequency between 3 and 3.5 kHz
are emitted irrespective of EOSs.  The effective amplitude of
gravitational waves can be $\agt 5 \times 10^{-21}$ at a distance of
50 Mpc, and hence, it may be detected by advanced
laser-interferometers. For the black hole formation case, the black
hole excision technique enables a longterm computation and extraction
of ring-down gravitational waves associated with a black hole
quasinormal mode.  It is found that the frequency and amplitude are
$\approx 6.5$--7 kHz and $\sim 10^{-22}$ at a distance of 50 Mpc for
the binary of mass $M \approx 2.7$-- $2.9M_{\odot}$.
\end{abstract}
\pacs{04.25.Dm, 04.30.-w, 04.40.Dg}

\maketitle

\section{Introduction}

Binary neutron stars in close orbits are strong emitters of
gravitational waves. A scenario based mainly on a recent discovery of
the binary system PSRJ0737-3039 \cite{NEW} suggests that the detection
rate of gravitational waves by the advanced laser-interferometric
gravitational wave observatory (LIGO) will be $\sim 40$--600 yr$^{-1}$
\cite{BNST}; i.e., one event per year is expected within the distance
of $\approx 35$--90 Mpc since the advanced LIGO is able to detect
gravitational waves from coalescing binary neutron stars within the
distance of about 300 Mpc. This indicates that a coalescing and
merging binary neutron star is one of the most promising sources for
kilometer-size laser-interferometric detectors \cite{KIP,Ando}.

Merger of binary neutron stars has been also proposed for many
years~\cite{NPT,GRB} as a possible formation scenario for a central
engine of short-hard gamma-ray bursts (SGRBs). Associations between
SGRBs and elliptical galaxies reported recently~\cite{short} make it
unlikely that SGRBs are related to supernova stellar core collapse
since elliptical galaxies have not produced massive stars in the past
$\sim 10^{10}$ yrs. In addition, recent observations of the afterglow
of the SGRB 050709 rule out the presence of a supernova light curve
and point to a binary compact object merger as the most likely central
engine~\cite{fox2005,sgrb724}.  The merger of compact-object binaries
(binary neutron stars or black hole-neutron star binaries) is thus the
most favored hypothesis for explaining SGRBs.  According to a standard
scenario based on the merger hypothesis, after the merger, a
stellar-mass black hole is formed with a surrounding accretion torus
of mass $\agt 0.01M_{\odot}$ \cite{RJ0}.  Energy extracted from this
system by either magnetohydrodynamic processes or neutrino radiation
powers the fireball and high-Lorentz factor jets for SGRBs, for
which the typical burst energy is between $10^{48}$ and $10^{49}$ ergs
(after a correction of a beaming factor) \cite{fox2005}.

Hydrodynamic simulations in the framework of full general relativity
provide the best approach for studying the merger of binary neutron
stars. Over the last several years, numerical methods for solving
coupled equations of the Einstein and hydrodynamic equations have been
developed (e.g.,
\cite{gr3d,bina,bina2,other,Font,shiba2d,STU,marks,illinois,Baiotti,STU2})
and now such simulations are feasible with an accuracy high enough for
yielding scientific results (e.g., \cite{STU,STU2}).

For many years, the fully general relativistic (GR) simulations for
the merger of binary neutron stars had been performed adopting an
ideal equation of state (EOS) $P=(\Gamma-1) \rho\varep$ where $P$,
$\rho$, $\varep$, and $\Gamma$ are pressure, rest-mass density,
specific internal energy, and adiabatic constant
\cite{bina,bina2,STU,marks,illinois}. However, simulations with
realistic EOSs are necessary for quantitative understanding for the
merger. From this motivation, in a previous paper, we performed
simulations taking into account realistic nuclear EOSs \cite{STU2} and
clarified that properties of the outcomes and gravitational waveforms
are significantly different from those obtained in the ideal EOS with
$\Gamma=2$. In particular, we found that a hypermassive neutron star
\cite{footnote1,BSS} of {\em elliptical shape} is formed for a
reasonable mass range of the merger progenitor because of the
stiffness of the realistic EOSs. As a consequence, hypermassive
neutron stars can be strong emitters of high-frequency, quasiperiodic
gravitational waves, which may be detected by the advanced
laser-interferometric gravitational wave detectors \cite{S2005}.

In this paper, we extend the previous work from the following
motivation. One is based on the recent possible discovery of a
high-mass neutron star with mass $2.1 \pm 0.2 M_{\odot}$ (one $\sigma$
error) \cite{nice}.  This measurement indicates that the maximum mass
of spherical neutron stars, $M_{\rm sph}$, may be larger than $\sim
2M_{\odot}$, and hence, very stiff EOSs such as those proposed in
\cite{EOS1,EOS2} are favored. In particular, if $M_{\rm sph}$ is
larger than $\sim 2.1M_{\odot}$, the nuclear EOS is so stiff that most
of the EOSs proposed so far will be rejected, resulting in that
restricted EOSs such as those derived by Akmal, Pandharipande, and
Ravenhall (hereafter APR) \cite{EOS1} survive. However, fully GR
simulations for the merger with such stiff EOSs have not been
performed to this time. This paper is mainly devoted to presenting new
numerical results obtained with the APR EOS.

The other motivation is to determine the final outcome for the case
that a black hole is formed promptly after the merger. Clarifying the
final state of the black hole system, in particular the surrounding
disk mass, is an important subject in exploring whether a remnant of
the merger can be a central engine of SGRBs. The previous simulations
\cite{STU,STU2} were performed only for a short time after the
formation of black hole because of the so-called grid stretching
around the black hole horizon. Under this situation, the geometry is so
steep near the horizon that the accuracy in the numerical computation
is lost. The popular approach for overcoming this difficulty is to
adopt excision techniques \cite{Unruh,AB,DLSS}. In the present
simulation, the growth of the black hole is followed until the system
approximately reaches a relaxed state by employing a simple excision
technique~\cite{AB,DLSS}. With this technique, the disk mass
surrounding a black hole can be determined.  Simulations are performed
for a wide variety of mass ratio for the black hole formation case as
well as in two EOSs.  We show that the disk mass surrounding a black
hole depends strongly on the mass ratio of the binary. In addition,
the excision technique enables a simulation long enough to extract 
gravitational waves emitted after the formation of black holes. 
We show for the first time that gravitational waves are
determined by a quasinormal mode of the formed black hole. 

To determine the disk mass around a formed black hole, Oechslin and
Janka recently performed a series of interesting simulations for
merger of binary neutron stars \cite{OJ} using a realistic EOS (the
so-called Shen's EOS) \cite{shen}, which is also so stiff that the
maximum mass of spherical neutron stars is $\approx 2.2M_{\odot}$.
They employ an approximate formulation of general relativity (the
conformal flatness approximation for the spatial three-geometry).  In
this approximation, gravitational radiation is neglected. They follow 
the merger process only in the formation of hypermassive neutron
stars. Assuming that the hypermassive neutron stars eventually
collapse to a black hole by a dissipation mechanism, they estimate the
disk mass (i.e., they compute the rest-mass of fluid elements with the
sufficiently large specific angular momentum which are expected to
escape falling into the black hole). They conclude that the disk mass
around the formed black hole will be always larger than $0.02
M_{\odot}$ because the angular momentum transport during the merger
process efficiently works. As shown in Sec. IV, the angular momentum
transport indeed plays an important role in the formation of disks
around the hypermassive neutron stars, and hence, our numerical results agree
partly with theirs. However, for the merger of high-mass binaries, our
results do not agree with theirs.

In their case, the merger results in a hypermassive neutron star even
for the total gravitational mass $\approx 3M_{\odot}$. This result
disagrees with ours. The reason may be partly due to different choice
of the EOS from ours, but is also likely due to the fact that they do
not take into account radiation reaction of gravitational waves.  As
shown in Sec. \ref{sec:gw} as well as in \cite{STU2}, angular momentum
is dissipated by a factor of $\sim 15\%$ in the first 3 ms from the
last one orbit. This significant dissipation induces prompt collapse
to a black hole for the high-mass case in our results. In this case, a
black hole is formed in much shorter than 1 ms after the onset of the
merger, and hence, the angular momentum transport does not work
efficiently before the black hole formation, unless the mass
difference of two stars is significant. We find that the disk mass is
much smaller than $0.01M_{\odot}$ in the prompt formation of a black
hole for the nearly equal-mass case, as shown in Sec. IV.

The paper is organized as follows. In Sec. II A--D, basic equations,
gauge conditions, excision scheme, methods for extracting
gravitational waves, and quantities used in the analysis for numerical
results are reviewed. Then, the hybrid EOSs adopted in this paper are
described in Sec. II E. In Sec. III, initial conditions and setting
for simulations are described. In Sec. IV, numerical results are
shown, paying attention to the merger process, the formed outcome, and
the disk mass surrounding a black hole.  Implication of our results to
formation of a central engine of SGRBs is also discussed in 
\ref{sec:grb}. In Sec. V, gravitational waveforms are presented.
Section VI is devoted to a summary. Throughout this paper, we adopt
the geometrical units in which $G=c=1$ where $G$ and $c$ are the
gravitational constant and the speed of light.  Latin and Greek
indices denote spatial components ($x, y, z$) and space-time
components ($t, x, y, z$), respectively: $r \equiv
\sqrt{x^2+y^2+z^2}$. $\delta_{ij}(=\delta^{ij})$ denotes the Kronecker
delta.

\section{Formulation}

\subsection{Summary of formulation}

Our formulation and numerical scheme for fully GR simulations in three
spatial dimensions are the same as in~\cite{STU,STU2}, to which the
reader may refer for details of basic equations.

The fundamental variables for geometry are $\alpha$: lapse function,
$\beta^k$: shift vector, $\gamma_{ij}$: metric in three-dimensional
spatial hypersurface, and $K_{ij}$: extrinsic curvature. In addition,
we define the conformal factor $\psi \equiv e^{\phi}\equiv
\gamma^{1/12}$, conformal three-metric $\tilde
\gamma_{ij}=e^{-4\phi}\gamma_{ij}$, three auxiliary functions
$F_i\equiv \delta^{jk}\pa_{j} \tilde \gamma_{ik}$, the trace of the
extrinsic curvature $K$, and a tracefree part of the extrinsic
curvature $\tilde A_{ij} \equiv e^{-4\phi}(K_{ij}-\gamma_{ij} K/3)$.
Here, $\gamma={\rm det}(\gamma_{ij})$. 

The Einstein evolution equations are solved using a version of the
BSSN formalism following previous papers \cite{SN,gr3d,bina2,STU}: We
evolve $\tilde \gamma_{ij}$, $\phi$, $F_i$, $\tilde A_{ij}$, and $K$
using an unconstrained free evolution code. The latest version of our
formulation and numerical method is described in \cite{STU}.  The
point worthy to note is that the equation for $\phi$ is written to a
conservative form similar to the continuity equation, and solving this
improves the accuracy of the conservation of the ADM
(Arnowitt-Deser-Misner) mass and the total angular momentum significantly.

The fundamental variables for the hydrodynamics are 
$\rho$: rest-mass density, 
$\varep$: specific internal energy, 
$P$: pressure, $u^{\mu}$: four velocity, and
\beqn
v^i = {dx^i \over dt}={u^i \over u^t}. 
\eeqn
For our numerical implementation of the hydrodynamic equations, 
we define a weighted density, 
a weighted four-velocity, and a specific energy defined, respectively, by
\beqn
&&\rho_* \equiv \rho \alpha u^t e^{6\phi}, \\
&&\hat u_i \equiv h u_i, \\
&& \hat e \equiv h\alpha u^t - {P \over \rho \alpha u^t},
\eeqn
where $h=1+\varepsilon+P/\rho$ denotes the specific enthalpy. 
GR hydrodynamic equations are written into the
conservative form for variables $\rho_*$, $\rho_* \hat u_i$, and
$\rho_* \hat e$, and solved using a high-resolution central (HRC) scheme
\cite{kurganov-tadmor,SFont}.
In this approach, the transport terms such as $\pa_i
(\cdots)$ are computed by Kurganov-Tadmor scheme with a third-order
(piecewise parabolic) spatial interpolation. In \cite{SFont}, we
illustrate that the results obtained in this scheme approximately
agree with those in a high-resolution shock-capturing scheme
based on the Roe-type reconstruction for the fluxes \cite{Font,shiba2d}. 
At each time step, $\alpha u^t$ is determined by solving an algebraic
equation derived from the normalization $u^{\mu}u_{\mu}=-1$, and then,
the primitive variables such as $\rho$, $\varepsilon$, and $v^i$ are updated.

A uniform atmosphere of small density $\rho \approx 10^7~{\rm g/cm^3}$
is added outside neutron stars at $t=0$, since the vacuum is not
allowed in any shock-capturing scheme. However, in the HRC scheme, the
density can be chosen to be much smaller than the previous values
\cite{STU,STU2}. This is a benefit in this scheme. The integrated
baryon rest-mass of the atmosphere is $\sim 10^{-4}M_{\odot}$ in the
present simulation with the largest grid size (see Sec. III). Hence,
the effect of the atmosphere for the evolution of binary neutron stars
and its contribution to the disk mass around a black hole eventually
formed is negligible. (However, for the nearly equal-mass merger, the
disk mass surrounding a black hole is small as several $\times
10^{-4}M_{\odot}$; see Sec.  \ref{sec:angmom_trans}. In such case, we
subtract the contribution of the atmosphere.)

As the time slicing condition, an approximate maximal slice (AMS)
condition $K \approx 0$ has been adopted following previous papers
\cite{gw3p2,gr3d,bina2} to this time. In this condition, $\alpha$ is
determined by approximately solving an elliptic-type equation. This
condition is also adopted in the case of hypermassive neutron star
formation in this paper. On the other hand, for the formation of a
black hole in which an excision evolution is necessary to follow its
growth, the AMS condition is not advantageous since it is
not easy to find an appropriate boundary condition for $\alpha$ at 
the excision surface. Thus, for the black hole formation, we adopt
a dynamical time slicing in which the lapse function is determined from
\beq
\pa_t \alpha = -\alpha K. \label{alphaK}
\eeq
In this condition, we need to determine the initial value of $\alpha$.
Here, we initially impose the maximal slicing condition for $\alpha$. 
A few simulations were performed in both slicing conditions until formation
of a black hole, and we confirmed that the results depend weakly on the
slicing conditions, indicating that Eq. (\ref{alphaK}) gives a similar
slicing to the maximal slicing. 

As the spatial gauge condition, we adopt the following hyperbolic
gauge condition as in \cite{S03,STU};
\beq
\pa_t \beta^k=\tilde \gamma^{kl}(F_l+ \Delta t \pa_t F_l).
\eeq
Here, $\Delta t$ denotes a time step of the simulation. 
Successful numerical results for the merger of binary neutron stars in
this gauge condition are presented in \cite{STU,STU2}.

In the presence of a black hole, the location is determined using an
apparent horizon finder for which our method is described in
\cite{AH}.  The location of the apparent horizons is used for
determining an excision surface (see the next subsection).

Following previous works, we adopt binary neutron stars in close 
quasiequilibrium circular orbits as the initial condition.  In
computing the quasiequilibrium state, we use the so-called conformally
flat formalism for the Einstein equation \cite{WM}. Solutions in this
formalism satisfy the constraint equations in general relativity,
and hence, it can be used for the initial condition. The irrotational
velocity field is assumed since it is considered to be a good
approximation for coalescing binary neutron stars in nature
\cite{CBS}. The coupled equations of the field and hydrostatic
equations \cite{irre} are solved by a spectral method developed
by Bonazzola, Gourgoulhon, and Marck \cite{GBM}. Detailed numerical
calculations have been done by Taniguchi and part of the numerical
results are presented in \cite{TG}. 

\subsection{Black hole excision}

Whenever an apparent horizon is found during the simulation, subsequent
evolution is followed using an excision technique \cite{Unruh}. We adopt
the so-called simple excision method originally proposed by Alcubierre
and Br\"ugmann \cite{AB}. In this method, one first determines a
two-surface inside an apparent horizon which is used for the inner
boundary in the numerical simulation. And then, at the boundary, we
impose the so-called copying boundary condition in which the time
derivative of geometric variables $\tilde \gamma_{ij}$, $\phi$,
$\tilde A_{ij}$, $K$, and $F_i$ is assumed to be spatially constant at
the boundary.

In our method, we choose a cube as the excision boundary of which the
length of each edge is $2\ell=2n \Delta x$ where $\Delta x$ denotes
the grid spacing. $n$ is a positive integer determined from the
equation
\beq
n={\rm integer}[{\rm min}[r_{\rm AH}(\theta,\varphi)]/(\sqrt{2}\Delta x)]-1
\eeq
where $r_{\rm AH}(\theta,\varphi)$
denotes the coordinate radii of the apparent horizon. At the surface
of the cube, we simply copy the time derivative of each variable at
the next cell along the normal line of each surface.  At the corners, 
we copy the time derivative at the next cell along the line connecting 
the corner and origin, and at the edges, at the next cell along a
perpendicular line connecting the edge and another edge in the
diagonal direction.

Since the computational resource is restricted, the radii of the
apparent horizon are covered only by 6--8 grid zones. As a result, the
value of $n$ is small; typically $n=2$. With such a small value, it is
difficult to maintain the accuracy for a long time. However, it is
still possible to continue the simulation for $\sim 0.5$--0.7 ms after
formation of a black hole, which is sufficiently long for
approximately determining the disk mass around the black hole, and for
computing ring-down gravitational waveforms associated with a black
hole quasinormal mode.

\subsection{Extracting gravitational waves}

Gravitational waves are computed in terms of the gauge-invariant
Moncrief variables in a flat spacetime \cite{moncrief} as in our
series of papers (e.g., \cite{gw3p2,STU,STU2,SS3}).  The detailed
equations are describe in \cite{STU,SS3} to which the reader may
refer. In this method, we split the metric in the wave zone into the
flat background and linear perturbation. Then, the linear part is
decomposed using the tensor spherical harmonics and gauge-invariant
variables are constructed for each mode of eigen values $(l,m)$.  The
gauge-invariant variables of $l \geq 2$ can be regarded as
gravitational waves in the wave zone, and hence, we focus on such
modes. In the merger of binary neutron stars of mass ratio larger than
$\sim 0.7$, we have found that the even-parity mode of $(l, |m|)=(2,
2)$ is much larger than other modes. Thus, in the following, we pay
attention only to this mode.

Using the gauge-invariant variables, the luminosity and the angular 
momentum flux of gravitational waves can be defined by
\beqn
&&{dE \over dt}={r^2 \over 32\pi}\sum_{l,m} |\pa_t R_{lm}|^2
\label{dedt} \\
&&{dJ \over dt}={r^2 \over 32\pi}\sum_{l,m}
|m(\pa_t R_{lm}) R_{lm}|, \label{dJdt} 
\eeqn
where $R_{lm}$ is the gauge-invariant variable. In this paper, 
we focus only on the even-parity mode with $l=2$ for $R_{lm}$.
The total radiated energy and angular momentum are
obtained by the time integration of $dE/dt$ and $dJ/dt$. 

To search for the characteristic frequencies of gravitational waves,
the Fourier spectra are computed by 
\beq
\bar R_{lm}(f)=\int e^{2\pi i f t} R_{lm}(t)dt,
\eeq
where $f$ denotes a frequency of gravitational waves. 
Using the Fourier spectrum, the energy power spectrum is defined by 
\beq
{dE \over df}={\pi \over 4}r^2 \sum_{l\geq 2, m\geq 0}
|\bar R_{lm}(f) f|^2 ~~~(f > 0), \label{power}
\eeq
where for $m\not=0$, we define 
\beq
\bar R_{lm}(f)
\equiv \sqrt{|\bar R_{l m}(f)|^2 + |\bar R_{l -m}(f)|^2}~~(m>0), 
\eeq
and use $|\bar R_{lm}(-f)|=|\bar R_{lm}(f)|$ for deriving Eq. (\ref{power}). 

Since we focus only on the $l=2$ even-parity mode, 
the gravitational waveforms are written as
\beqn
&&h_+=
{1 \over r}\biggl[ \sqrt{{5 \over 64\pi}}
\{ R_{22+}(1+\cos^2\theta)\cos(2\varphi) \nonumber \\
&&~~~~~~~~~~~~~~~~~~+R_{22-}(1+\cos^2\theta)\sin(2\varphi) \} \nonumber \\
&&~~~~~~~~~~
+ \sqrt{{15 \over 64\pi}}R_{20} r \sin^2\theta \biggr],\label{eq34} \\
&&h_{\times}={2 \over r} \sqrt{{5 \over 64\pi}}
\Bigl[ -R_{22+} \cos\theta \sin(2\varphi) \nonumber \\
&&~~~~~~~~~~~~~~~~~~~+R_{22-} \cos\theta \cos(2\varphi)\Bigr]. \label{eq35}
\eeqn
In Eqs. (\ref{eq34}) and (\ref{eq35}), the variables are defined by 
\beqn
R_{22\pm} \equiv {R_{22} \pm R_{2~-2} \over \sqrt{2}}r. 
\eeqn
In the following, we present
\beqn
R_+=\sqrt{{5 \over 16\pi}}R_{22+},~~~
R_{\times}=\sqrt{{5 \over 16\pi}}R_{22-}.  
\eeqn
These have the unit of length and provide the amplitude of a given mode
measured by an observer located in the most optimistic direction. 
The amplitude of gravitational waves, $h_{\rm gw}$,
observed at a distance of $r$
along the optimistic direction ($\theta=0$) is written as 
\beqn
h_{\rm gw} \approx 10^{-22} \biggl( {\sqrt{R_{+}^2+R_{\times}^2}
\over 0.31~{\rm km}}\biggr)
\biggl({100~{\rm Mpc} \over r}\biggr). \label{hamp}
\eeqn

\subsection{Definitions of quantities and methods for calibration}

In numerical simulations, we refer to the total baryon rest-mass, 
the ADM mass, and the angular momentum of the system, which are given by 
\beqn
M_* &&\equiv \int \rho_* d^3x={\rm const}, \\
M &&\equiv -{1 \over 2\pi} 
\oint_{r\rightarrow\infty} \pa_i \psi dS_i \nonumber \\
&&=\int \biggl[ \rho_{\rm H} e^{5\phi} +{e^{5\phi} \over 16\pi}
\biggl(\tilde A_{ij} \tilde A^{ij}-{2 \over 3}K^2 \nonumber \\
&&~~~~~~~~~~~~~~~~~~~~~~~~~~~~~~-\tilde R_k^{~k} 
e^{-4\phi}\biggr)\biggr]d^3x, \label{eqm00}\\
J &&\equiv {1 \over 8\pi}\oint_{r\rightarrow\infty} 
\varphi^ i \tilde A_i^{~j} e^{6\phi} dS_j \nonumber \\
&&=\int e^{6\phi}\biggl[J_i \varphi^i  
+{1 \over 8\pi}\biggl( \tilde A_i^{~j} \pa_j \varphi^i
-{1 \over 2}\tilde A_{ij}\varphi^k\pa_k \tilde \gamma^{ij}
\nonumber \\
&&~~~~~~~~~~~~~~~~~~~~~~~~~~
+{2 \over 3}\varphi^j \pa_j K \biggr) \biggr]d^3x,
\label{eqj00}
\eeqn
where $dS_j=r^2 \pa_j r d(\cos\theta)d\varphi$, $\varphi^j=-y(\pa_x)^j
+ x(\pa_y)^j$, $\rho_{\rm H}=\rho \alpha u^t \hat e$,
$J_i=\rho \hat u_i$, and $\tilde R_k^{~k}$ denotes the Ricci scalar
with respect to $\tilde \gamma_{ij}$. To derive the expressions for
$M$ and $J$ in the form of volume integral, the Gauss law is used.

The notations $M_{*1}$ and $M_{*2}$ are used to denote the baryon
rest-mass of the primary and secondary neutron stars, respectively. In
terms of them, the baryon rest-mass ratio is defined by
$Q_M \equiv M_{*2}/M_{*1} (\leq 1)$.

In numerical simulation, $M$ and $J$ are computed using the volume
integral shown in Eqs. (\ref{eqm00}) and (\ref{eqj00}).  Since the
computational domain is finite, they are not constant and decrease
after gravitational waves propagate away from the computational
domain.  Therefore, for $t > 0$, $M$ and $J$ are not equal to the ADM
mass and the total angular momentum defined at spatial infinity, but
quasilocal quantities. However, in this paper, we refer to them simply
as the ADM mass and the total angular momentum.

The decrease rates of $M$ and $J$ should be equal to the emission 
rates of the energy and the angular momentum by gravitational
radiation according to the conservation law.  Denoting the radiated
energy and angular momentum from the beginning of the simulation to
the time $t$ as $\Delta E(t)$ and $\Delta J(t)$, the conservation
relations are 
\beqn
&&M(t) + \Delta E(t) =M_0,\label{eqm01}\\
&&J(t) + \Delta J(t) =J_0,\label{eqj01}
\eeqn
where $M_0$ and $J_0$ are the initial values of $M$ and $J$.  We check 
that these conservation laws approximately hold during the simulation. 

During merger of binary neutron stars (from the last one orbit to a
relaxed state formed after the merger), the angular momentum is
dissipated by 15--$30\%$ (cf. Sec. \ref{sec:gw}). Obviously, the
dissipation effect plays a crucial role in determining the final
outcome. Therefore, checking that Eq. (\ref{eqj01}) holds in a
simulation is one of the most important procedures to confirm that the
numerical results are reliable.

In addition to checking the conservation of the mass and the angular
momentum, we monitor the violation of the Hamiltonian constraint in
the same manner as in \cite{shiba2d,STU,STU2} in the absence of black
hole. In its presence, the violation is defined for a region outside
the excision surface. As demonstrated in \cite{STU,STU2}, the typical
magnitude of the violation is of order 1\% throughout the simulation
in the absence of black hole. In its presence, the violation is
amplified. However, the typical magnitude is still $\sim 10\%$ in
the early phase of the excision run. The magnitude gradually increases 
until the crash of computation. 

\subsection{Equations of state}

\begin{table}[t]
\vspace{0mm}
\begin{center}
\begin{tabular}{lllllll} \hline
& ~$i$~ & ~$p_i$ (SLy)~ & ~$p_i$ (APR)~ &
~~~$i$~ & ~$p_i$ (SLy)~ & ~$p_i$ (APR)~ \\ \hline
&1 & 0.1037 & 0.0889 & ~~9 & $9\times 10^5$ & $9 \times 10^5$ \\ 
&2 & 0.1956 & 0.1821 & ~~10 & 4 & 4 \\ 
&3 & 39264  & $5.945 \times 10^5$  & ~~11 & 0.75 & 0.75 \\ 
&4 & 1.9503 & 2.4265  & ~~12 & 0.057 & 0.057 \\ 
&5 & 254.83 & 1600.0  & ~~13 & 0.138 & 0.138 \\ 
&6 & 1.3823 & 2.165   & ~~14 & 0.84 & 0.84 \\ 
&7 & $-1.234$ & $-6.96$  & ~~15 & 0.338 & 0.338 \\ 
&8 & $1.2 \times 10^5$ & $1.2 \times 10^5$     & & & \\ \hline
\end{tabular}
\caption{The values of $p_{i}$ in units of $c=G=M_{\odot}=1$. 
}
\end{center}
\end{table}

\begin{figure}[thb]
\begin{center}
\includegraphics[width=7cm]{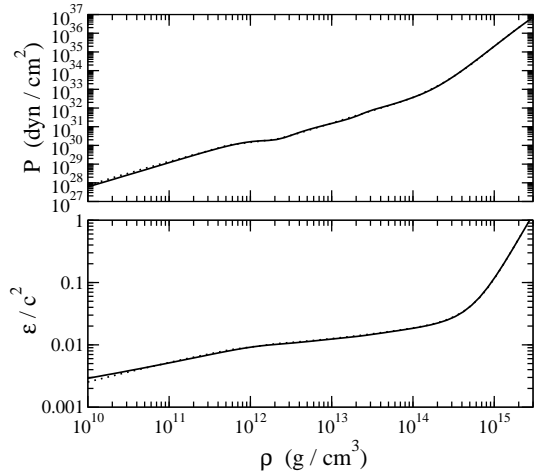}
\end{center}
\vspace{-4mm}
\caption{Pressure and specific internal energy as a function of baryon
rest-mass density $\rho$ for the APR EOS.  The solid and 
dotted curves denote the results by fitting formulae and numerical
data tabulated, respectively. 
\label{FIG1} }
\end{figure}

Following \cite{STU2}, we adopt a hybrid EOS for modeling neutron
stars' EOS; namely, we write the pressure and the specific internal
energy in the form
\beqn
P=P_{\rm cold} + P_{\rm th},\label{eosp}\\
\varep=\varep_{\rm cold} + \varep_{\rm th} \label{eose}. 
\eeqn
Here $P_{\rm cold}$ and $\varep_{\rm cold}$ are the cold
(zero-temperature) parts, and are written as a function of $\rho$. For
them, we assign realistic EOSs for zero-temperature nuclear matter. In
this paper, we adopt the APR \cite{EOS1} and SLy (Skyrme-Lyon) EOSs
\cite{EOS2}. These are tabulated as a function of the baryon rest-mass
density for a wide density range from $\sim 10~{\rm g/cm^3}$ to $\sim
10^{16}~{\rm g/cm^3}$. To simplify numerical implementation for the 
simulation, we make fitting formulae from the EOS tables as in
\cite{HP}.

In our approach, we first make a fitting formula for $\varep_{\rm cold}$
in the form 
\beqn
&&\varep_{\rm cold}(\rho)
=[(1+p_1\rho^{p_2}+p_3\rho^{p_4})(1+p_5 \rho^{p_6})^{p_7}
-1]\nonumber \\
&&~~~~~~~~~~~~~ \times f(-p_8\rho+p_{10})\nonumber \\
&&~~~~~~~~~~~~~ +p_{12} \rho^{p_{13}}f(p_8\rho-p_{10})f(-p_9\rho+p_{11})
\nonumber \\
&&~~~~~~~~~~~~~  +p_{14}\rho^{p_{15}}f(p_9\rho-p_{11}),
\eeqn
where $f(x)=1/(e^x +1)$, and the constant coefficients $p_i~(i=$1--15)
are listed in Table I. In making the formulae, we focus only on the
density for $\rho \geq 10^{10}~{\rm g/cm^3}$, since the matter of lower
density does not play an important role in the merger. Then, the
pressure is computed from the thermodynamic relation in the
zero-temperature limit \beqn P_{\rm cold} =\rho^2 {d \varep_{\rm cold}
\over d\rho}. \eeqn With this approach, the accuracy of the fitting
for the pressure is not as good as that in \cite{HP}, but the
first law of the thermodynamics is completely satisfied in contrast to
the work of \cite{HP}. 

In Fig. 1, we compare $P_{\rm cold}$ and $\varep_{\rm cold}$
calculated by the fitting formulae (solid curves) with the numerical
data tabulated (dotted curves) for the APR EOS \cite{footnote2}.
The same figures for the SLy EOS are shown in
\cite{STU2}. It is found that the fitting formulae agree approximately
with the tabulated data sets.  The relative error between two results is
within $\sim 10\%$ for $\rho > 10^{10}~{\rm g/cm^3}$ and less than 2\% 
for supranuclear density with $\rho \agt 2 \times 10^{14}~{\rm g/cm^3}$.

\begin{figure*}[tb]
\vspace{-2mm}
\begin{center}
\epsfxsize=2.8in
\leavevmode
\epsffile{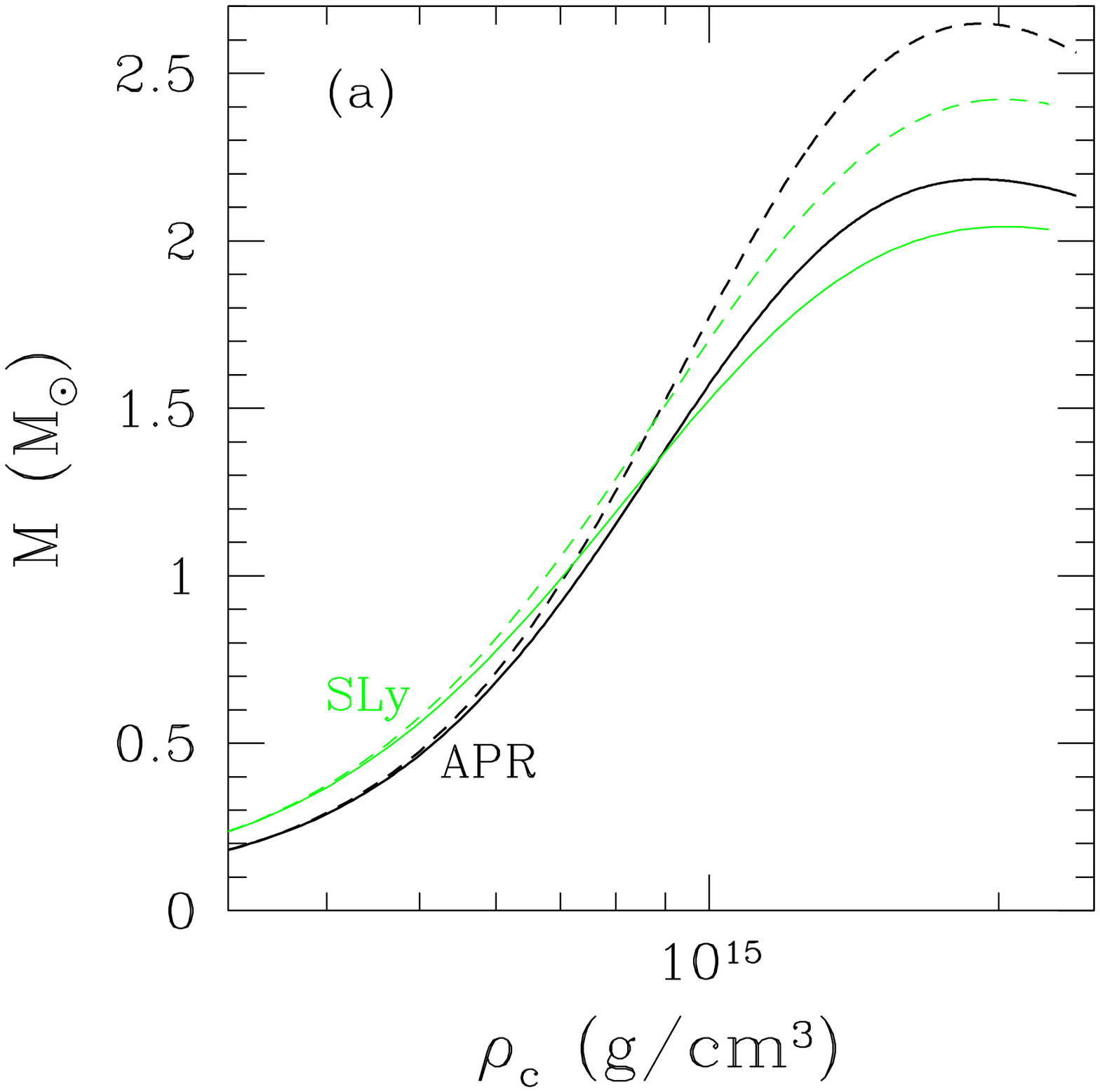} 
\epsfxsize=2.8in
\leavevmode
\hspace{1cm}\epsffile{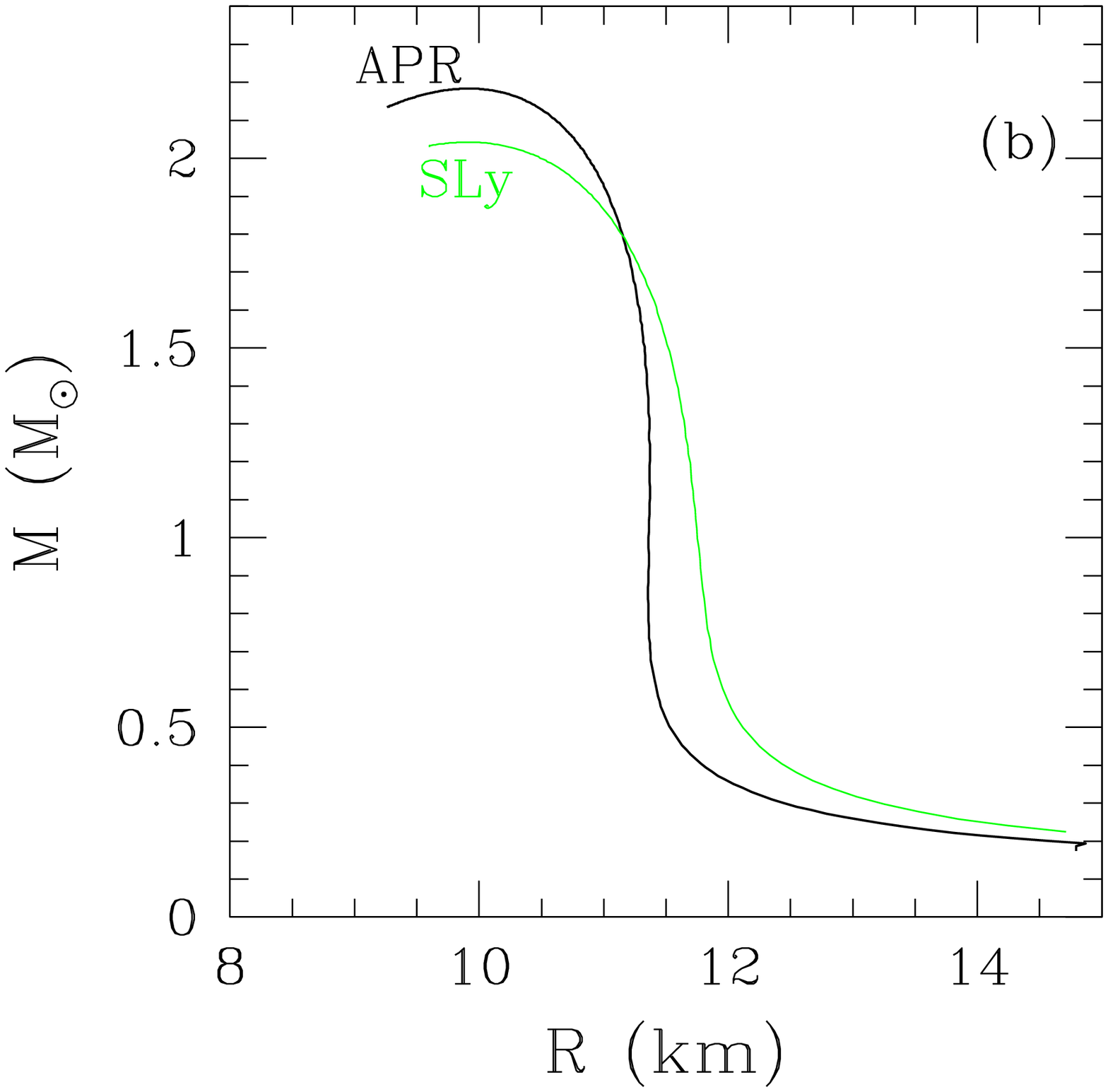}
\end{center}
\vspace{-10mm}
\caption{(a) ADM mass (solid curves) and total baryon rest-mass
(dashed curves) as a function of central baryon rest-mass density
$\rho_c$ and (b) relation between the circumferential radius and the
ADM mass for cold and spherical neutron stars in equilibrium. ``APR''
and ``SLy'' denote the sequences for the APR and SLy EOSs, respectively.
\label{FIG2} }
\end{figure*}

In Fig. \ref{FIG2}, we show the relations among the ADM mass $M$, the
total baryon rest-mass $M_*$, the central density $\rho_c$, and the
circumferential radius $R$ for cold and spherical neutron stars in the
APR and SLy EOSs. The maximum ADM mass (baryon rest-mass) for the APR
and SLy EOSs is about 2.18 and $2.04 M_{\odot}$, respectively. It is
worthy to note that for the APR EOS, the radius is in a narrow range
(11.2--11.4 km) for the ADM mass from $\sim 0.5M_{\odot}$ to $\sim
1.7M_{\odot}$, while for the SLy EOS, it decreases with increasing the
ADM mass. This results in the difference on disk formation in the
merger of unequal-mass neutron stars. 

\begin{table*}[tb]
\vspace{0mm}
\begin{center}
\begin{tabular}{ccccccccccc} \hline
\hspace{2mm} Model \hspace{2mm} & ~~$M_{\infty} (M_{\odot})$~~ &
~~$\rho_{\rm max}~(10^{14}{\rm g/cm^3})$~~ & ~~~$Q_M$~~~
& ~$M_* (M_{\odot})$~ &
~$M_0 (M_{\odot})$~ &  ~~~~~$q_0$~~~~~  & 
~$P_{0}$ (ms)~&
~~~$C_0$~~~ &  
~~~$Q_{*}$~~~  \\ \hline
APR1313 &1.30, 1.30 & 8.62, 8.62 & 1.00
& 2.858 & 2.568 & 0.918 & 2.064 & 0.114 & 1.075   \\ 
APR1214 &1.20, 1.40 & 8.28, 9.10 & 0.842 
& 2.861 & 2.569 & 0.920 & 2.158 & 0.111 & 1.076   \\ 
APR135135 &1.35, 1.35 & 8.85, 8.85 & 1.00
& 2.981 & 2.665 & 0.906 & 1.992 & 0.120 & 1.125   \\ 
APR1414 &1.40, 1.40 & 9.09, 9.09 & 1.00
& 3.106 & 2.762 & 0.896 & 1.923 & 0.125 & 1.173   \\ 
APR1515 &1.50, 1.50 & 9.59, 9.56 & 1.00
& 3.360 & 2.957 & 0.879 & 1.838 & 0.135 & 1.269   \\ 
APR145155 &1.45, 1.55 & 9.34, 9.86 & 0.927
& 3.360 & 2.959 & 0.886 & 1.969 & 0.129 & 1.269   \\ 
APR1416 &1.40, 1.60 & 9.09, 10.14 & 0.862
& 3.363 & 2.960 & 0.892 & 1.969 & 0.129 & 1.270   \\ 
APR135165 &1.35, 1.65 & 8.85, 10.43 & 0.800
& 3.366 & 2.960 & 0.888 & 1.968 & 0.129 & 1.271   \\ 
APR1317 &1.30, 1.70 & 8.62, 10.74 & 0.743
& 3.370 & 2.960 & 0.883 & 2.057 & 0.126 & 1.272   \\ 
APR125175 &1.25, 1.75 & 8.40, 11.09 & 0.690
& 3.377 & 2.962 & 0.885 & 2.145 & 0.122 & 1.275   \\
APR1218 &1.20, 1.80 & 8.17, 11.44 & 0.639
& 3.378 & 2.957 & 0.861 & 2.189 & 0.120 & 1.275   \\ \hline
SLy1313 &1.30, 1.30 & 8.57, 8.57 & 1.00
& 2.847 & 2.568 & 0.922 & 2.110 & 0.112 & 1.175   \\ 
SLy1414 &1.40, 1.40 & 9.16, 9.16 & 1.00
& 3.093 & 2.763 & 0.902 & 2.012 & 0.122 & 1.277   \\ 
SLy135145 &1.35, 1.45 & 8.85, 9.48 & 0.923
& 3.094 & 2.763 & 0.901 & 2.013 & 0.122 & 1.277   \\ 
SLy1315 &1.30, 1.50 & 8.56, 9.80 & 0.851
& 3.096 & 2.764 & 0.904 & 2.104 & 0.118 & 1.278   \\ 
SLy125155 &1.25, 1.55 &8.42, 10.15 & 0.786
& 3.099 & 2.765 & 0.904 & 2.150 & 0.117 & 1.278   \\ 
SLy1216 &1.20, 1.60 &8.02, 10.54 & 0.726 
& 3.103 & 2.765 & 0.903 & 2.242 & 0.113 & 1.279   \\
\hline
\end{tabular}
\caption{ List of several quantities for initial data (binary neutron
stars in quasicircular orbits).  The ADM mass of each star when they
are in isolation $M_{\infty}$, the maximum density for each star, the
baryon rest-mass ratio $Q_M \equiv M_{*2}/M_{*1}$, the total baryon
rest-mass, the total ADM mass $M_{0}$, non-dimensional spin parameter
$q_0=J_0/M_{0}^2$, orbital period $P_{0}$, the orbital compactness
[$C_0\equiv (M_{0}\Omega)^{2/3}$], and the ratio of the total baryon
rest-mass to the maximum allowed mass for a spherical and cold neutron
star ($Q_{*}\equiv M_*/M_{*~\rm max}^{\rm sph}$). }
\end{center}
\end{table*}

$P_{\rm th}$ and $\varep_{\rm th}$ in Eqs. (\ref{eosp})
and (\ref{eose}) are the thermal (finite-temperature) parts. During the
simulation, $\rho$ and $\varep$ are computed from hydrodynamic variables
$\rho_*$ and $\hat e$. Thus, $\varep_{\rm th}$ is determined by
$\varep-\varep_{\rm cold}$. The thermal part of the pressure $P_{\rm th}$
is related to the specific thermal energy 
$\varepsilon_{\rm th}\equiv \varepsilon-\varepsilon_{\rm cold}$ as
\beq
P_{\rm th}=(\Gamma_{\rm th}-1)\rho \varepsilon_{\rm th},
\eeq
where $\Gamma_{\rm th}$ is an adiabatic constant for which we set
$\Gamma_{\rm th}=2$ taking into account the fact that the EOSs for
high-density nuclear matter are stiff. Since $\Gamma_{\rm th} \approx 5/3$
for the ideal non-relativistic Fermi gas 
\cite{Chandra}, it is reasonable to consider that it is much larger
than 5/3 for the nuclear matter.  We note that in \cite{STU2},
we also chose $\Gamma_{\rm th}=1.3$ and 1.65 and found that the
numerical results depend only weakly on its value.

\begin{table*}[tb]
\vspace{0mm}
\begin{center}
\begin{tabular}{cccccccc} \hline
~~~~~ Model ~~~~~ & 
~~Grid size~~ & ~~$L$ (km)~~ & ~~$\lambda_{0}$ (km)~~
& ~$f_{\rm merger}$ (kHz)~
& ~$\lambda_{\rm merger}$ (km)~& ~Product~
& Disk mass $(M_{\odot})$ \\ \hline
APR1313    & (665, 665, 333)  & 139 & 309 & 3.20 & 94 &NS& ---\\ 
APR1313b   & (377, 377, 189)  & 78.7& 309 & 3.18 & 94 &NS& ---\\ 
APR1214    & (377, 377, 189)  & 74.5& 323 & 3.23 & 93 &NS& ---\\ 
APR135135  & (377, 377, 189)  & 77.8& 299 & 3.35 & 89 &NS& ---\\
APR1414    & (665, 665, 333)  & 135 & 288 & 3.79 & 79 &NS& ---\\ 
APR1515    & (633, 633, 317)  & 125 & 276 & 6.5 & 46 &BH&$4\times 10^{-4}$\\ 
APR1515b   & (377, 377, 189)  & 74.5& 276 & --- & ---&BH&$2\times 10^{-4}$\\ 
APR145155  & (633, 633, 317)  & 125 & 295 & 6.5 & 46 &BH&$5\times 10^{-4}$\\ 
APR1416    & (633, 633, 317)  & 125 & 295 & 6.5 & 46 &BH&$1.0\times 10^{-3}$\\ 
APR1416b   & (377, 377, 189)  & 74.5& 295 & --- & ---&BH&$7  \times 10^{-4}$\\ 
APR135165  & (633, 633, 317)  & 125 & 295 & 6.5 & 46 &BH&$2.7\times 10^{-3}$\\ 
APR135165b & (377, 377, 189)  & 74.5& 295 & --- & ---&BH&$1.9\times 10^{-3}$\\ 
APR1317    & (633, 633, 317)  & 125 & 308 & 6.5 & 46 &BH&$6.9\times 10^{-3}$\\
APR1317b   & (377, 377, 189)  & 74.5& 308 & --- & ---&BH&$5.0\times 10^{-3}$\\
APR125175  & (377, 377, 189)  & 74.5& 276 & --- & ---&BH&$1.2\times 10^{-2}$\\
APR1218    & (377, 377, 189)  & 74.5& 276 & --- & ---&BH&$1.5\times 10^{-2}$\\
\hline
SLy1313    & (633, 633, 317)  & 131  & 316& 3.20& 94 &NS& ---\\ 
SLy1414    & (633, 633, 317)  & 131  & 302& 6.7 & 45 &BH&$4\times 10^{-3}$\\ 
SLy1414b   & (377, 377, 189)  & 77.8 & 302& --- & ---&BH&$3\times 10^{-3}$\\ 
SLy135145  & (377, 377, 189)  & 77.8 & 302& --- & ---&BH&$5\times 10^{-3}$\\ 
SLy1315    & (377, 377, 189)  & 77.8 & 315& --- & ---&BH&$1.4\times 10^{-2}$\\ 
SLy125155  & (633, 633, 317)  & 131  & 322& 6.7 & 45 &BH&$4.9\times 10^{-2}$\\ 
SLy125155b & (377, 377, 189)  & 77.8 & 322& --- & ---&BH&$3.6\times 10^{-2}$\\ 
SLy1216    & (377, 377, 189)  & 77.8 & 336& --- & ---&BH&$5.7\times 10^{-2}$\\
\hline
\end{tabular}
\caption{List of setting for simulation and summary of the outcome.
$L$, $\lambda_0$, and $f_{\rm merger}$ denote the location of outer
boundaries along each axis, the wave length of gravitational waves at
$t=0$, and the frequency of gravitational waves from the formed
hypermassive neutron stars or of ring-down gravitational waves of a
quasinormal mode of black holes, respectively. $\lambda_{\rm merger}$
denotes the wave length of gravitational waves $\lambda_{\rm
merger}=c/f_{\rm merger}$, which is shown to be $\alt L$ for the large
grid sizes.  In the last two columns, the product at $t=10$ ms and
disk mass for the black hole formation case are shown. ``NS'' implies
that a hypermassive neutron star is the outcome at $t=10$ ms, and
``BH'' implies that a black hole is promptly formed.  The disk mass is
evaluated at $t - t_{\rm AH}=0.5$ ms where $t_{\rm AH}$ is the time at
apparent horizon formation. Note that we do not compute gravitational
waves for the small grid size (377, 377, 189) in the case of prompt
black hole formation. }
\end{center}
\end{table*}

\section{Initial condition and setting for simulation}\label{sec:init}

In Table II, we summarize several quantities that characterize
irrotational binary neutron stars in quasiequilibrium circular
orbits used as initial conditions for the present simulations.
Since the lifetime of binary neutron stars from the birth to the merger
is longer than $\sim 10^8$ yrs for the observed systems \cite{Stairs},
the temperature of each neutron star will be very low ($\alt 10^5$ K)
\cite{Tsuruta} at the onset of merger; i.e., the thermal energy
per nucleon is much smaller than the Fermi energy of neutrons. 
Hence, cold nuclear EOSs are employed in giving the initial condition. 

We choose binaries of an orbital separation which is slightly larger
than that for an innermost orbit.  Here, the innermost orbit is
defined as a close orbit for which Lagrange points appear at the inner
edge of at least one of two neutron stars \cite{USE,GBM}. If the
orbital separation becomes smaller than that of the innermost orbit,
mass transfer sets in and a dumbbell-like structure is formed.  Until
the innermost orbit is reached, the circular orbit is stable, and
hence, the innermost stable circular orbit (ISCO) does not exist
outside the innermost orbit for the present cases. However, the ISCO
seems to be close to the innermost orbit since the decrease rates of
the energy and the angular momentum as a function of the orbital
separation along the quasiequilibrium sequences are close to zero near
the innermost orbit.

The ADM mass of each neutron star, when it is in isolation (i.e., when
the orbital separation is infinity), is denoted by $M_{\infty}$, and
chosen in the range between $1.2M_{\odot}$ and $1.8M_{\odot}$. The
labels APR and SLy denote the binary models constructed in the APR and
SLy EOSs, respectively. We select the models for a wide range of rest-mass
ratio, $0.6 \alt Q_M \leq 1$, to find the dependence of the disk mass
around a black hole on $Q_M$ for the high-mass case with $M
\sim 2.76$--$2.96M_{\odot}$. 

The simulations are performed using a fixed uniform grid in the reflection
symmetry with respect to the equatorial plane (the orbital plane). In
this paper, the typical grid size is (633, 633, 317) or (665, 665,
333) for $(x, y, z)$.  The grid covers the region $-L \leq x \leq L$,
$-L \leq y \leq L$, and $0 \leq z \leq L$ where $L$ is a constant. The
grid spacing $\Delta x$ is $\approx 0.4$ km with which the major
diameter of each star is covered with about 45 grid points
initially. We have shown that with this grid spacing, a convergent
numerical result is obtained in the merger simulations \cite{STU}.

Accuracy in the computation of gravitational waveforms (in particular
amplitude) and the radiation reaction depends on the location of the
outer boundaries if the wavelength, $\lambda$, is larger than $L$
\cite{STU}.  For $L \alt 0.4 \lambda$, the amplitude and the radiation
reaction of gravitational waves are significantly overestimated
\cite{STU,SU01}. Due to the restriction of the computational power, it
is difficult to take a huge grid size in which $L$ is much larger than
$\lambda$. As a consequence of the present restricted computational
resources, $L$ has to be chosen as $\sim 0.4$--$0.45 \lambda_0$ where
$\lambda_0$ denotes the value of $\lambda$ at $t=0$. Hence, the error
associated with the small value of $L$ is inevitable; the amplitude
and radiation reaction of gravitational waves are overestimated in the
early phase of the simulation. The overestimation of the radiation
reaction leads to slight spurious shortening of the late inspiraling
phase.

However, the typical wavelength of gravitational waves just before the
merger quickly becomes shorter due to a short radiation-reaction time
scale, and hence, the accuracy of the wave extraction is improved with
the evolution of the system. In particular, the wavelength of
quasiperiodic gravitational waves emitted from the formed hypermassive
neutron star and of ring-down gravitational waves associated with a
quasinormal mode of the formed black hole (denoted by
$\lambda_{\rm merger}$) is much shorter than $\lambda_0$ and satisfies
the condition $\lambda_{\rm merger} < L$ for the grid size
(633, 633, 317) (see Table III). Therefore,
the waveforms in the merger stage are computed accurately (within
$\sim 10\%$ error) as confirmed in \cite{STU}.

For some models, the simulations are performed for a smaller grid size
(377, 377, 189) but with the same grid spacing $\Delta x$. In this
case, gravitational waveforms cannot be accurately computed for the
inspiraling phase because of the small value of $L/\lambda_0$.
However, if we pay attention only to the outcome after the merger,
such small value for $L$ may be allowed since the properties of the
outcome depend weakly on $L$.  This fact is indeed confirmed in the
simulations for several models (cf. Table III; about the reason for
the systematic underestimation of the disk mass, see
\ref{sec:BHdisk}). Thus, we perform simulations for models APR135135,
APR1214, APR125175, APR1218, SLy135145, SLy1315, and SLy1216 only with
the small grid size. In this case, the focus is only on the outcome,
not on computing gravitational waveforms.

With the (633, 633, 317) grid size, about 240 GBytes computational
memory is required. For the case of the hypermassive neutron star
formation, the simulations are performed for about 30,000 time steps
(until $t \sim 10$ ms) and then stopped to save the computational
time.  The computational time for one model in such a simulation is
about 180 CPU hours using 32 processors on FACOM VPP5000 in the data
processing center of National Astronomical Observatory of Japan
(NAOJ). For the case of the black hole formation, the simulations are
continued until mass accretion rate from the surrounding disk to a formed
black hole relaxes approximately to a constant. In this case, the
computational time is about 60 CPU hours for about 12,000 time steps.

\section{Numerical results: Merger process}
\subsection{Summary of the outcome}\label{sec:outcome}

For all the cases, the binary orbit is stable at $t=0$, but after
slight decrease of the orbital separation due to gravitational
radiation reaction, merger sets in. In the present simulations, the
merger starts in one orbit (in $t \sim 2$ ms) irrespective of the
models. If the total ADM mass of the system is high enough, a black
hole is formed within $\sim 1$ ms after two stars come into contact.
On the other hand, for models with the ADM mass smaller than a
threshold mass $M_{\rm thr}$, a hypermassive neutron star is formed
and survives for more than 10 ms. However, it will collapse to a black
hole eventually due to gravitational radiation reaction or angular
momentum transport by other effects (see discussion in
Sec. \ref{sec:angmom_trans}).  In the case of black hole formation
(models APR1515, APR1416, APR135165, APR1317, APR125175, APR1218,
SLy1414, SLy135145, SLy1315, SLy125155, and SLy1216), the evolution of the
black holes is followed using a black hole excision technique \cite{AB} until
the mass accretion rate from the disk to the formed black hole becomes
approximately constant.

\begin{figure}[thb]
\vspace{-4mm}
\begin{center}
\epsfxsize=3.2in
\leavevmode
\epsffile{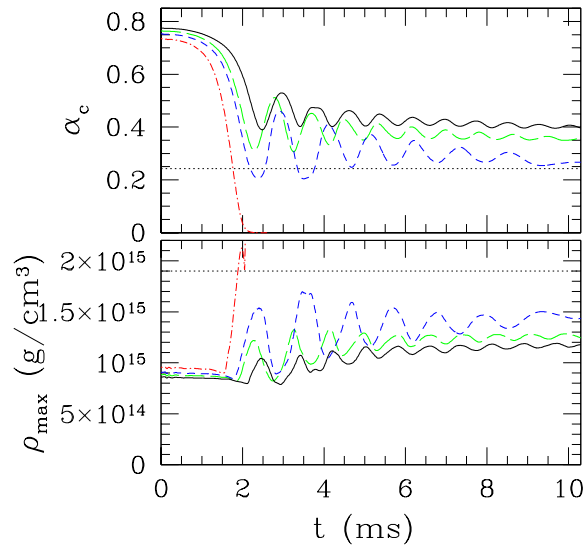}
\end{center}
\vspace{-10mm}
\caption{Evolution of the central value of the lapse function
$\alpha_c$ and the maximum values of the rest-mass density $\rho_{\rm
max}$ for models APR1313 (solid curves), APR135135 (long-dashed
curves), APR1414 (dashed curves), and APR1515 (dot-dashed
curves). The dotted horizontal lines denote the central values of the
lapse and rest-mass density for the marginally-stable, spherical 
neutron star in equilibrium with the cold APR EOS. 
\label{FIG3}}
\end{figure}

In Fig. 3, we show the evolution of the central value of the lapse
function $\alpha_c$ and the maximum of the baryon rest-mass density
$\rho_{\rm max}$ for models APR1313, APR135135, APR1414, and
APR1515. For model APR1515, $\alpha_c$ collapses to zero and
$\rho_{\rm max}$ quickly increases at $t \sim 1.5$ ms, implying that a
black hole is formed promptly soon after the onset of the merger (at
$t \approx 2.04$ ms).  For other cases, $\alpha_c$ and $\rho_{\rm
max}$ settle down to relaxed values, implying that a quasistationary
neutron star is formed. Since the mass is larger than the maximum
allowed limit of rigidly rotating neutron stars, these neutron stars
are hypermassive \cite{BSS}.

For model APR1414, $\alpha_c$ and $\rho_{\rm max}$ oscillate with an
amplitude larger than that for models APR1313 and APR135135. This behavior
results from the fact that the self-gravity is large enough for the
merged object to deeply shrink surmounting the centrifugal force. This
indicates that the total ADM mass of this model ($M \approx
2.76M_{\odot}$) is slightly smaller than the threshold value for the
prompt black hole formation.  Since a black hole is formed for $M
\approx 2.96M_{\odot}$, the threshold mass for black hole formation is
$M_{\rm thr}\approx 2.8$--$2.9M_{\odot}$.  In a previous paper
\cite{STU2}, we adopted the SLy and FPS
(Friedman-Pandharipande-Skyrme) EOSs and found that $M_{\rm thr}
\approx 2.7M_{\odot}$ and $2.5M_{\odot}$, respectively. The high value
of $M_{\rm thr}$ for the APR EOS is reasonable since it is stiffer
than the SLy and FPS EOSs, and the maximum allowed mass for spherical,
cold neutron star $M_{\rm sph}$ is also larger.  The present result
combined with the previous one \cite{STU2} suggests an empirical
relation $M_{\rm thr}/M_{\rm sph} = 1.30$--1.35 for these stiff
nuclear EOSs.

\begin{figure*}[thb]
\begin{center}
\epsfxsize=2.2in
\leavevmode
\epsffile{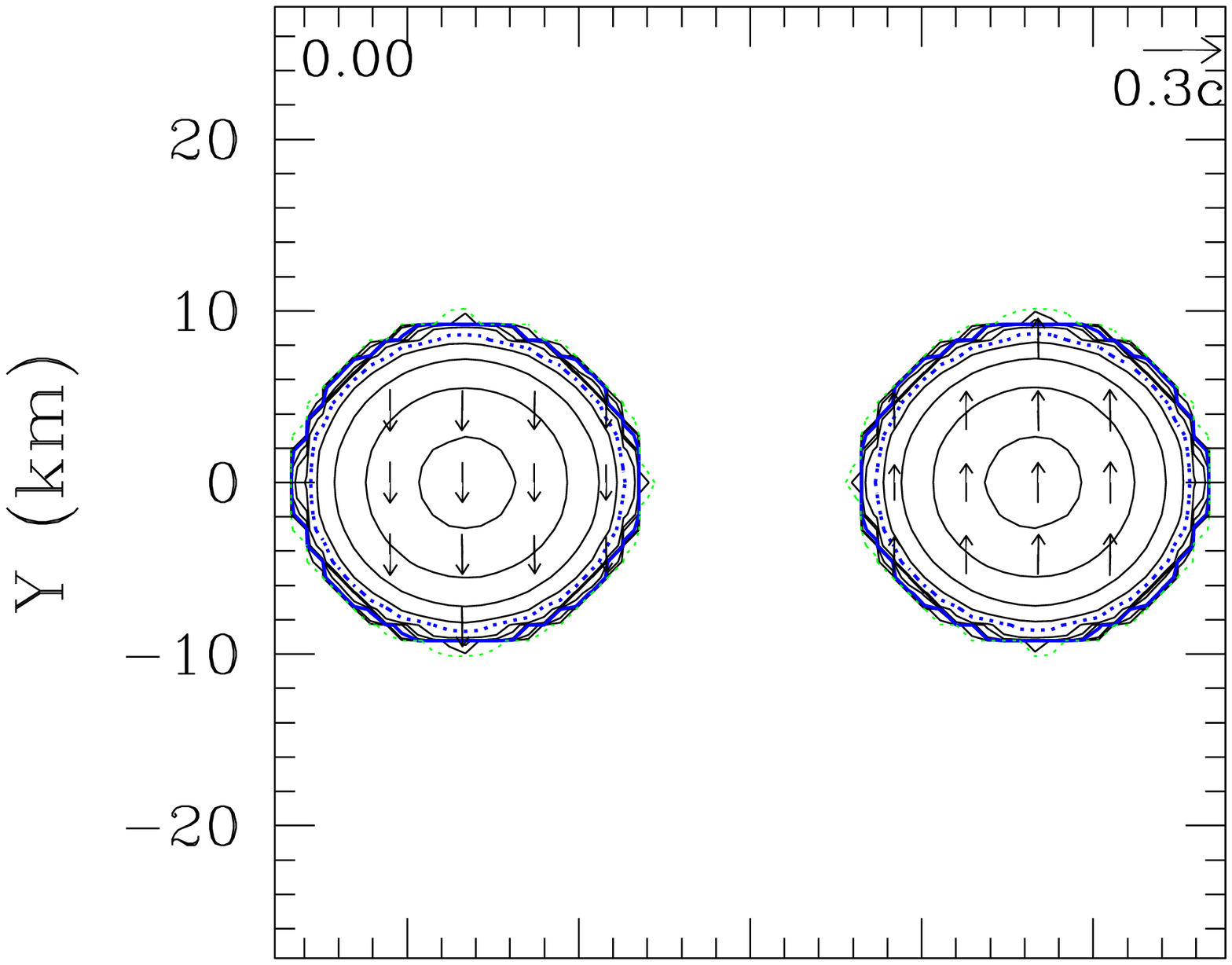}
\epsfxsize=2.2in
\leavevmode
\hspace{-1.73cm}\epsffile{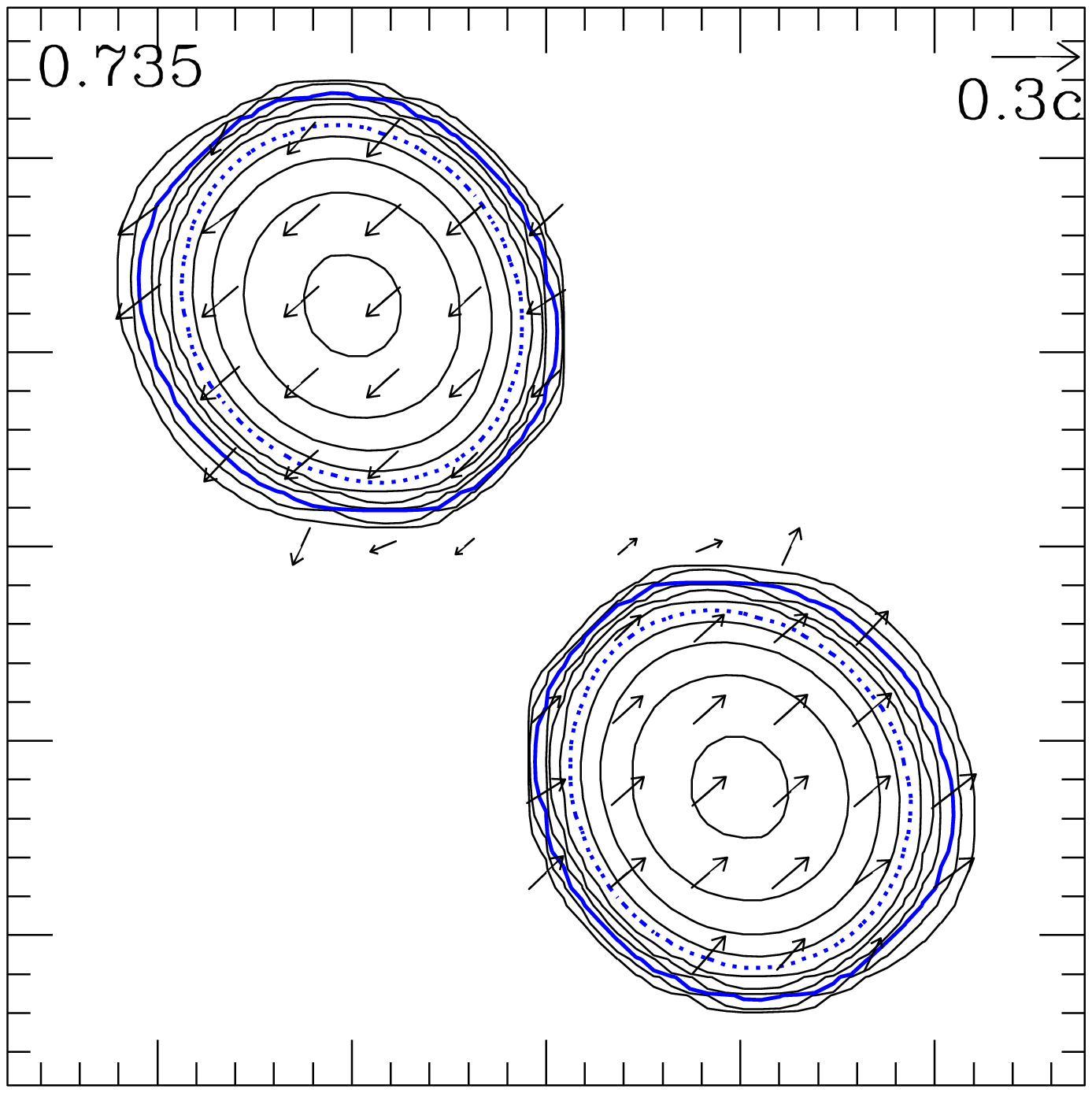} 
\epsfxsize=2.2in
\leavevmode
\hspace{-1.73cm}\epsffile{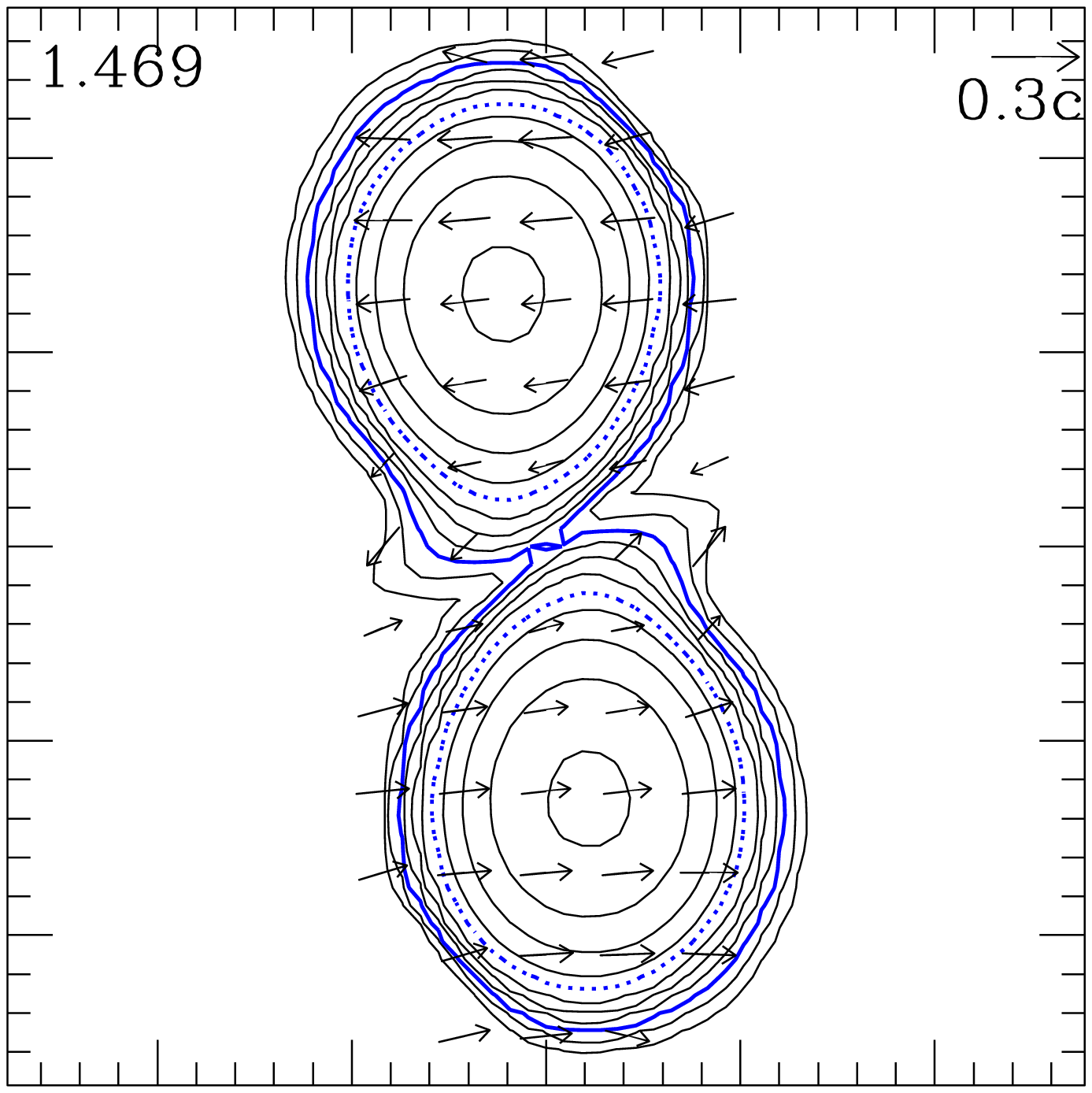} 
\epsfxsize=2.2in
\leavevmode
\hspace{-1.73cm}\epsffile{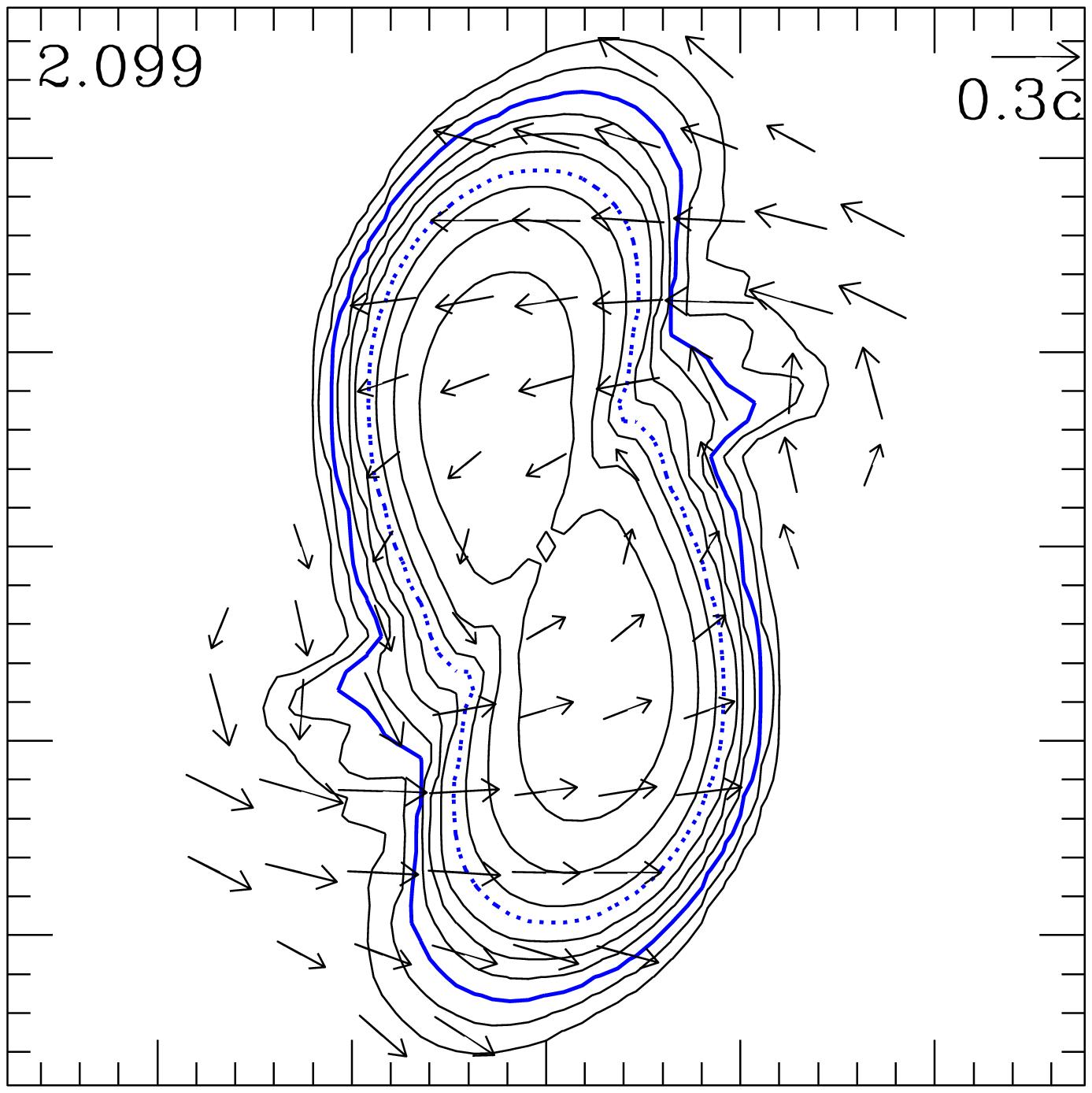} \\
\vspace{-1.76cm}
\epsfxsize=2.2in
\leavevmode
\epsffile{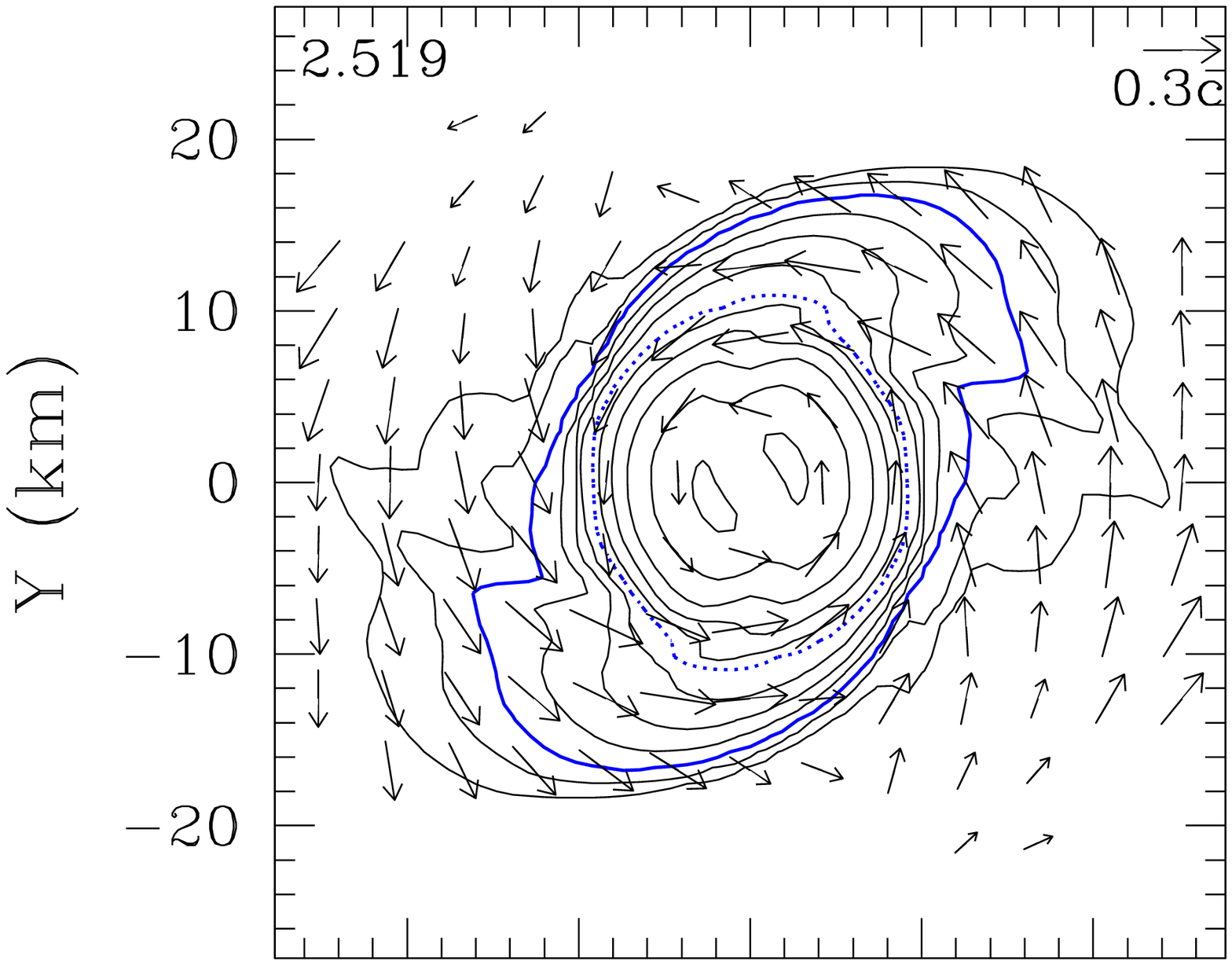} 
\epsfxsize=2.2in
\leavevmode
\hspace{-1.73cm}\epsffile{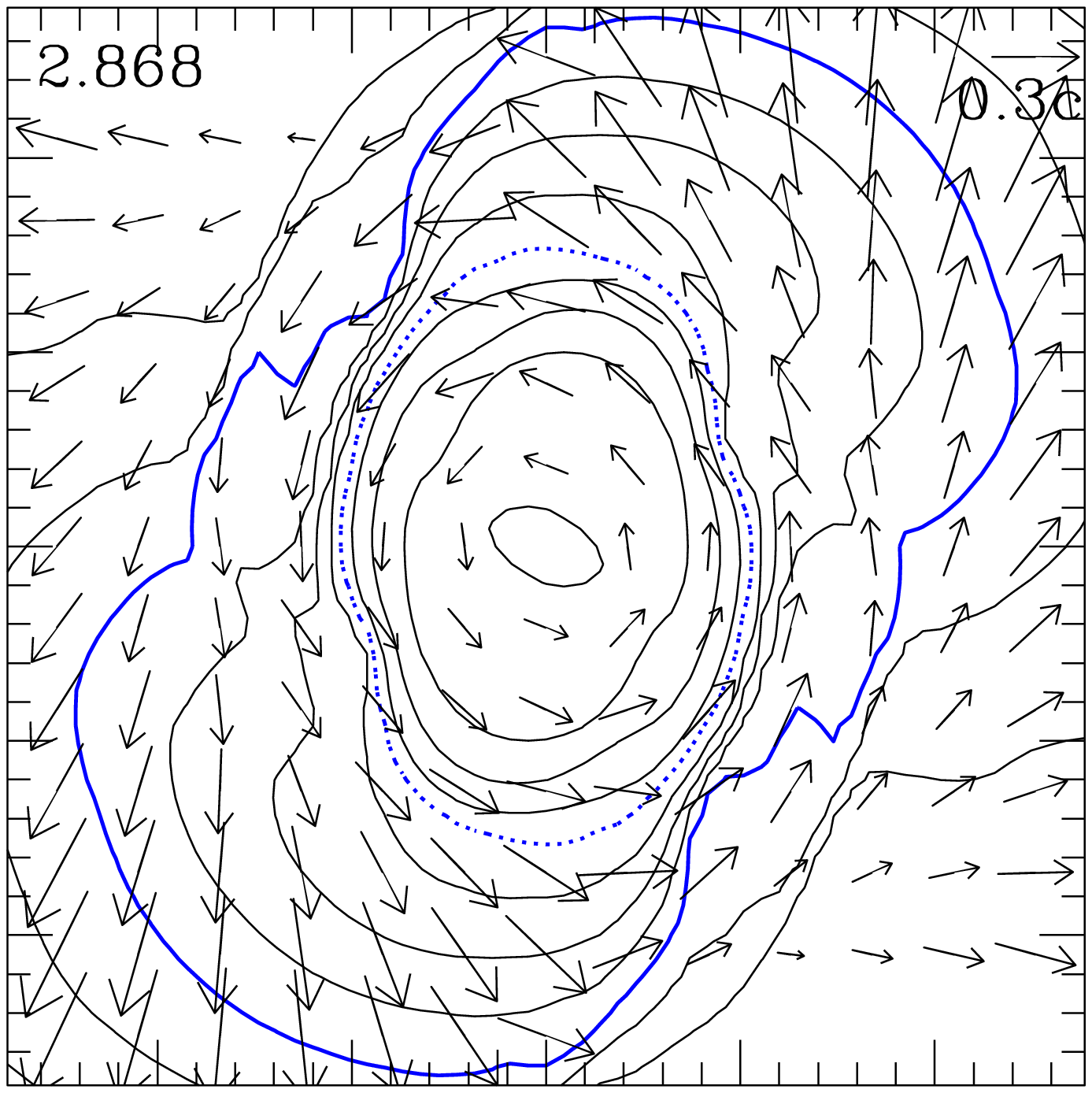}
\epsfxsize=2.2in
\leavevmode
\hspace{-1.73cm}\epsffile{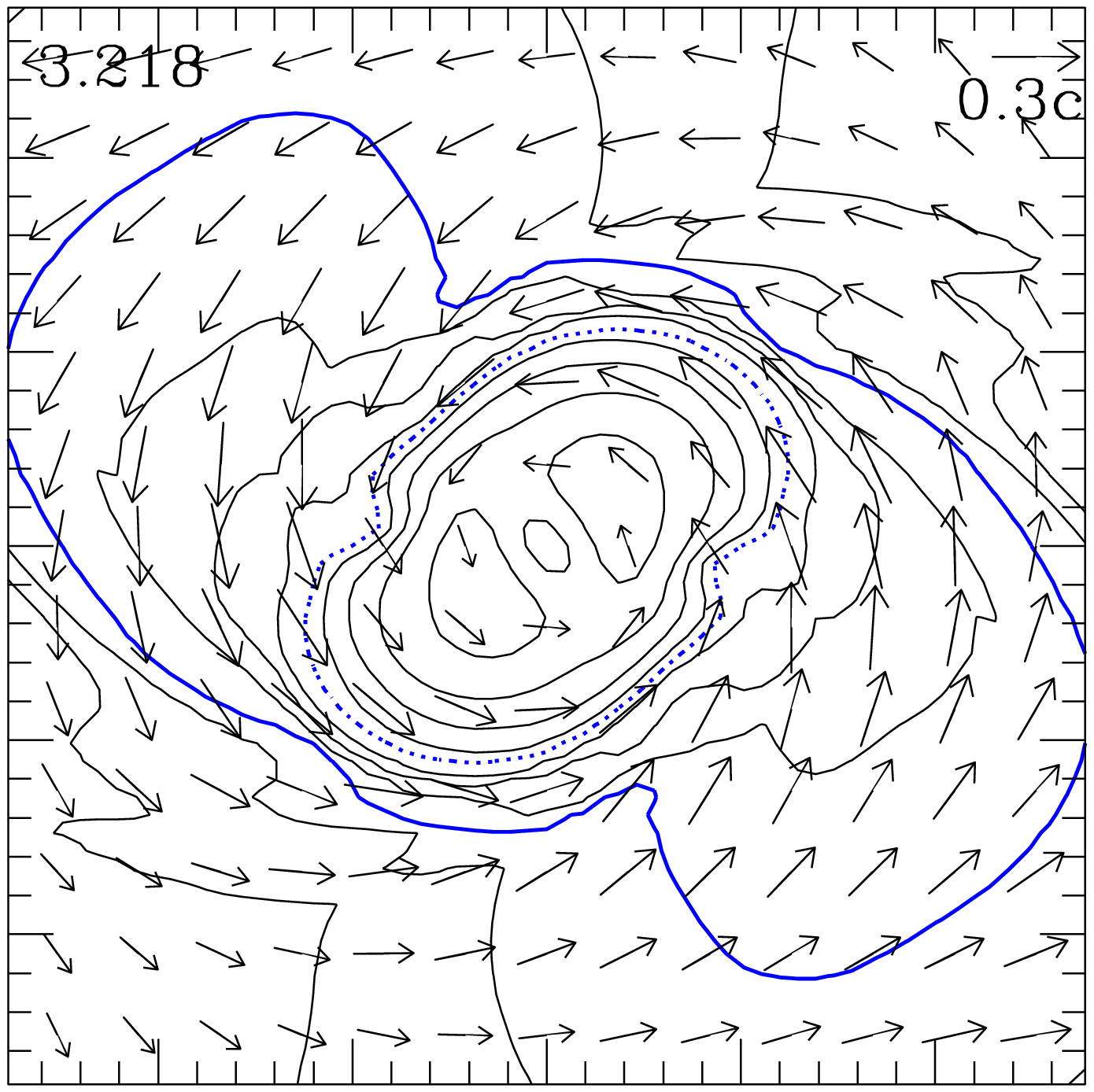}
\epsfxsize=2.2in
\leavevmode
\hspace{-1.73cm}\epsffile{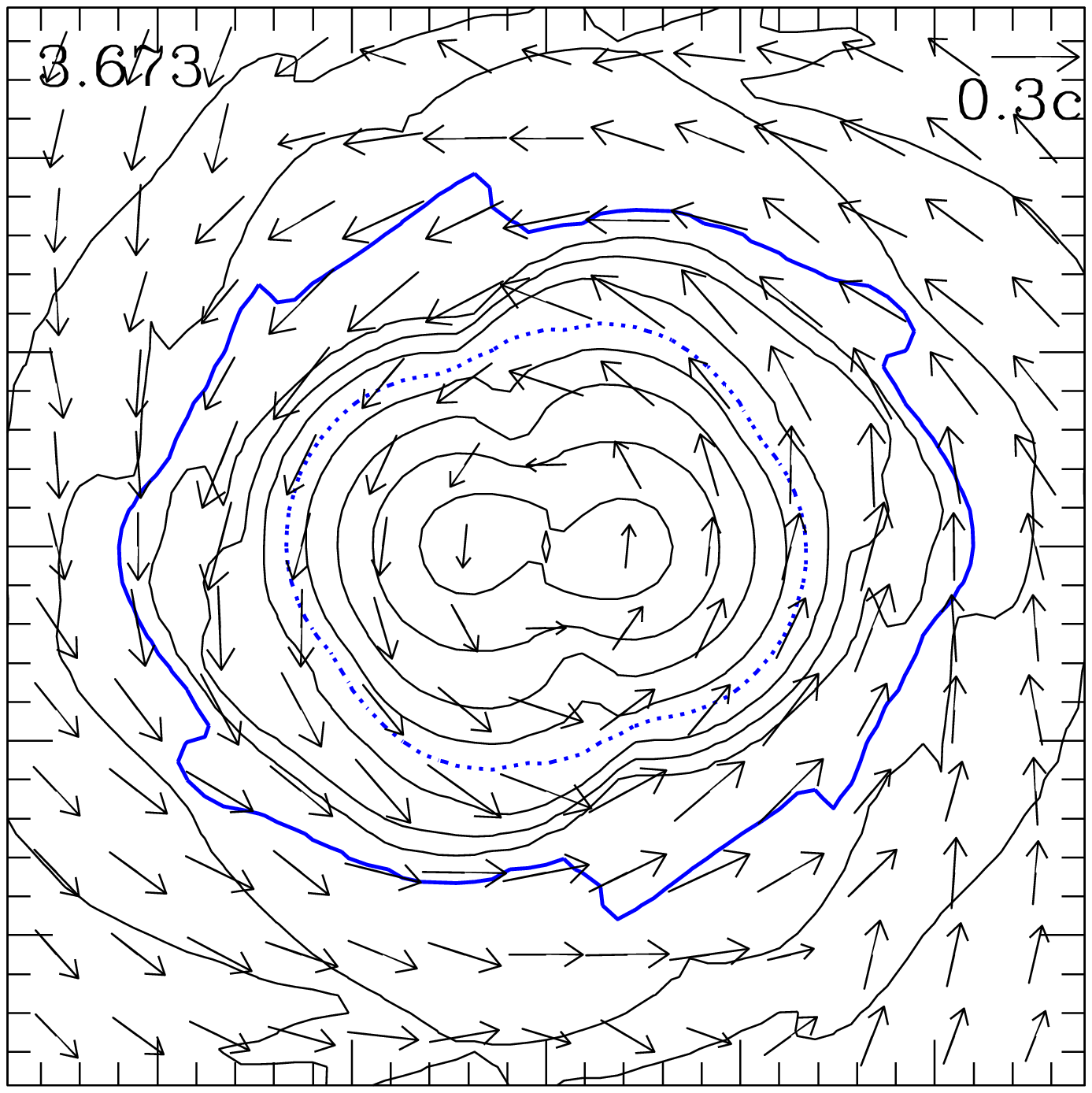} \\
\vspace{-1.76cm}
\epsfxsize=2.2in
\leavevmode
\epsffile{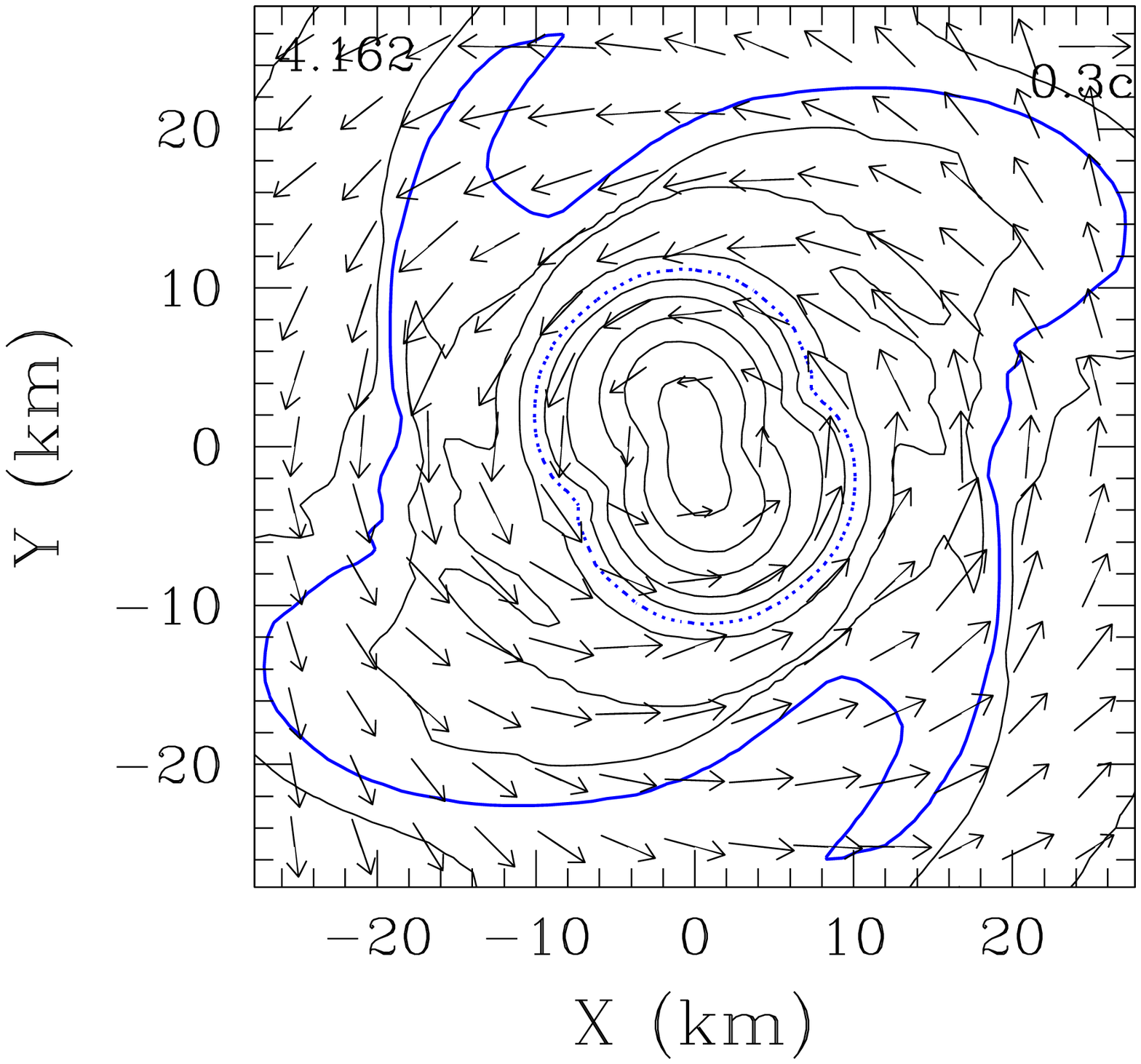} 
\epsfxsize=2.2in
\leavevmode
\hspace{-1.73cm}\epsffile{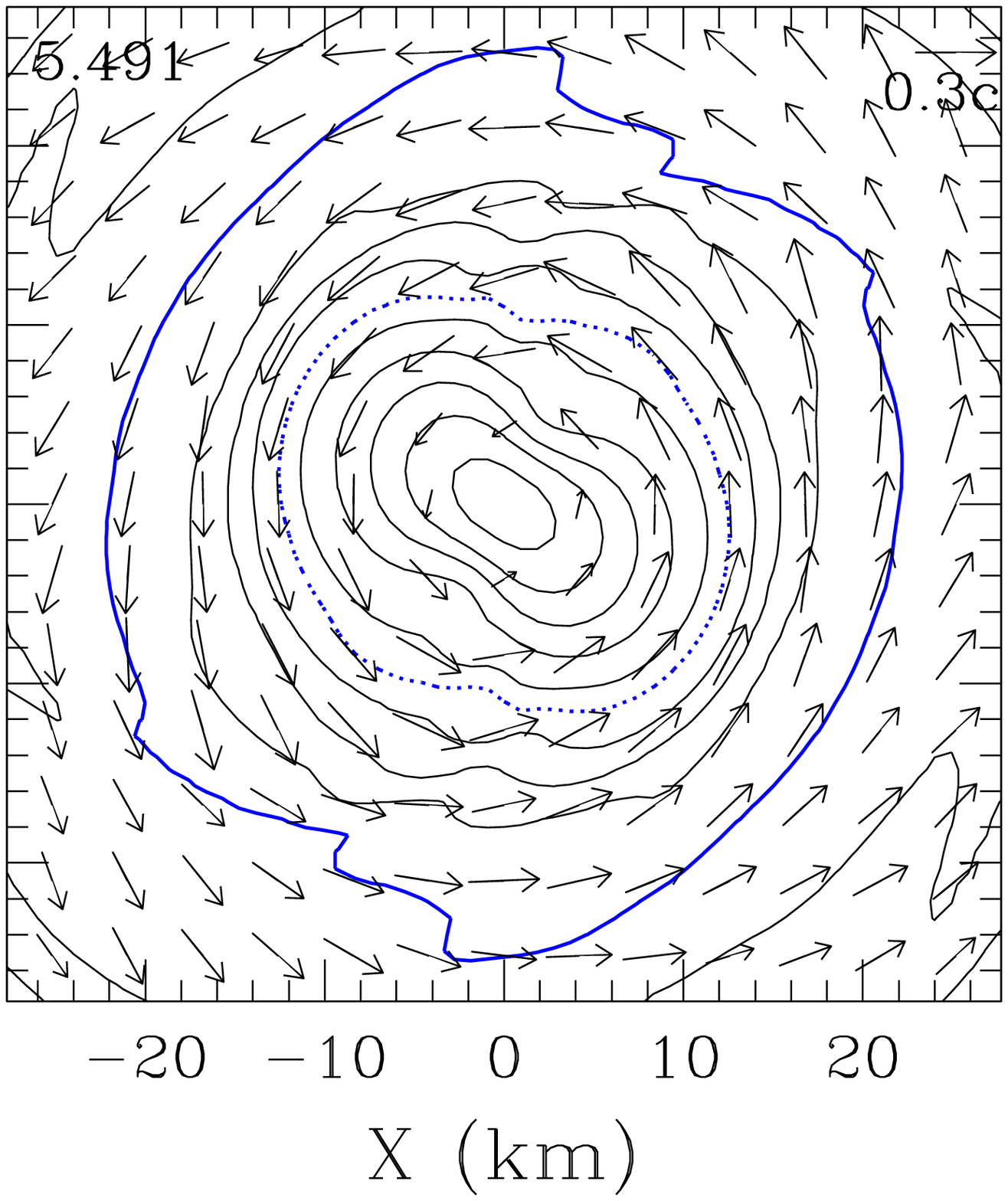}
\epsfxsize=2.2in
\leavevmode
\hspace{-1.73cm}\epsffile{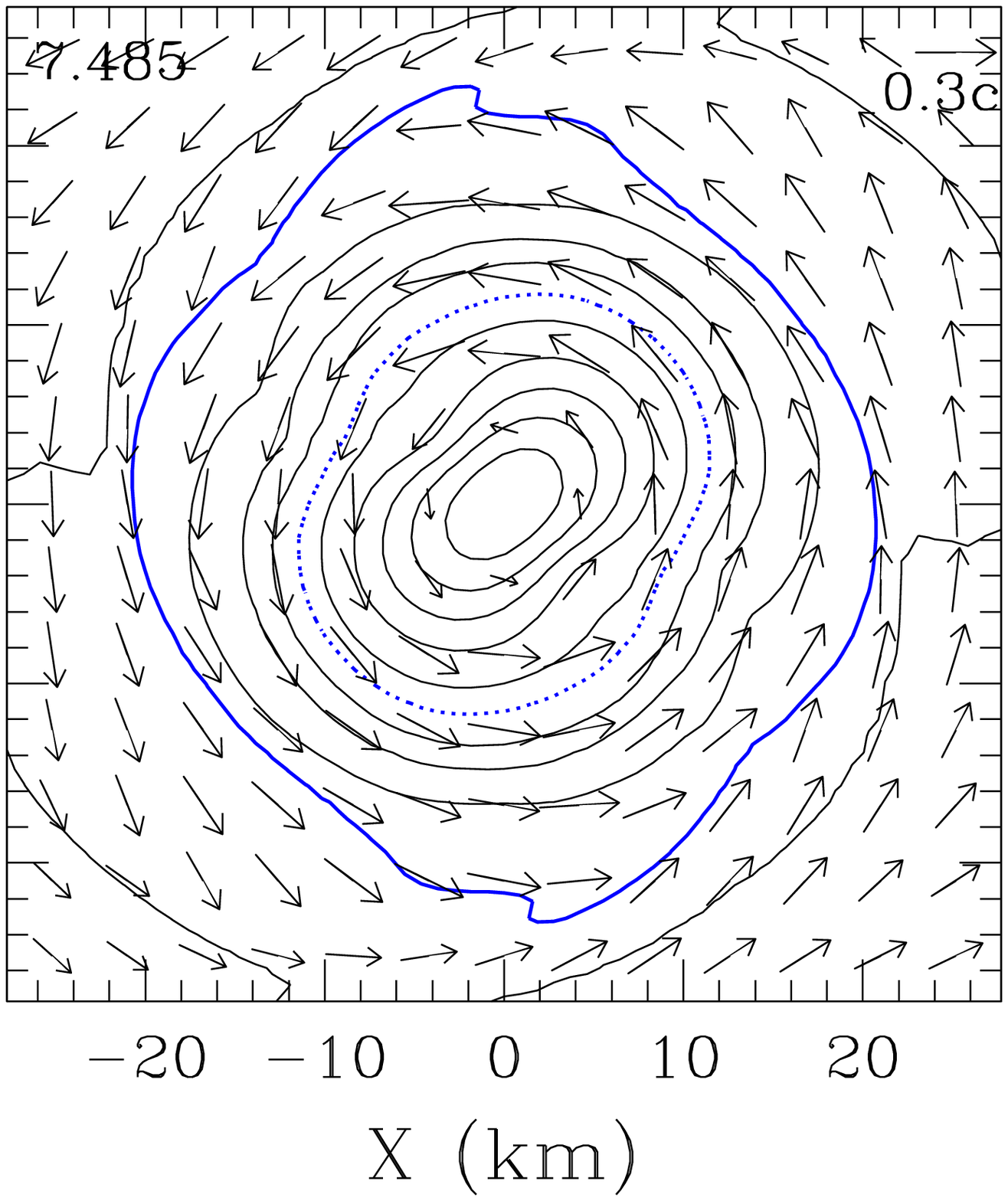}
\epsfxsize=2.2in
\leavevmode
\hspace{-1.73cm}\epsffile{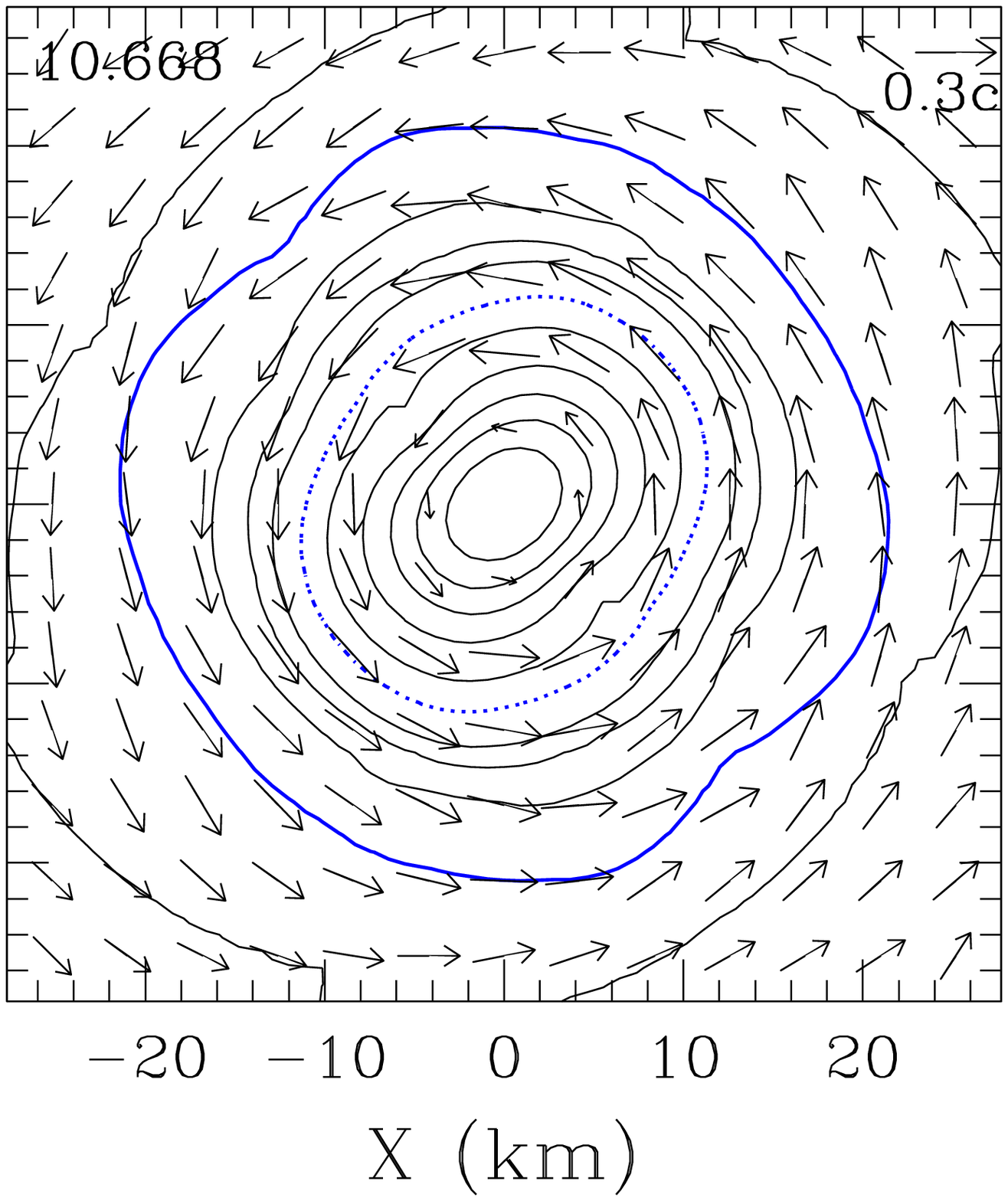} \\
\vspace{-4mm}
\caption{Snapshots of the density contour curves for $\rho$ in
the equatorial plane for model APR1313.  The solid contour curves are
drawn for $\rho= 2\times 10^{14} \times i ~{\rm g/cm^3}~(i=1, 2, 3, \cdots)$
and for $ 1\times 10^{14-0.5 i}~{\rm g/cm^3}~(i=1 \sim
6)$. The (blue) thick dotted and solid curves denote $1 \times 10^{14}~{\rm
g/cm^3}$ and $1 \times 10^{12}~{\rm g/cm^3}$, respectively. The number
in the upper left-hand side denotes the elapsed time from the
beginning of the simulation in units of ms. Vectors indicate the local
velocity field $(v^x,v^y)$, and the scale is shown in the upper
right-hand corner.
\label{FIG4}}
\end{center}
\end{figure*}

We note that this relation is highly different from that for the
$\Gamma$-law EOS with $\Gamma=2$ \cite{STU} for which $M_{\rm
thr}/M_{\rm sph} \approx 1.7$. As discussed in \cite{STU2}, the
compactness of each neutron star in the stiff nuclear EOSs is larger
than that with the $\Gamma=2$ EOS for a given mass. Accordingly, for a
given total mass, the binary system at the onset of the merger is more
compact. This implies that the angular momentum is dissipated more
before the merger sets in with the stiff nuclear EOSs.  The
dissipation of the angular momentum by $\sim 0.1 J_0$ before the merger
helps the prompt black hole formation, since the hypermassive neutron
star formation requires a substantial centrifugal force to sustain the
self-gravity.

In recent papers \cite{LBS}, ratio of the maximum mass of
hypermassive neutron stars in equilibrium to $M_{\rm sph}$ has been
investigated in detail. These works coincidently show a similar
dependence of the ratio $M_{\rm thr}/M_{\rm sph}$ on the EOSs.  Hence,
the small value of $M_{\rm thr}/M_{\rm sph}$ may be partly due to the
absence of high-mass differentially rotating neutron stars in
equilibrium.

\begin{figure*}[thb]
\begin{center}
\epsfxsize=2.2in
\leavevmode
\epsffile{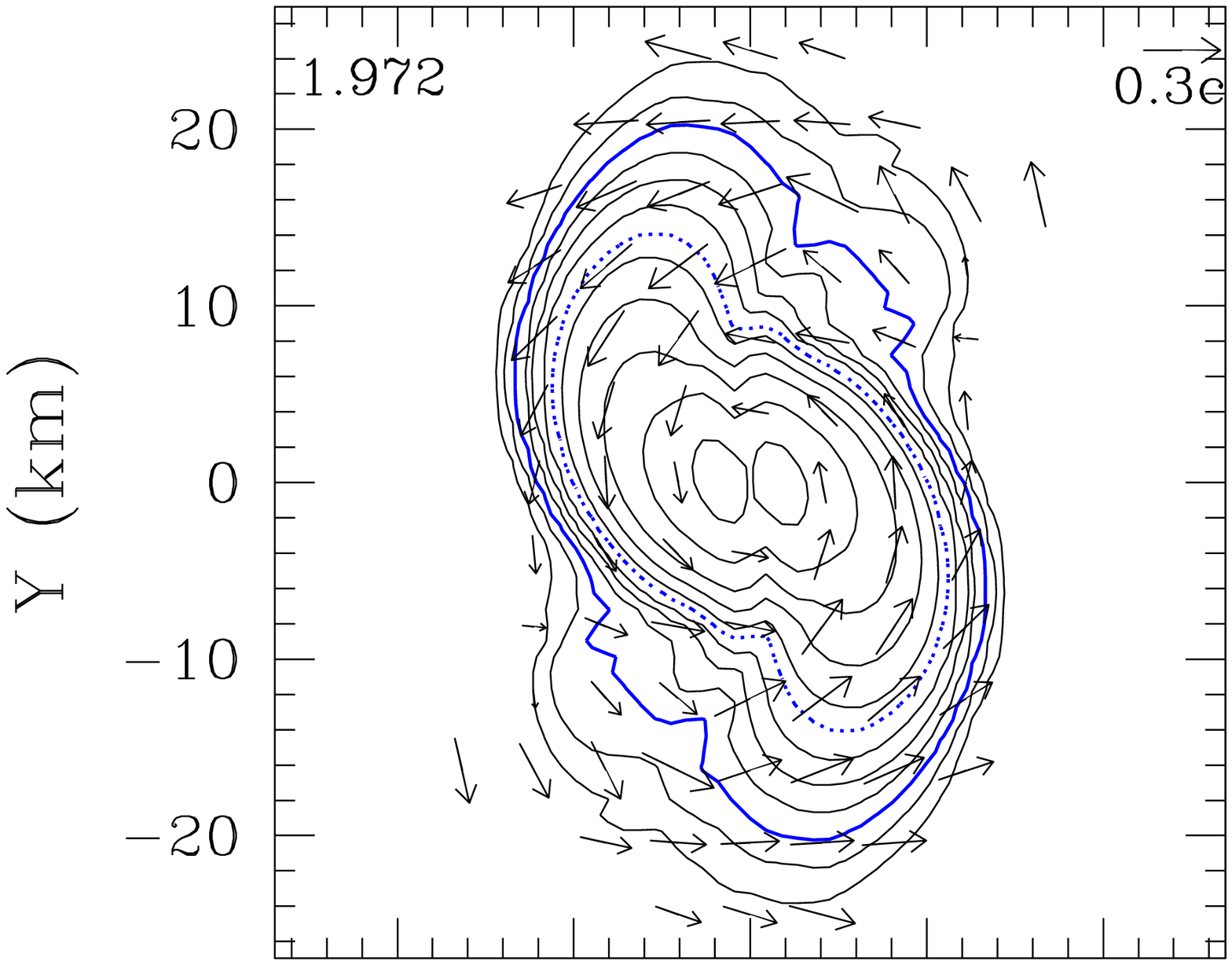}
\epsfxsize=2.2in
\leavevmode
\hspace{-1.73cm}\epsffile{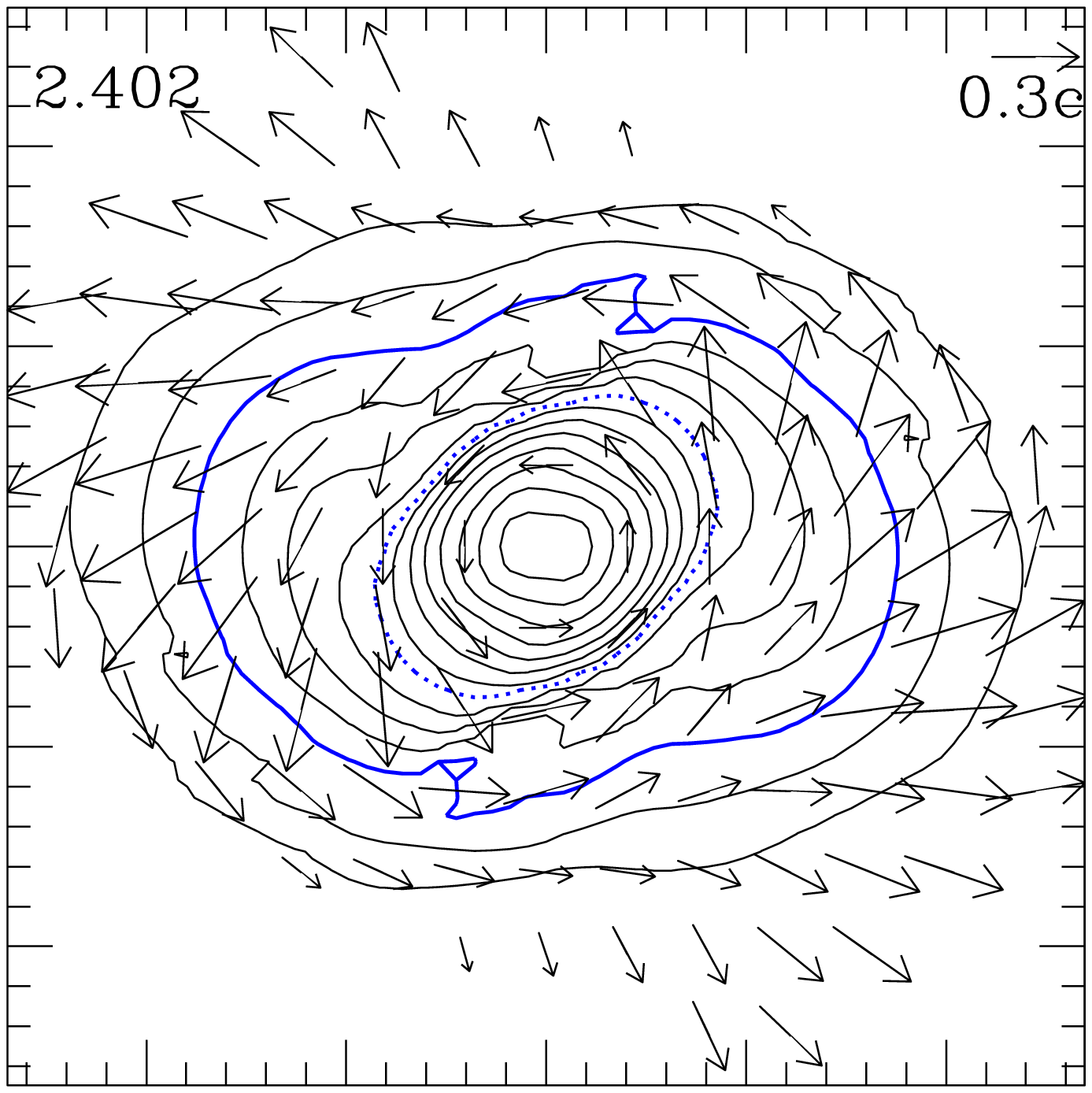} 
\epsfxsize=2.2in
\leavevmode
\hspace{-1.73cm}\epsffile{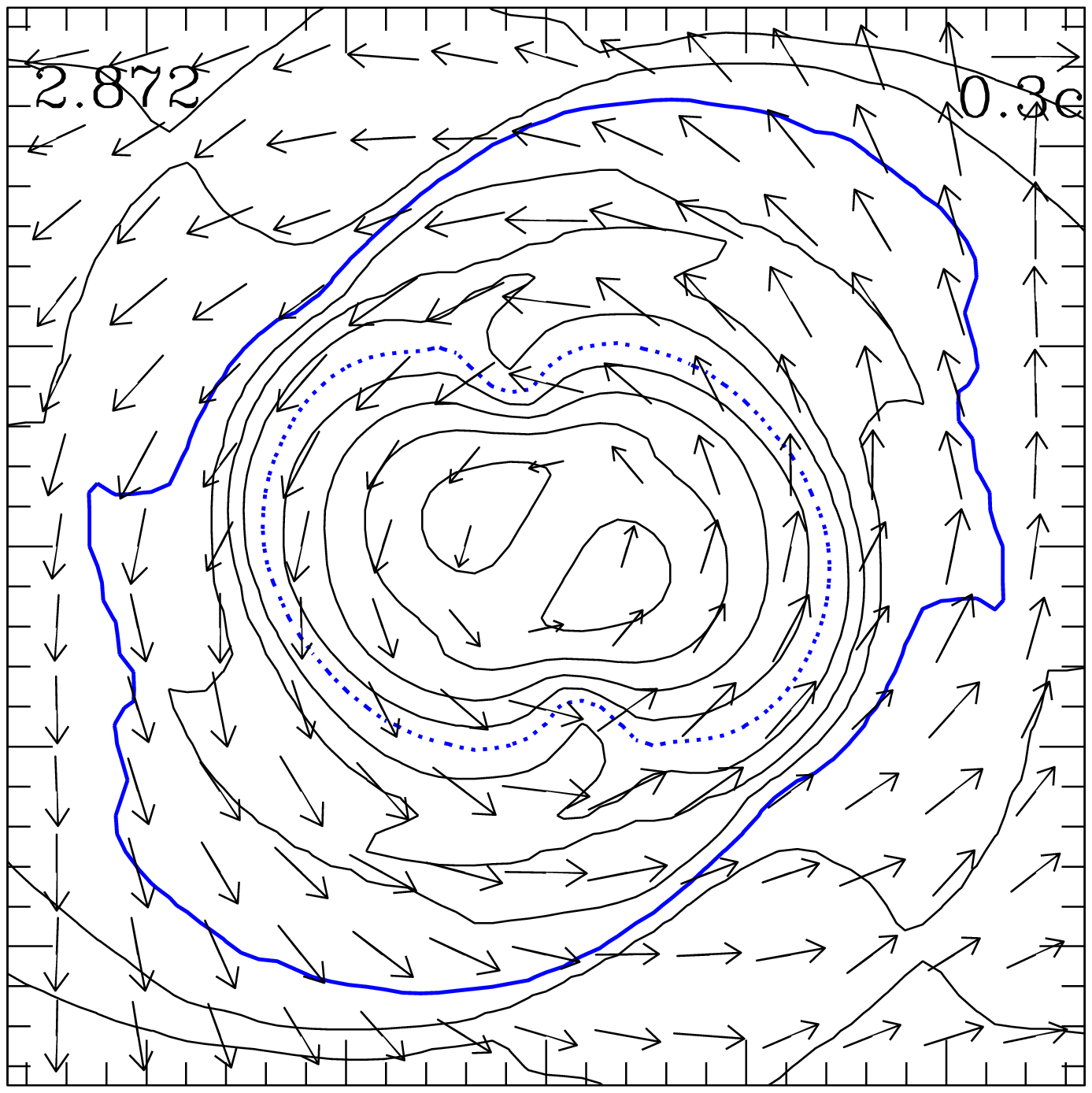}
\epsfxsize=2.2in
\leavevmode
\hspace{-1.73cm}\epsffile{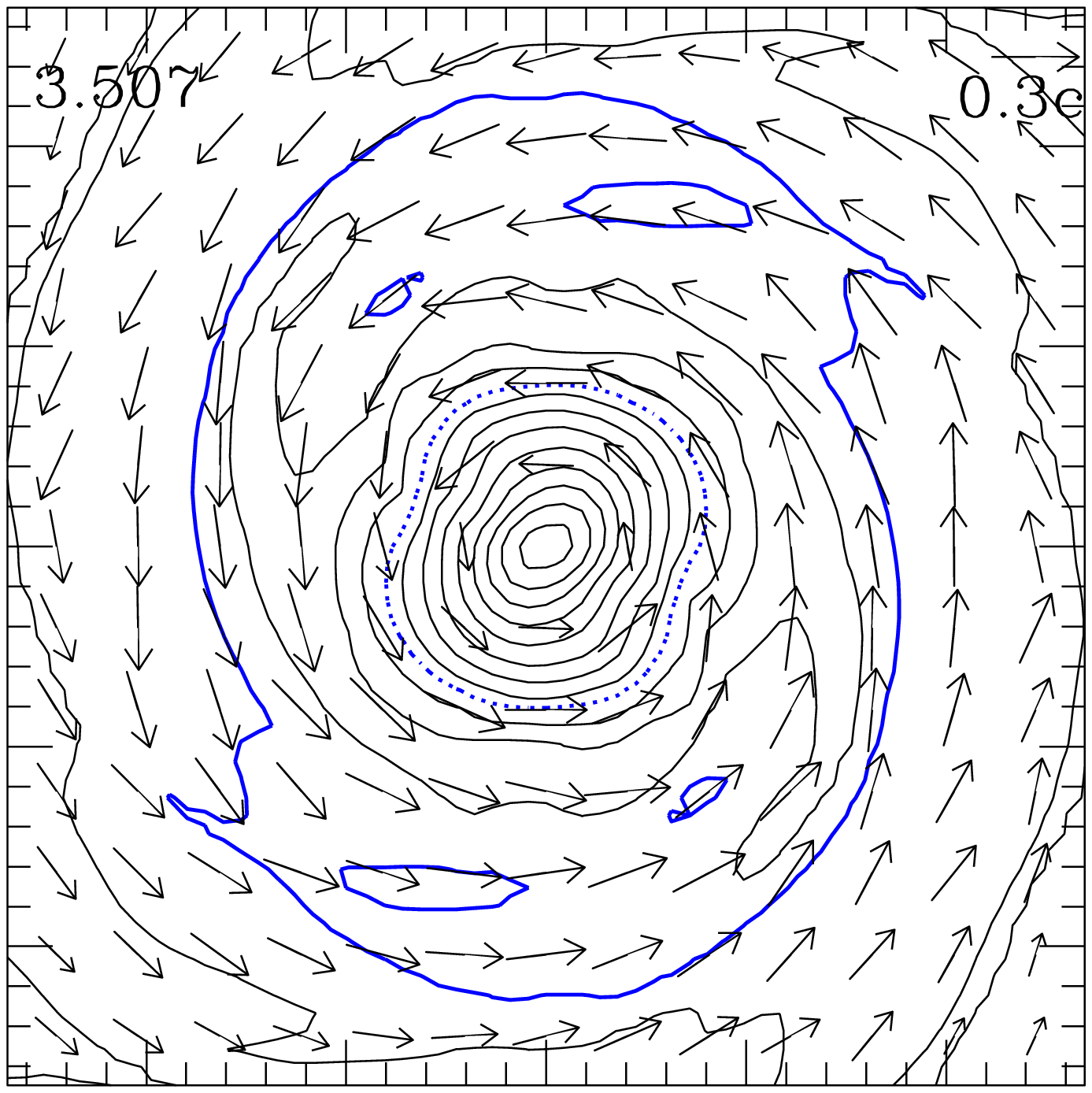} \\
\vspace{-1.76cm}
\epsfxsize=2.2in
\leavevmode
\epsffile{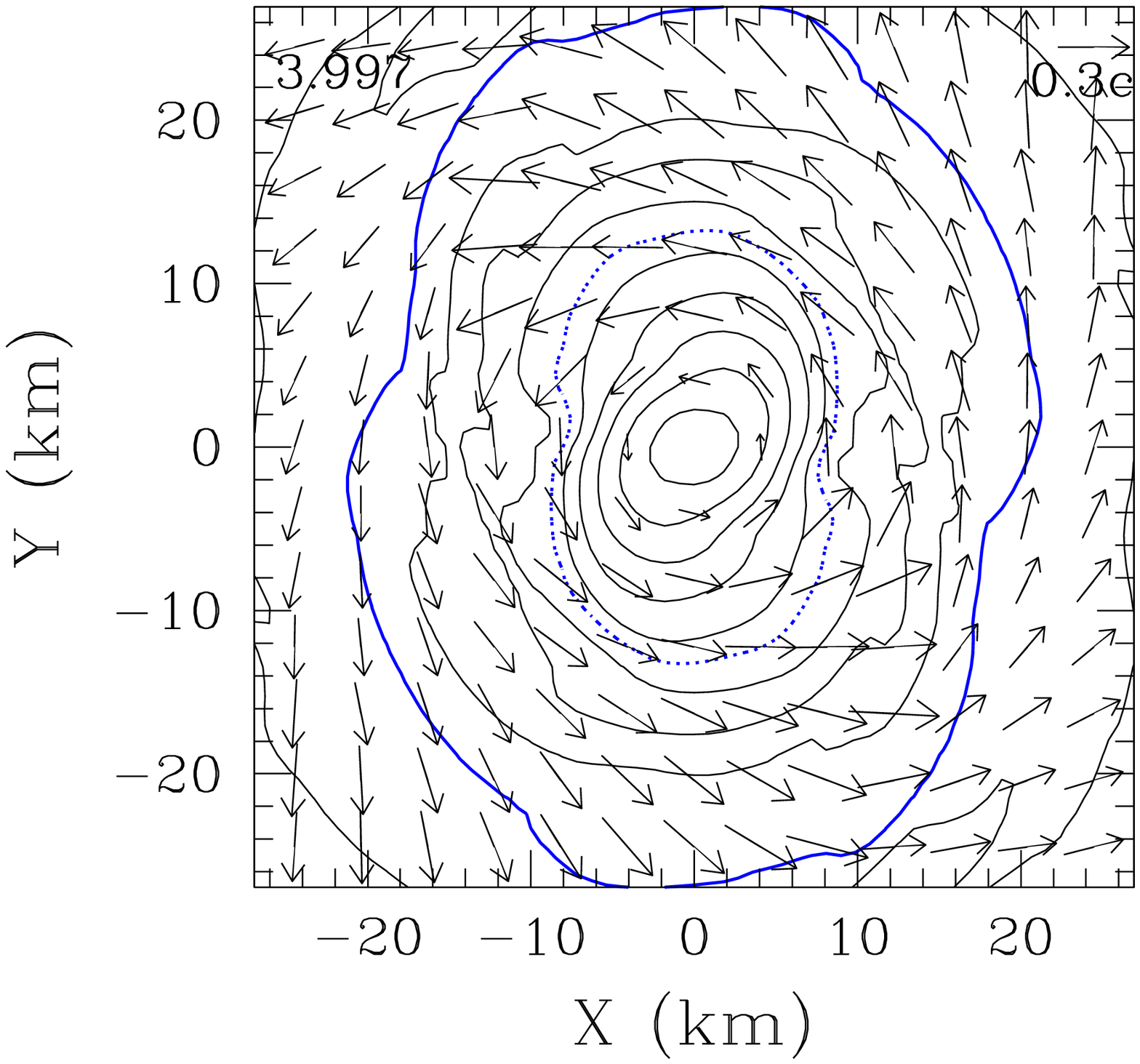}
\epsfxsize=2.2in
\leavevmode
\hspace{-1.73cm}\epsffile{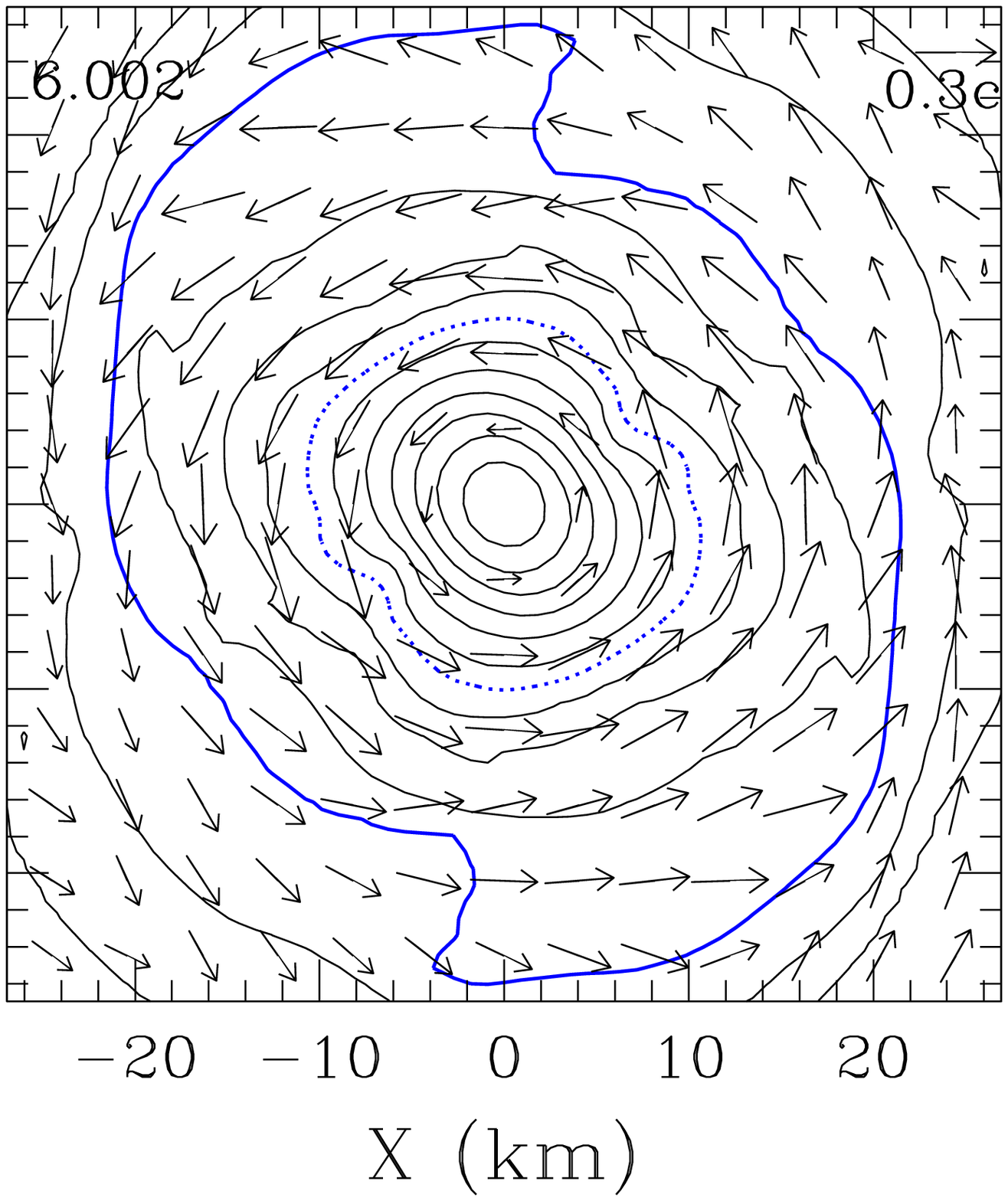} 
\epsfxsize=2.2in
\leavevmode
\hspace{-1.73cm}\epsffile{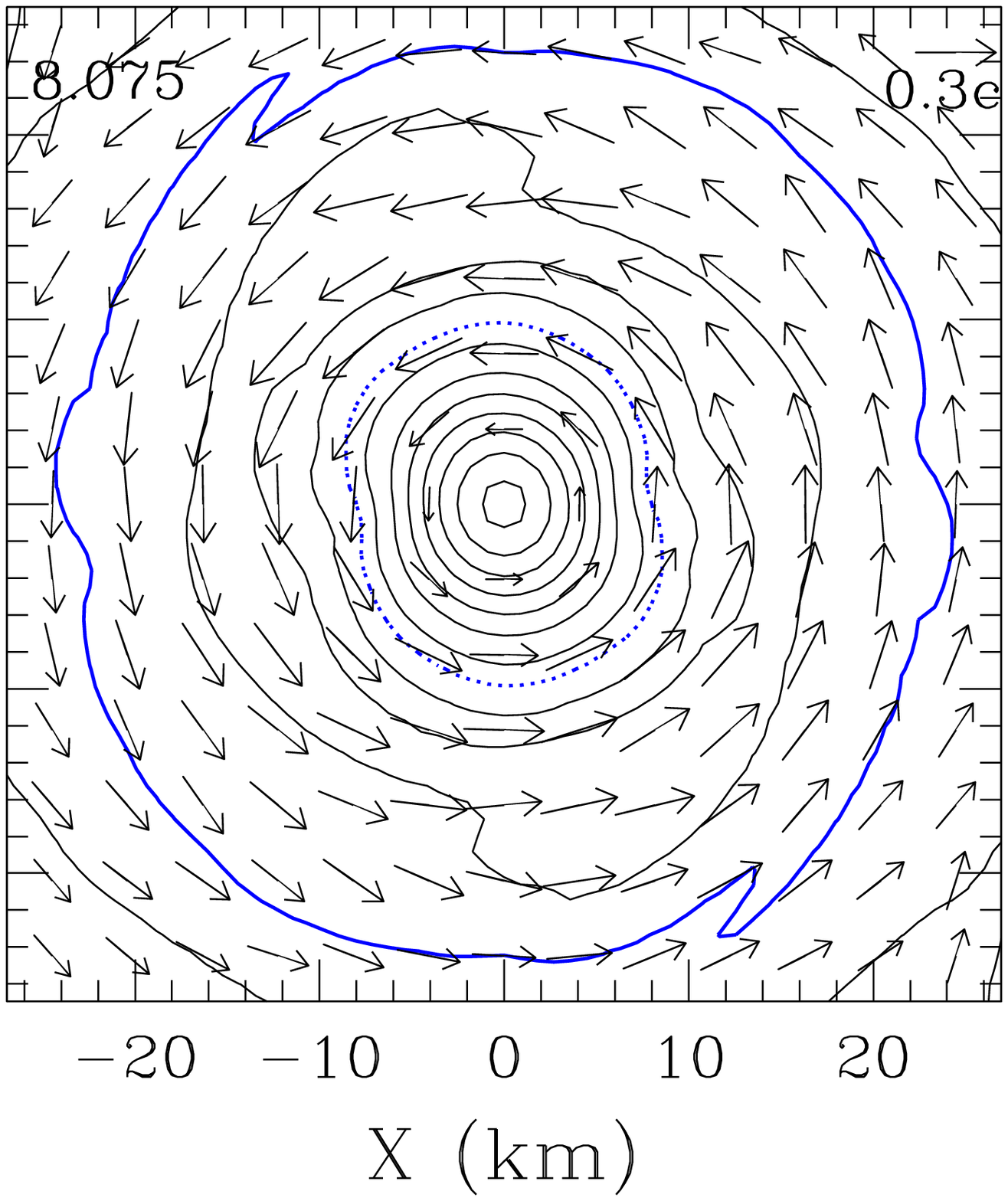}
\epsfxsize=2.2in
\leavevmode
\hspace{-1.73cm}\epsffile{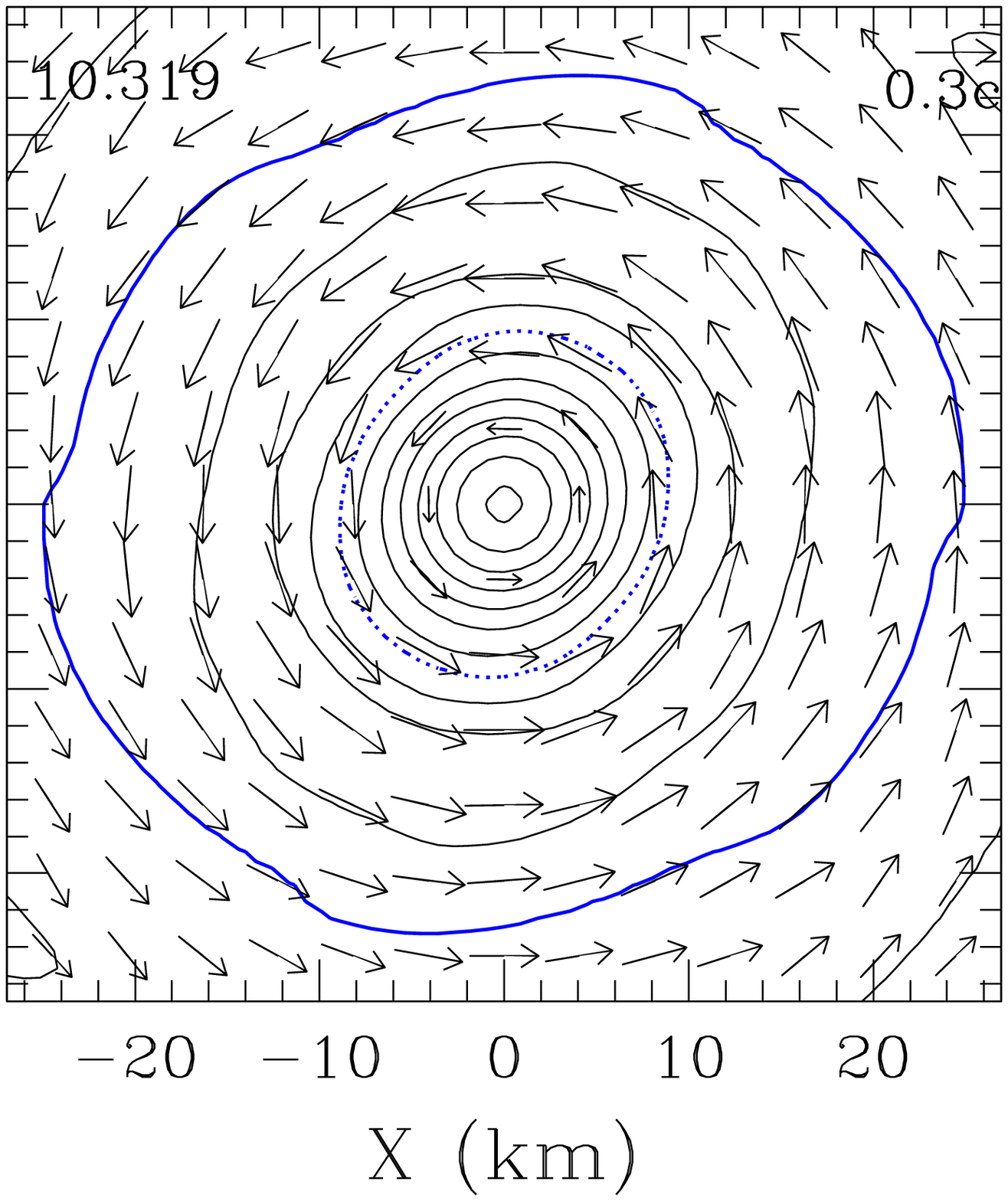}
\vspace{-6mm}
\caption{\small
The same as Fig. \ref{FIG4} but for model APR1414. 
\label{FIG5}}
\end{center}
\end{figure*}

\begin{figure*}[thb]
\vspace{-3mm}
\begin{center}
\epsfxsize=2.2in
\leavevmode
\epsffile{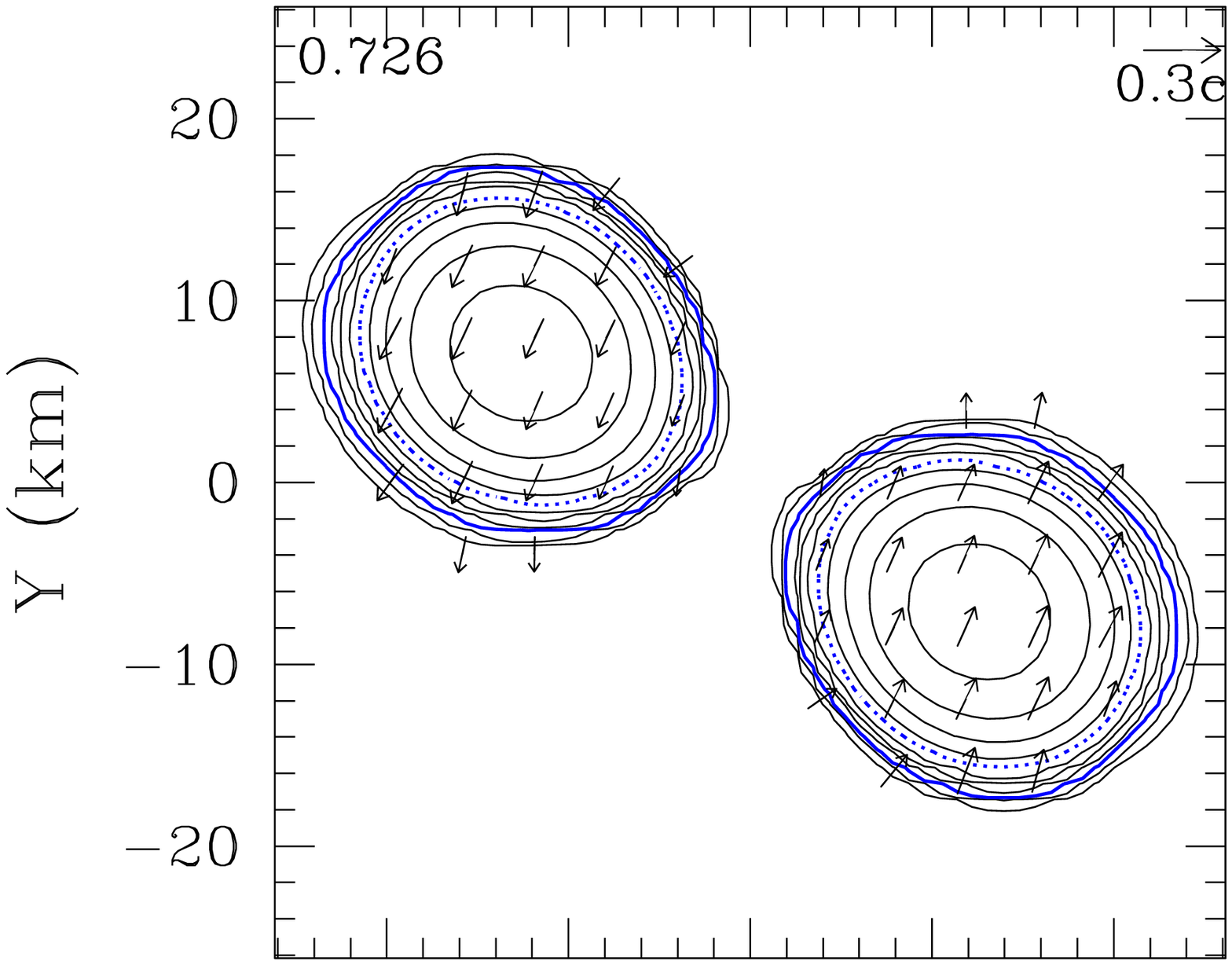}
\epsfxsize=2.2in
\leavevmode
\hspace{-1.73cm}\epsffile{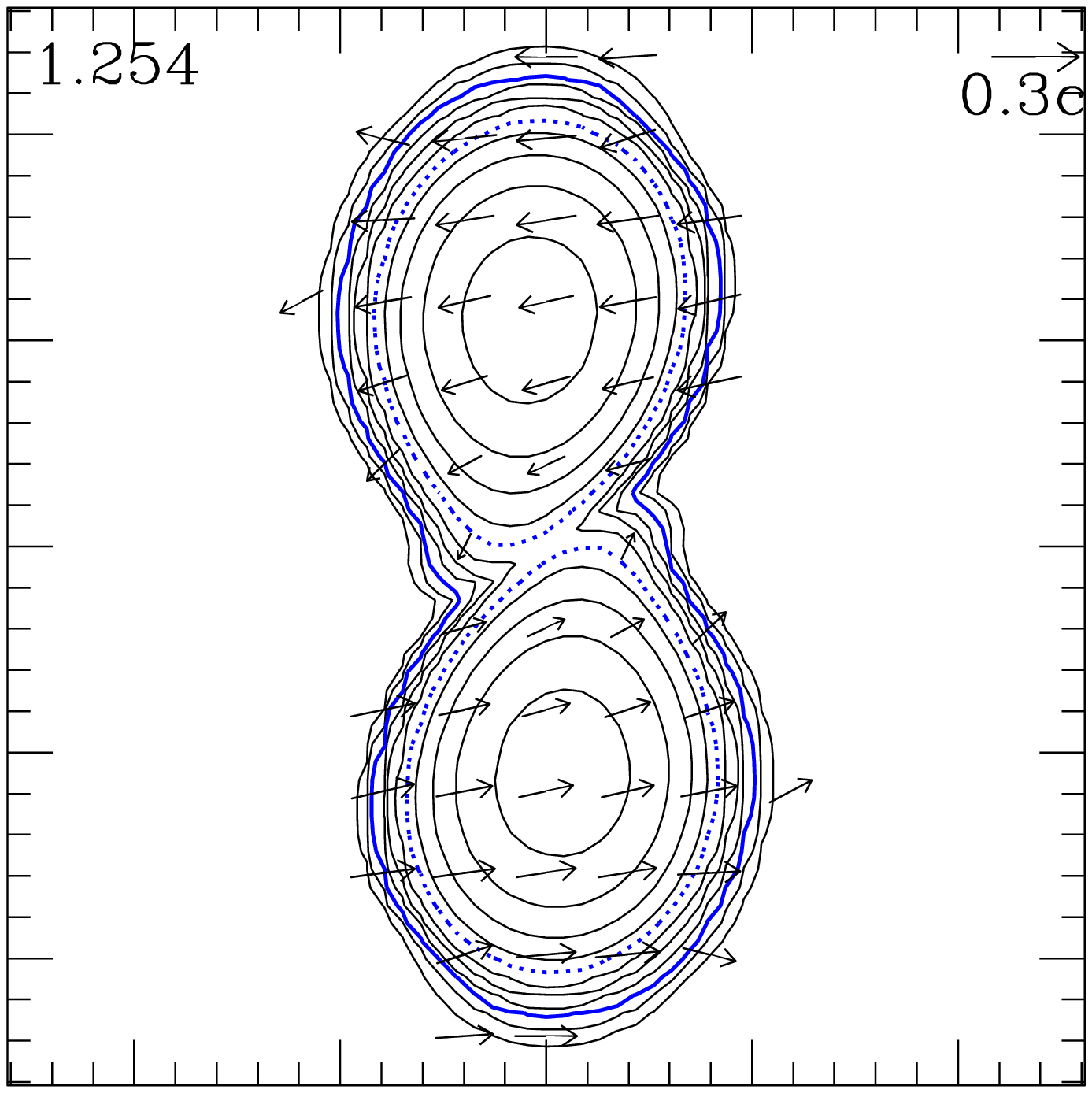} 
\epsfxsize=2.2in
\leavevmode
\hspace{-1.73cm}\epsffile{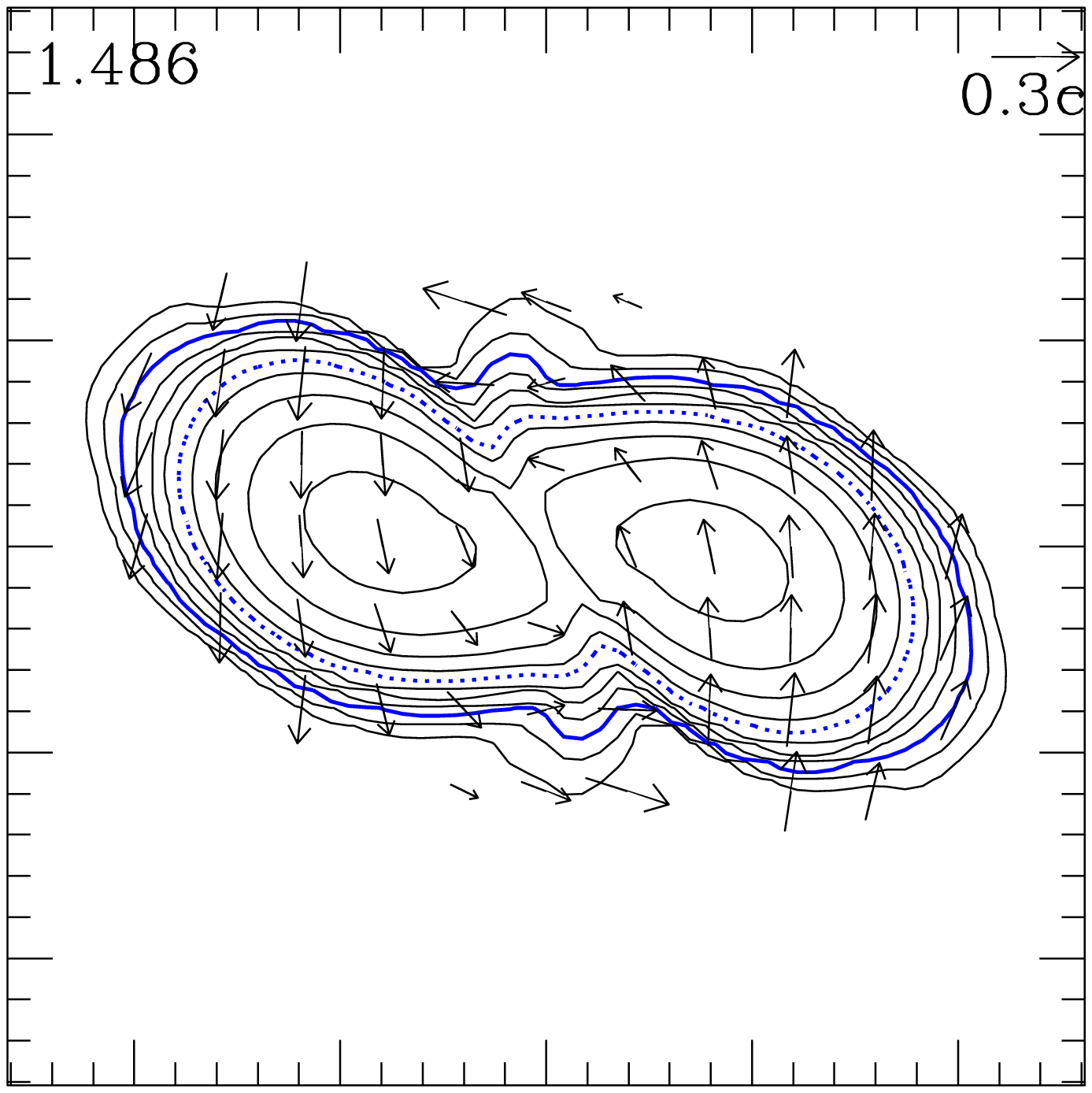} 
\epsfxsize=2.2in
\leavevmode
\hspace{-1.73cm}\epsffile{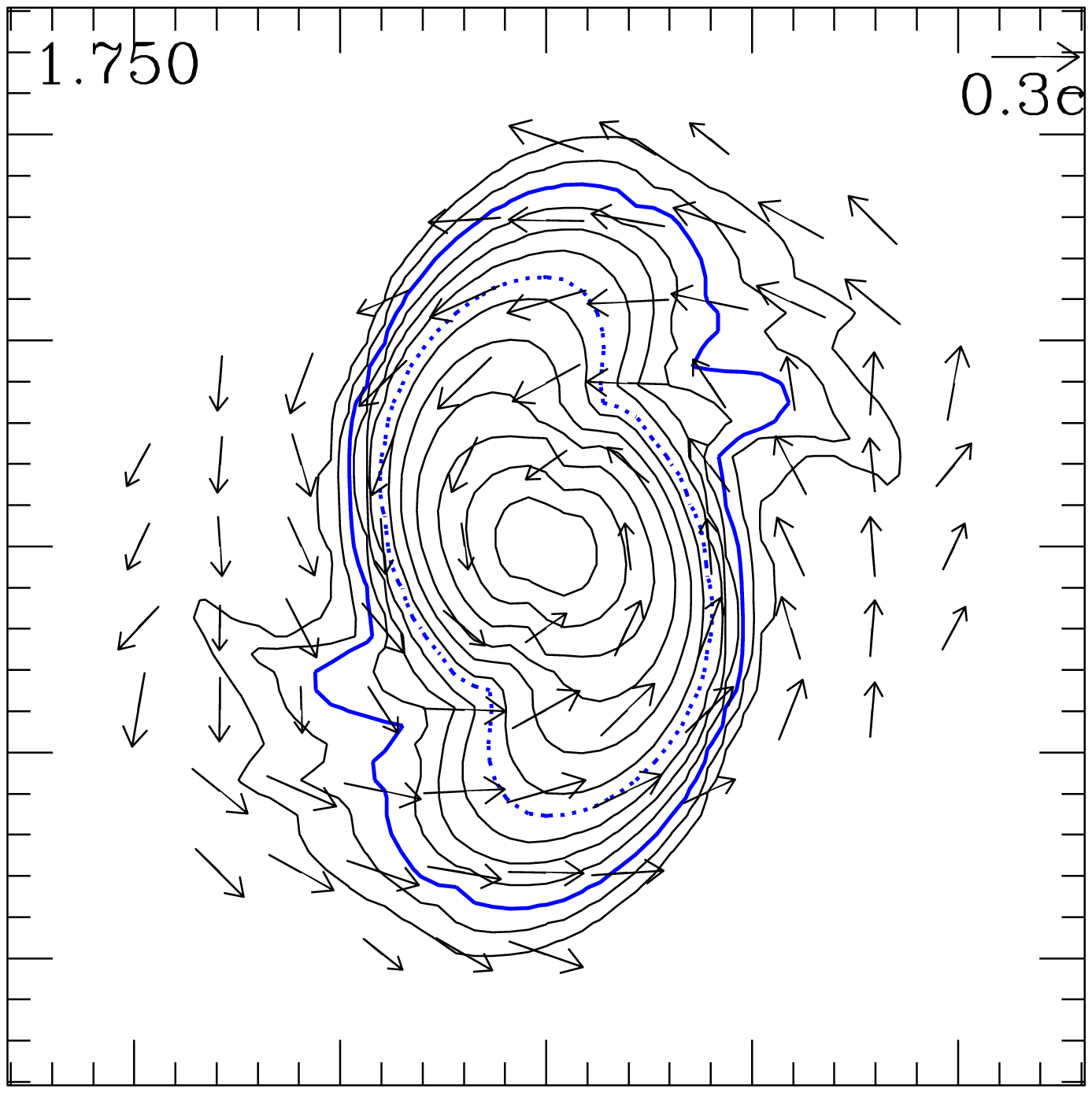} \\
\vspace{-1.76cm}
\epsfxsize=2.2in
\leavevmode
\epsffile{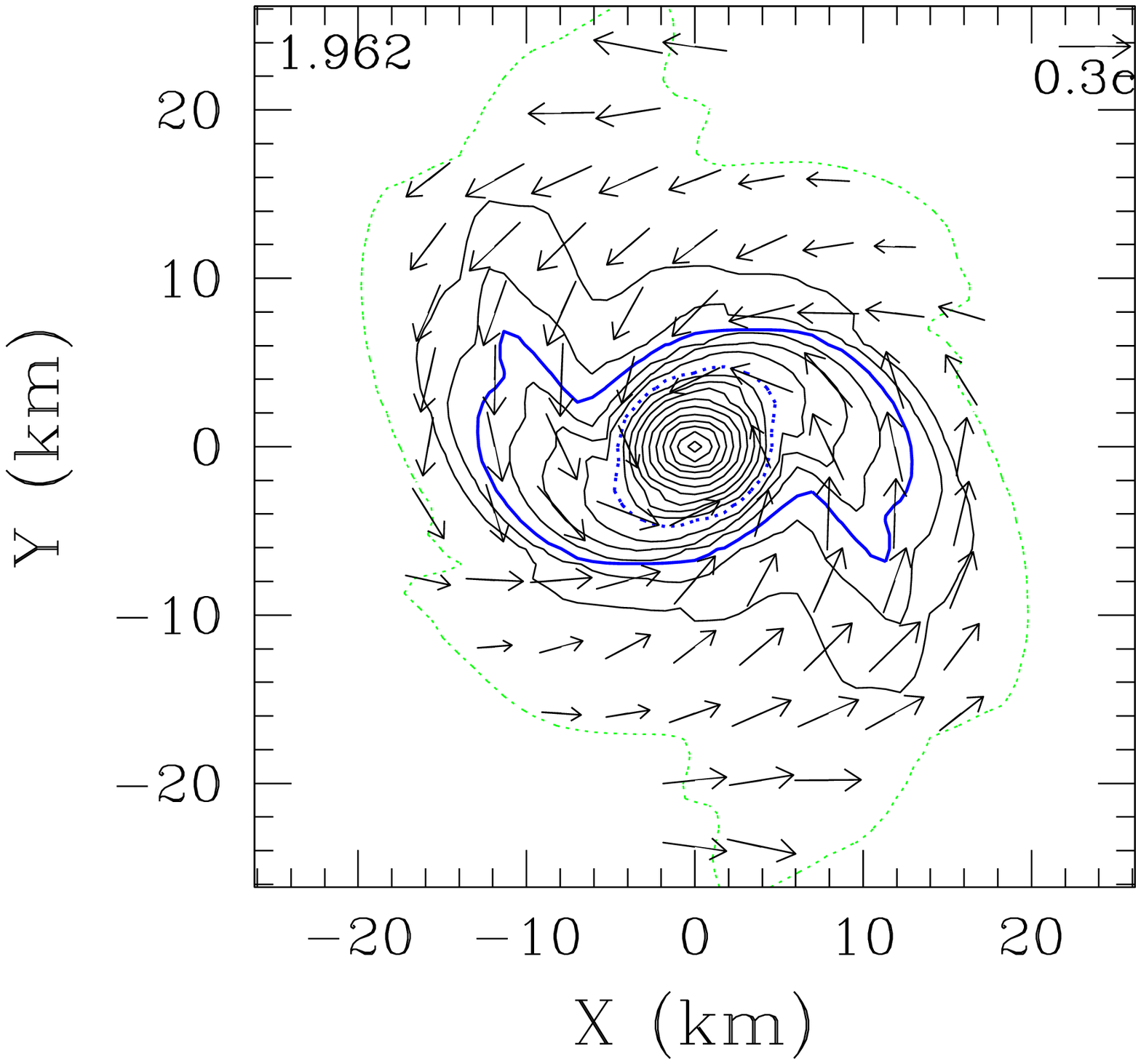}
\epsfxsize=2.2in
\leavevmode
\hspace{-1.73cm}\epsffile{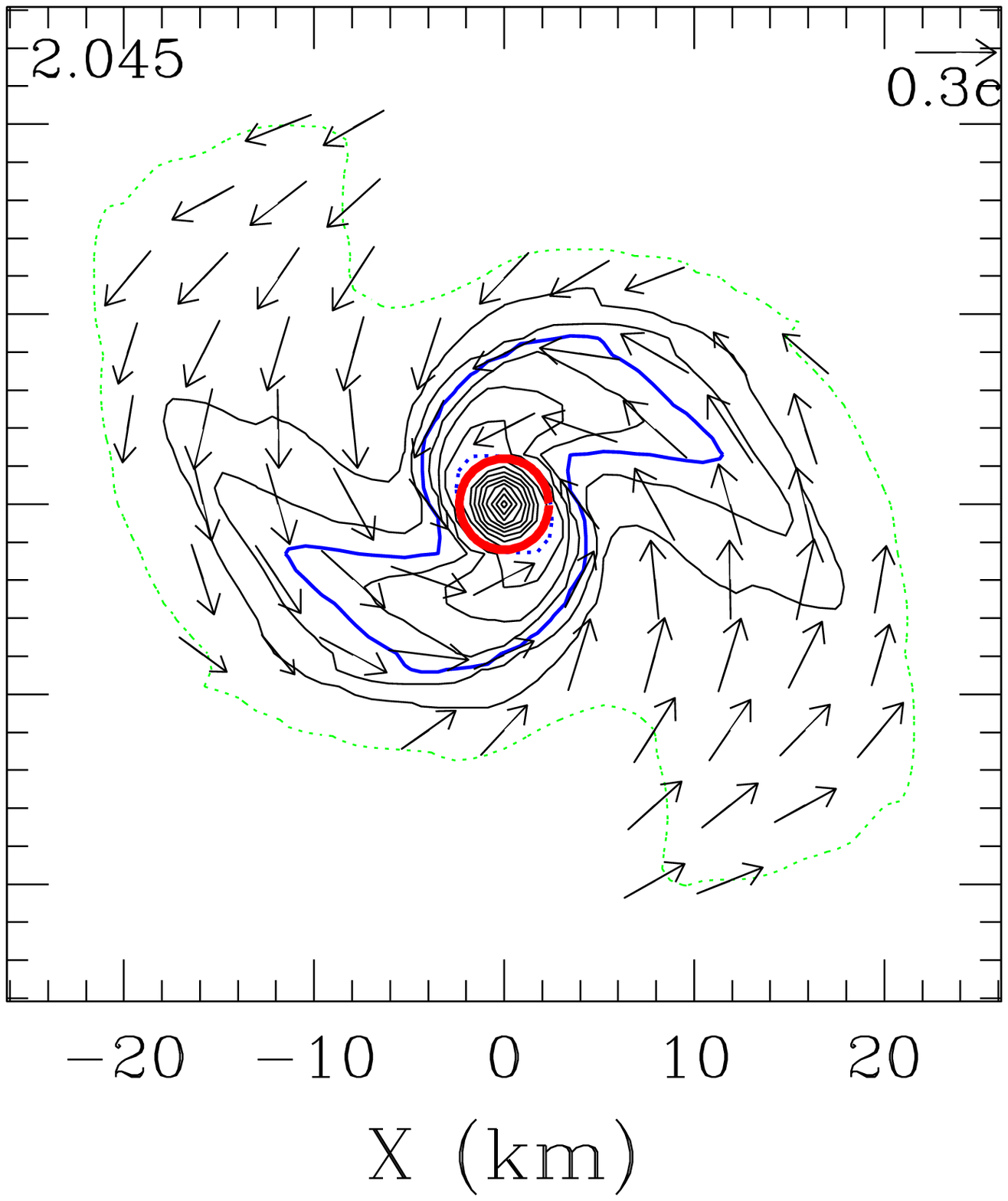} 
\epsfxsize=2.2in
\leavevmode
\hspace{-1.73cm}\epsffile{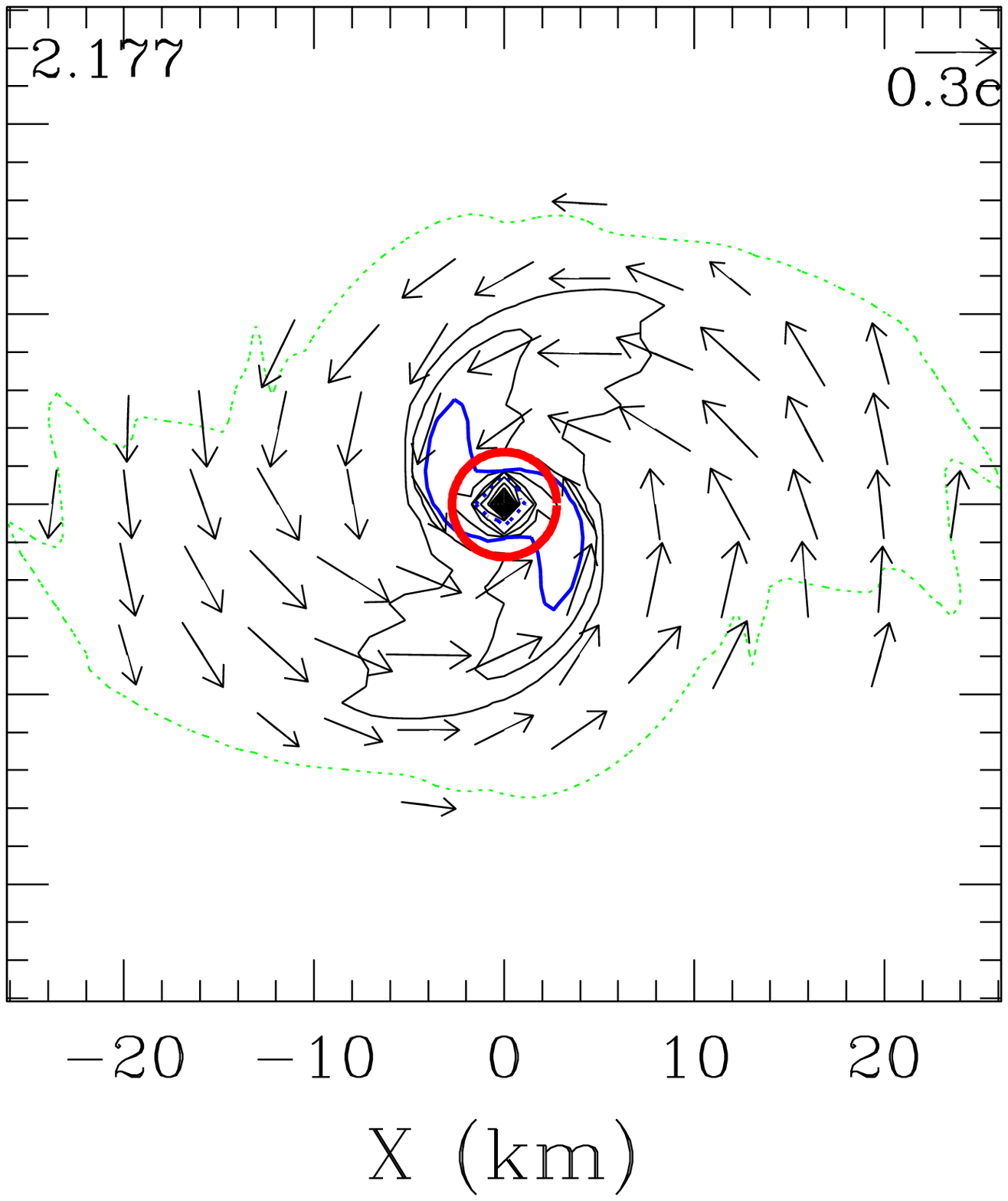} 
\epsfxsize=2.2in
\leavevmode
\hspace{-1.73cm}\epsffile{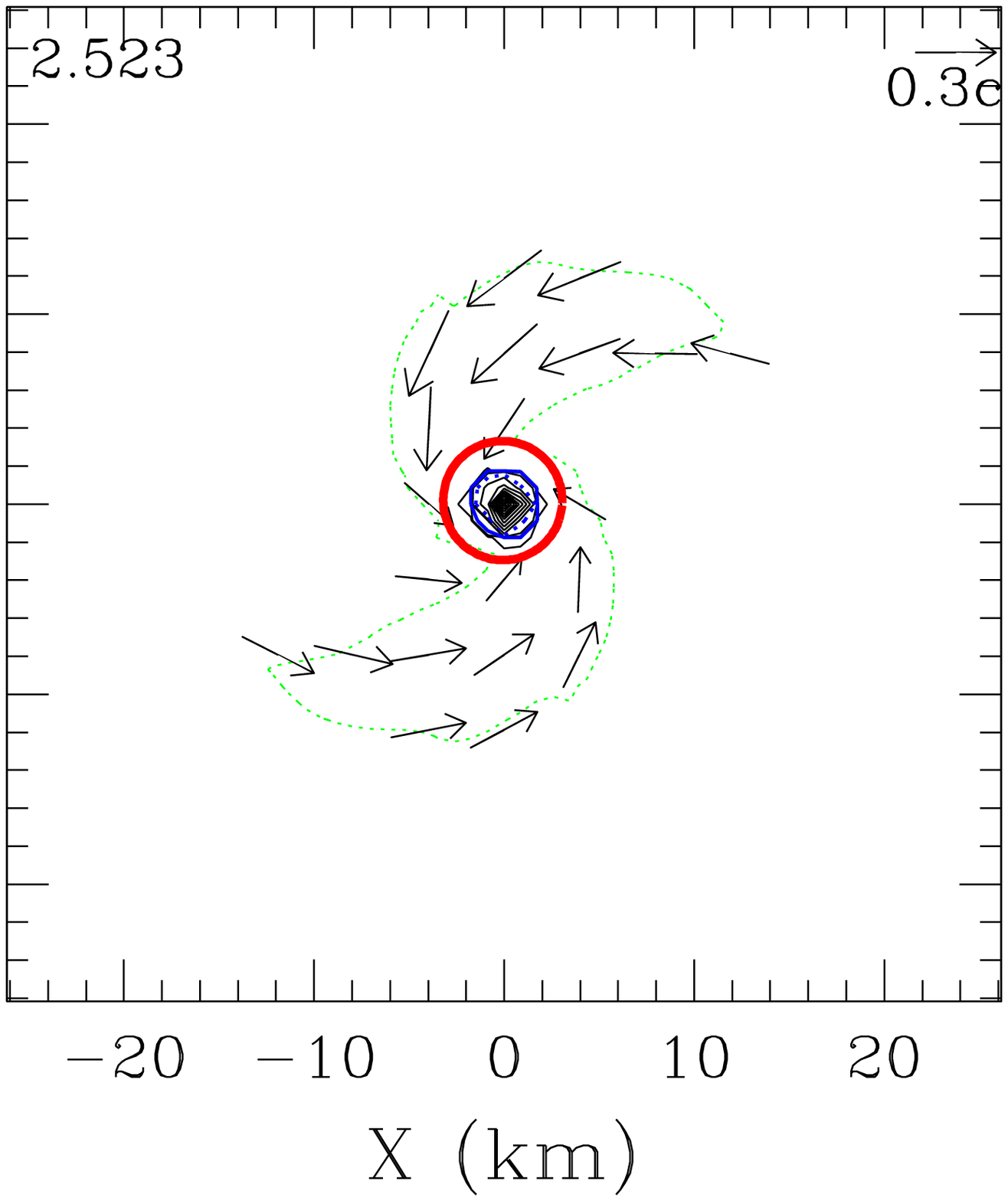} 
\vspace{-6mm}
\caption{The same as Fig. \ref{FIG4} but for model APR1515. 
The outermost (green) dotted curves denote $1 \times 10^{10}~{\rm g/cm^3}$. 
The thick circles around the origin in the last three panels 
denote the location of the apparent horizon.
\label{FIG6}}
\end{center}
\end{figure*}

\begin{figure*}[thb]
\begin{center}
\epsfxsize=2.2in
\leavevmode
\epsffile{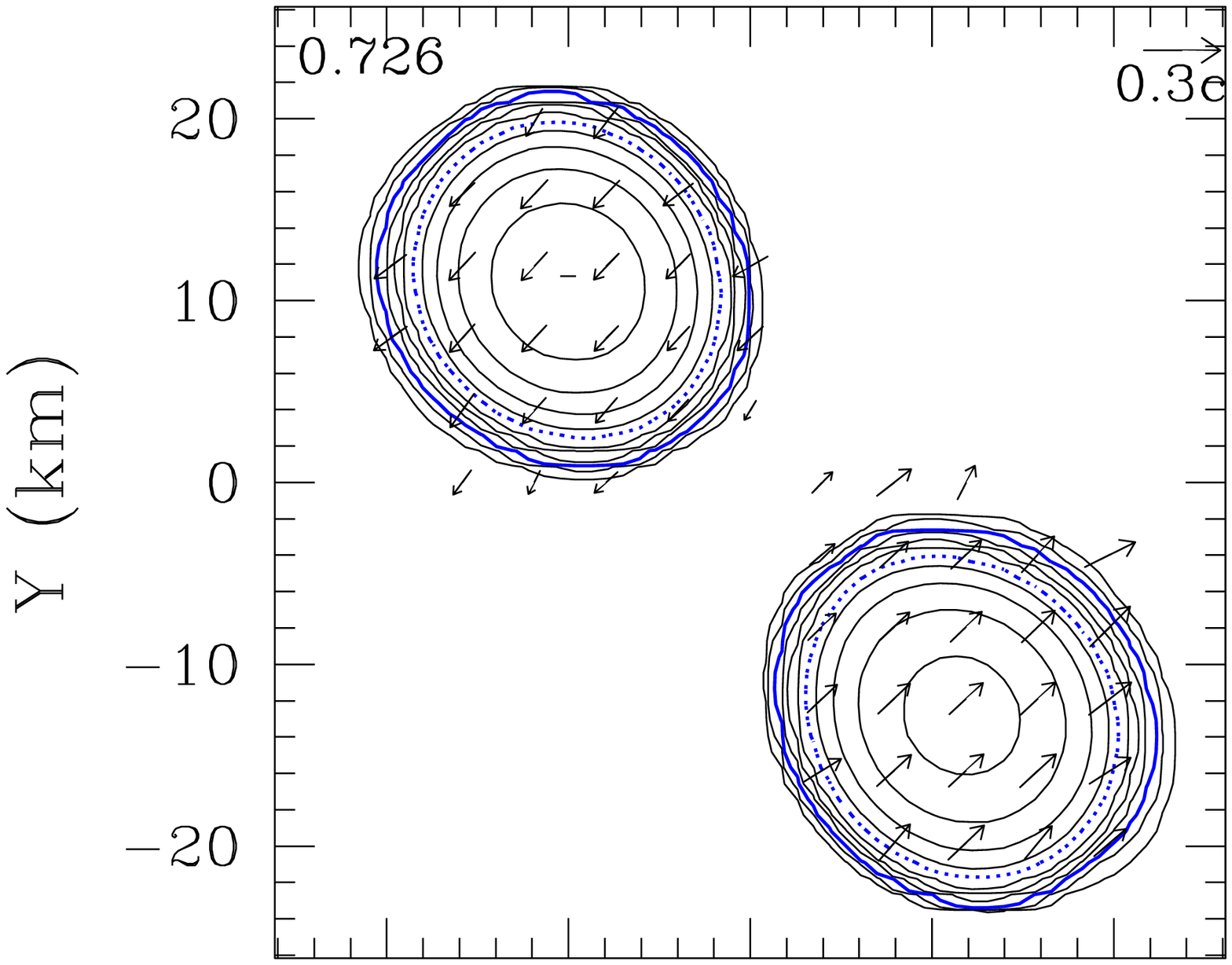}
\epsfxsize=2.2in
\leavevmode
\hspace{-1.73cm}\epsffile{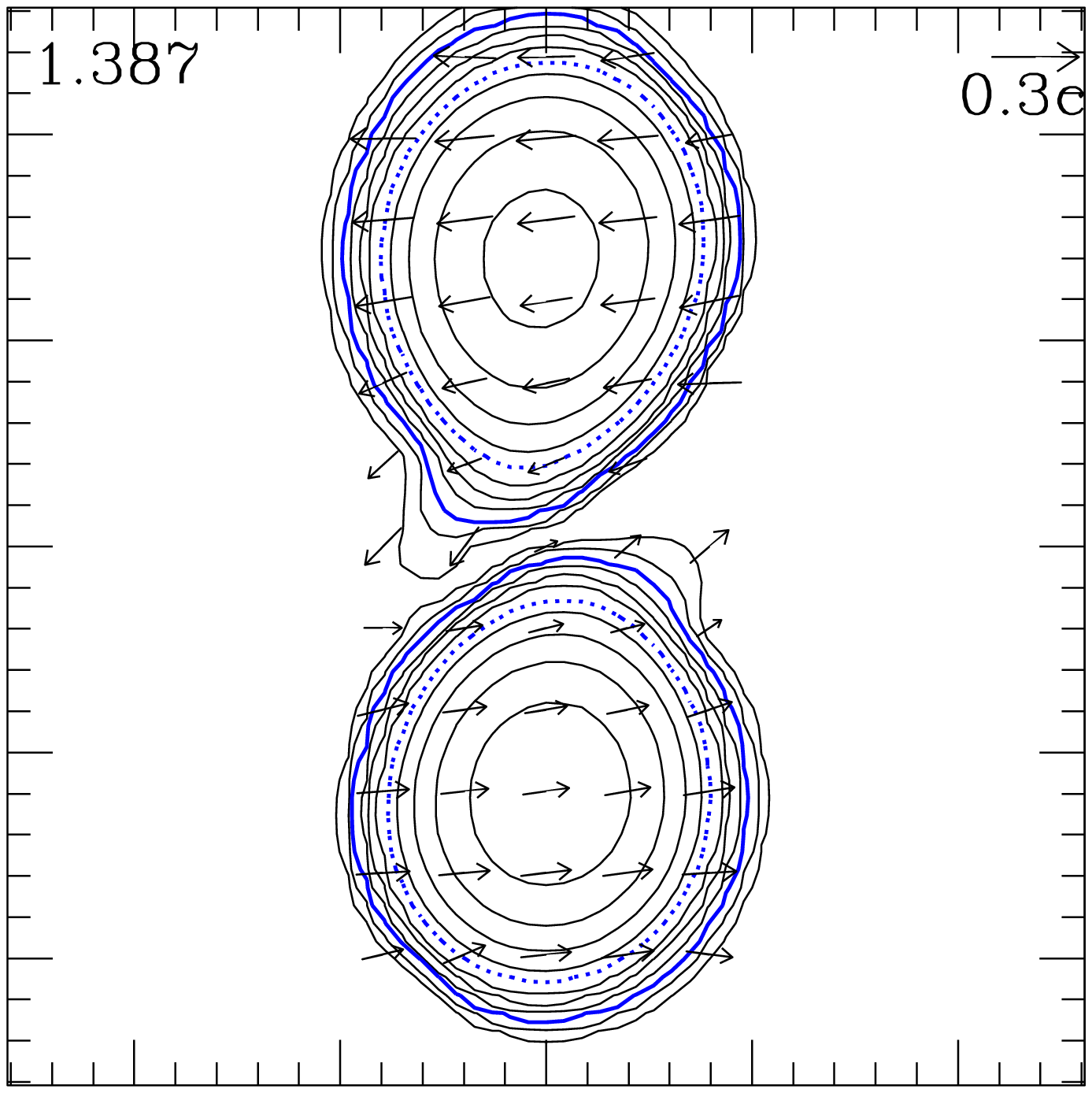} 
\epsfxsize=2.2in
\leavevmode
\hspace{-1.73cm}\epsffile{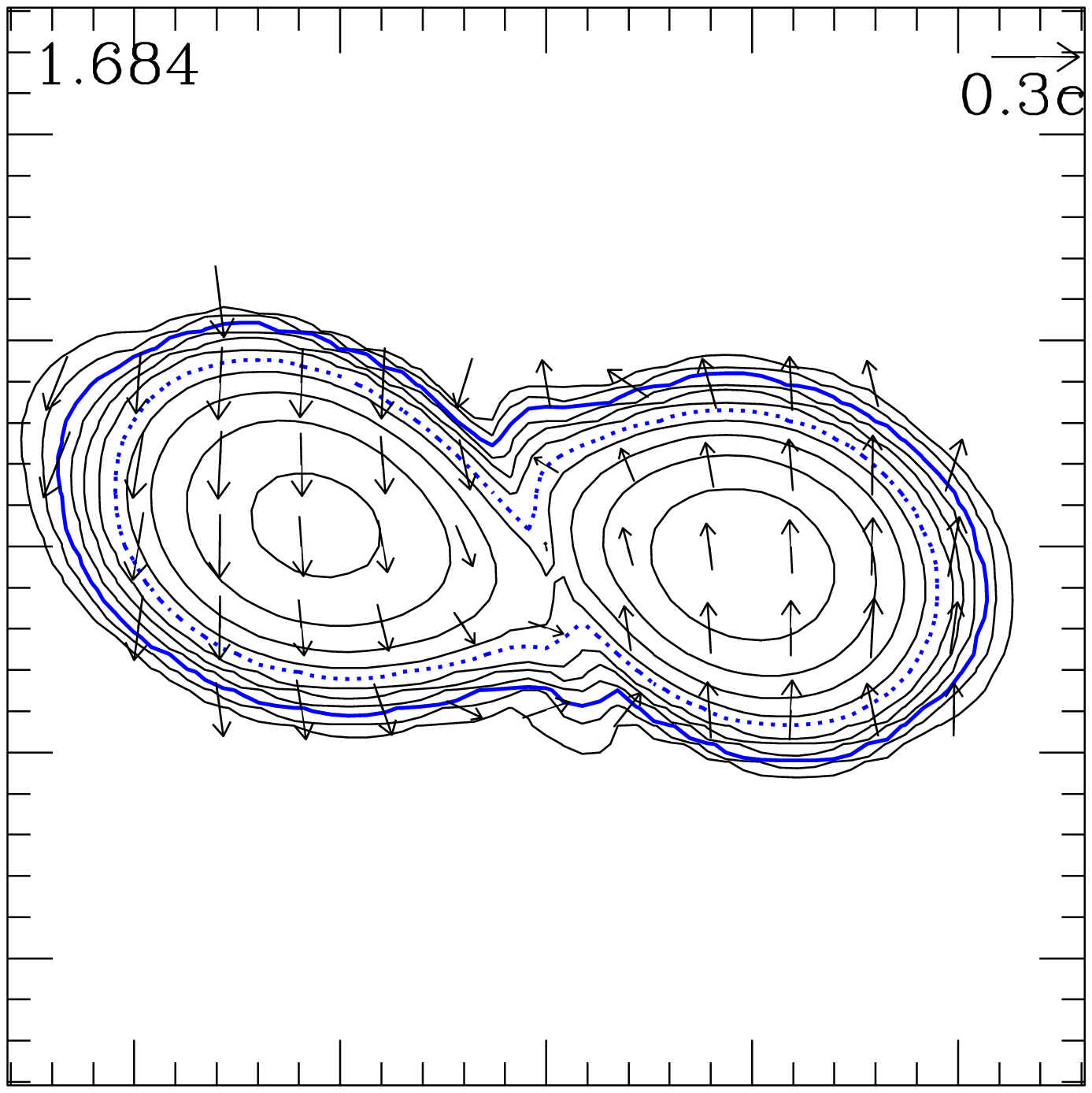}
\epsfxsize=2.2in
\leavevmode
\hspace{-1.73cm}\epsffile{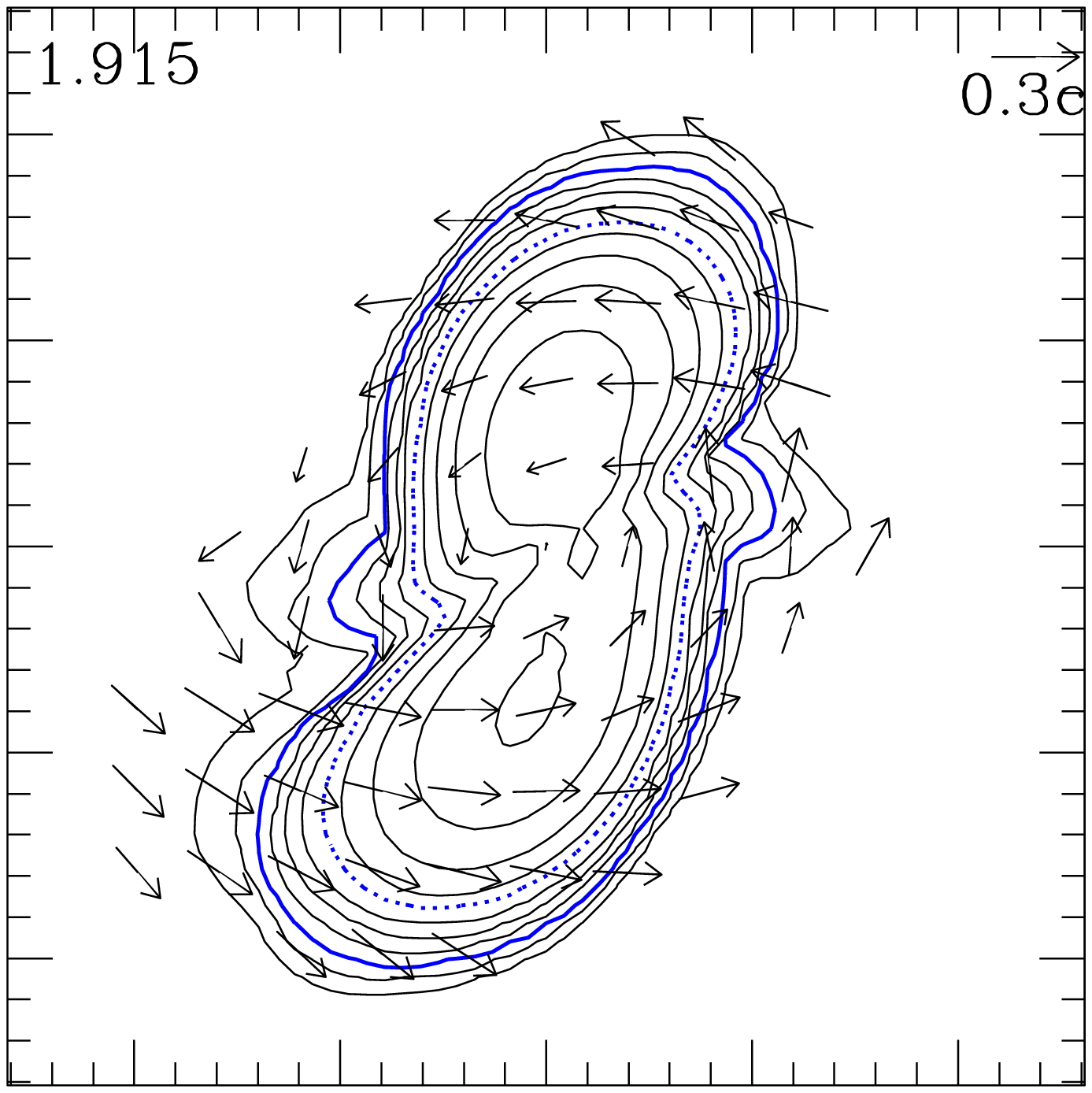} \\
\vspace{-1.76cm}
\epsfxsize=2.2in
\leavevmode
\epsffile{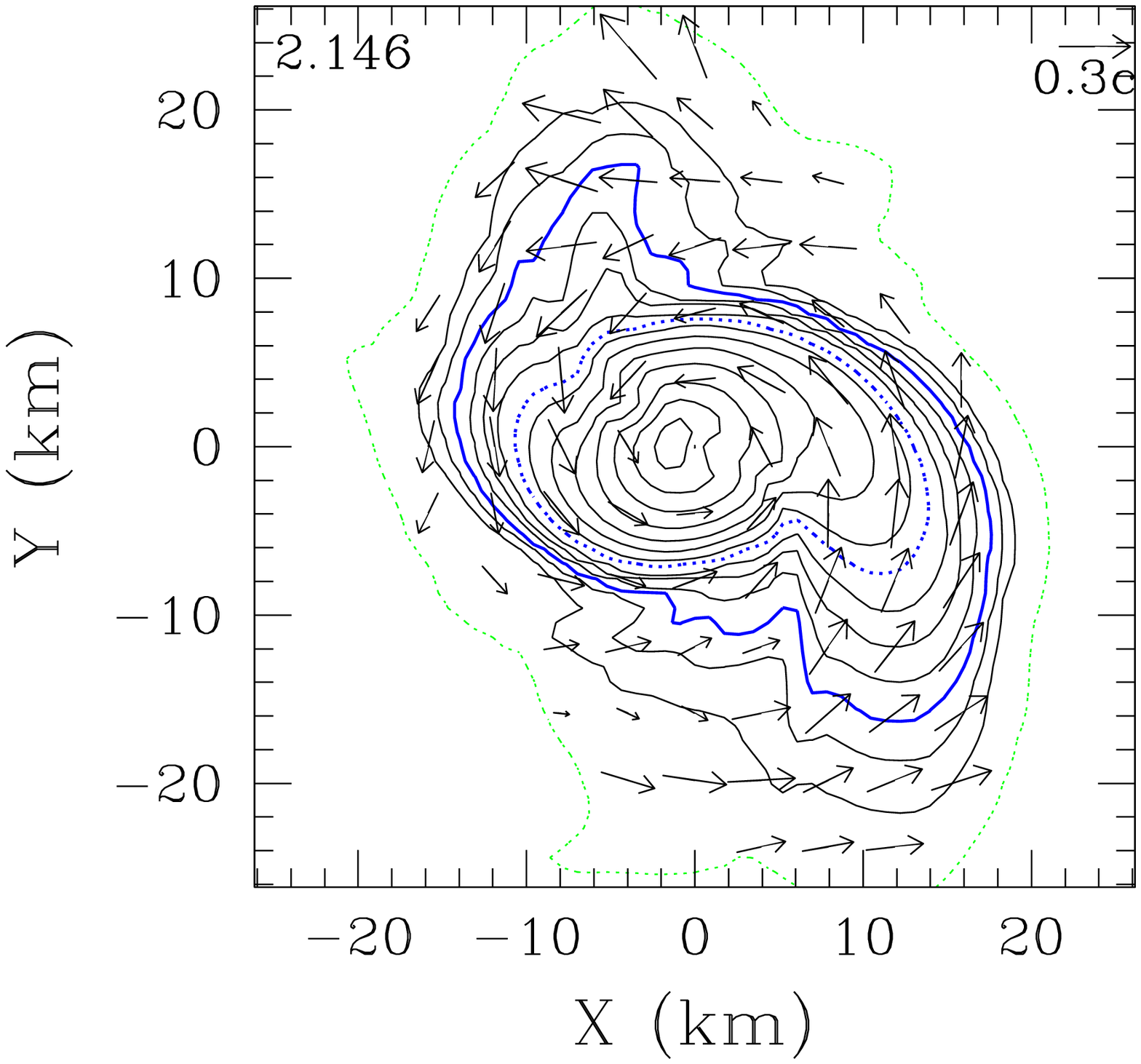}
\epsfxsize=2.2in
\leavevmode
\hspace{-1.73cm}\epsffile{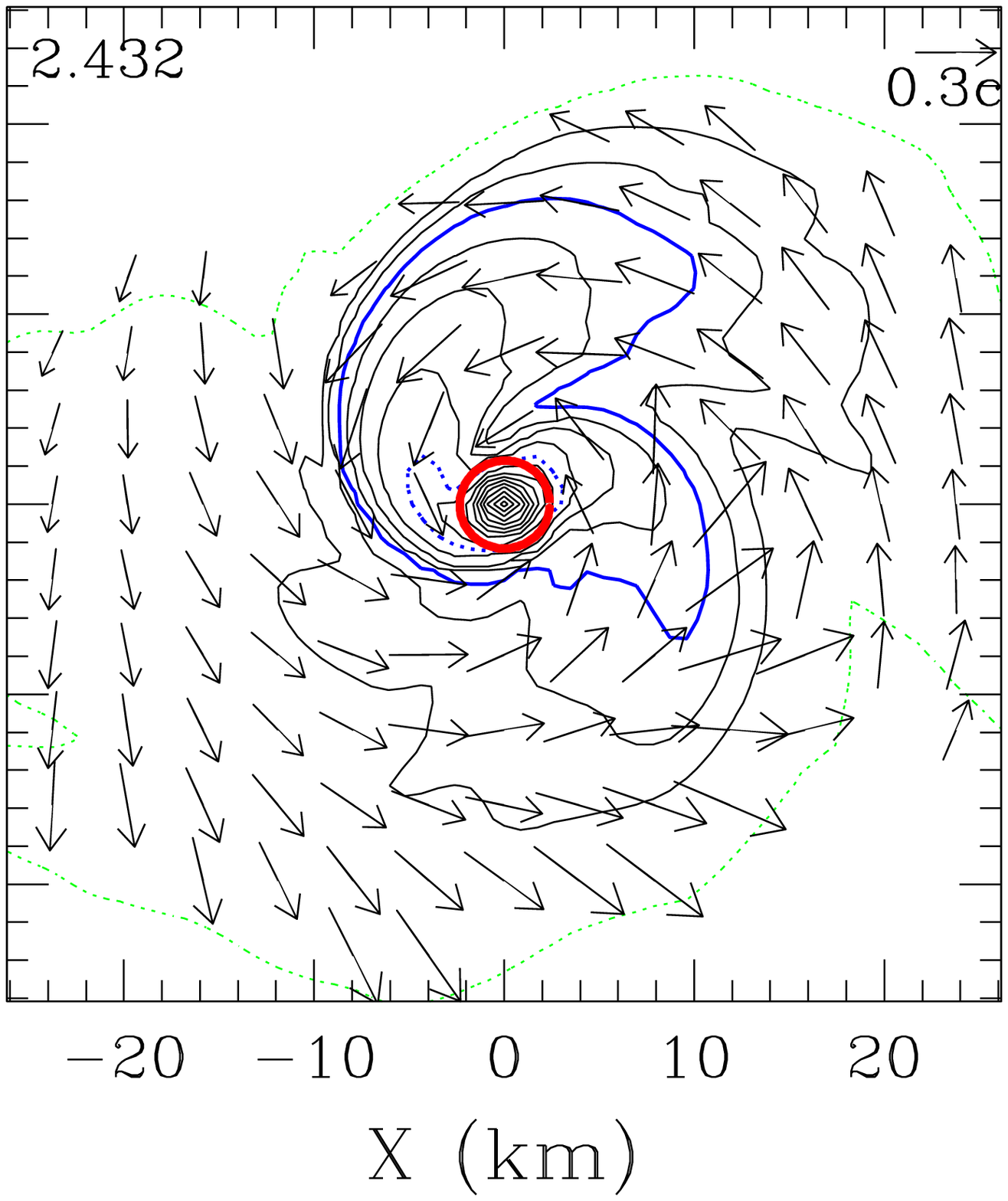} 
\epsfxsize=2.2in
\leavevmode
\hspace{-1.73cm}\epsffile{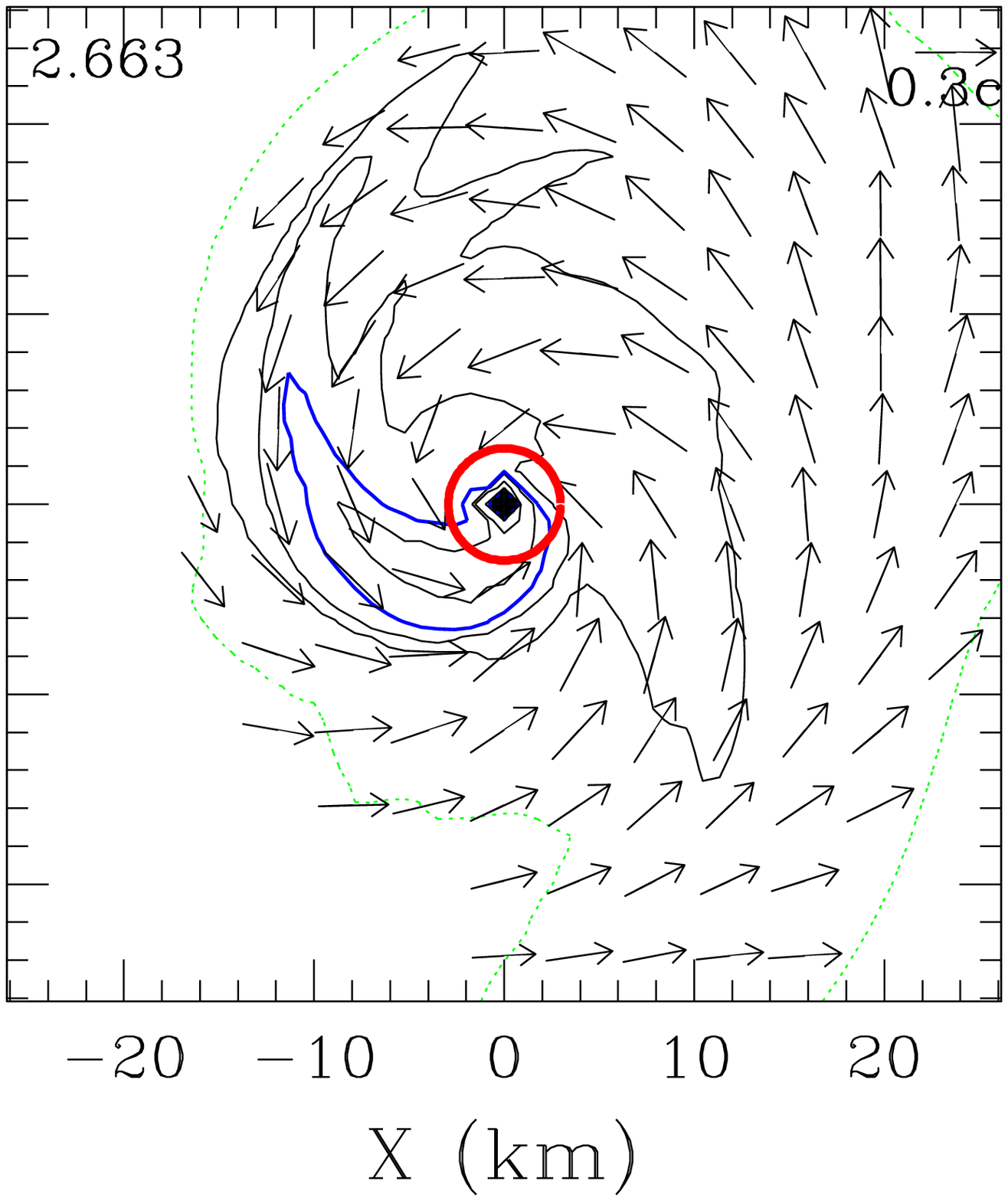}
\epsfxsize=2.2in
\leavevmode
\hspace{-1.73cm}\epsffile{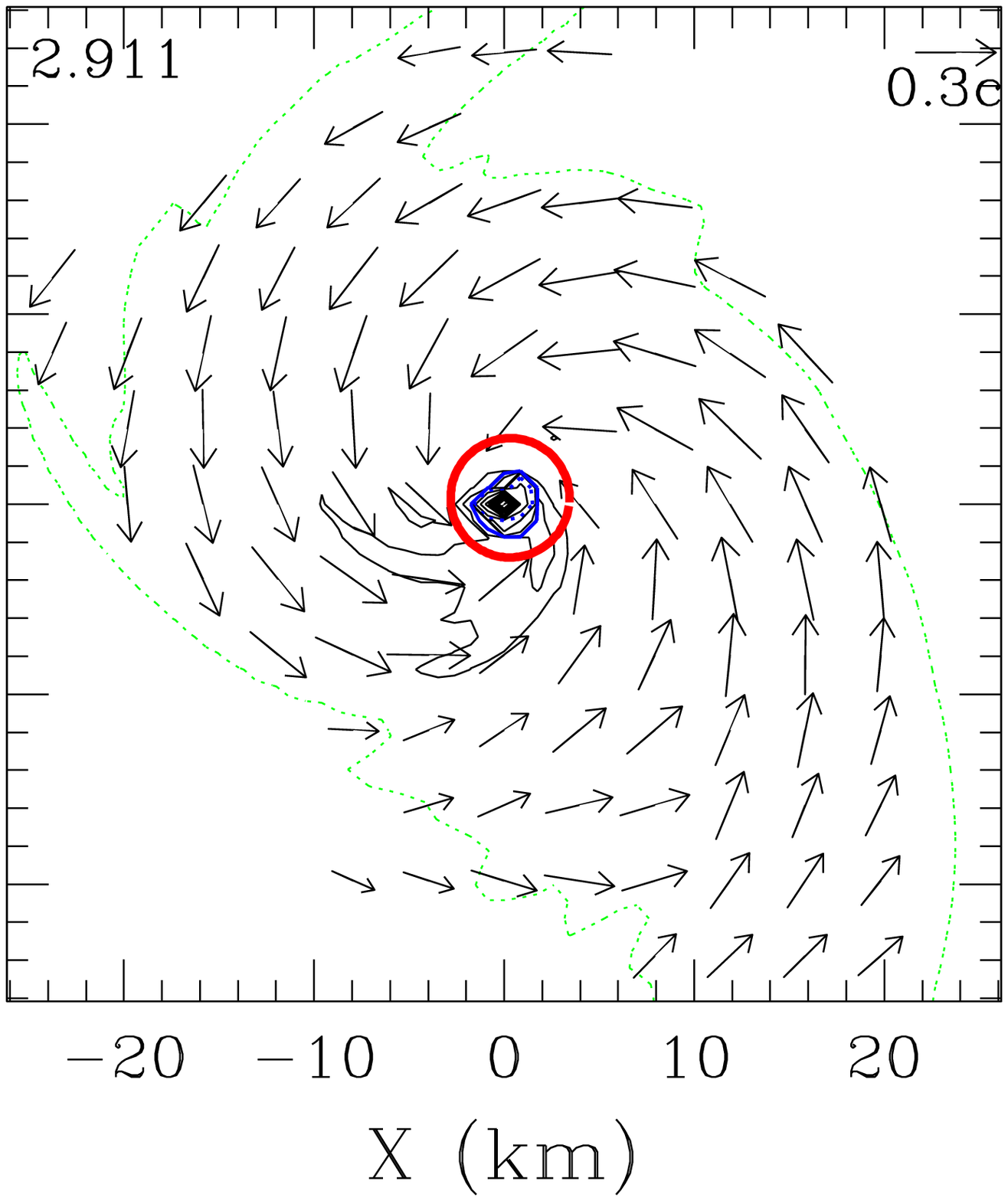}
\vspace{-6mm}
\caption{\small
The same as Fig. \ref{FIG6} but for model APR1416. 
\label{FIG7}}
\end{center}
\end{figure*}

\begin{figure*}[thb]
\vspace{-3mm}
\begin{center}
\epsfxsize=2.2in
\leavevmode
\epsffile{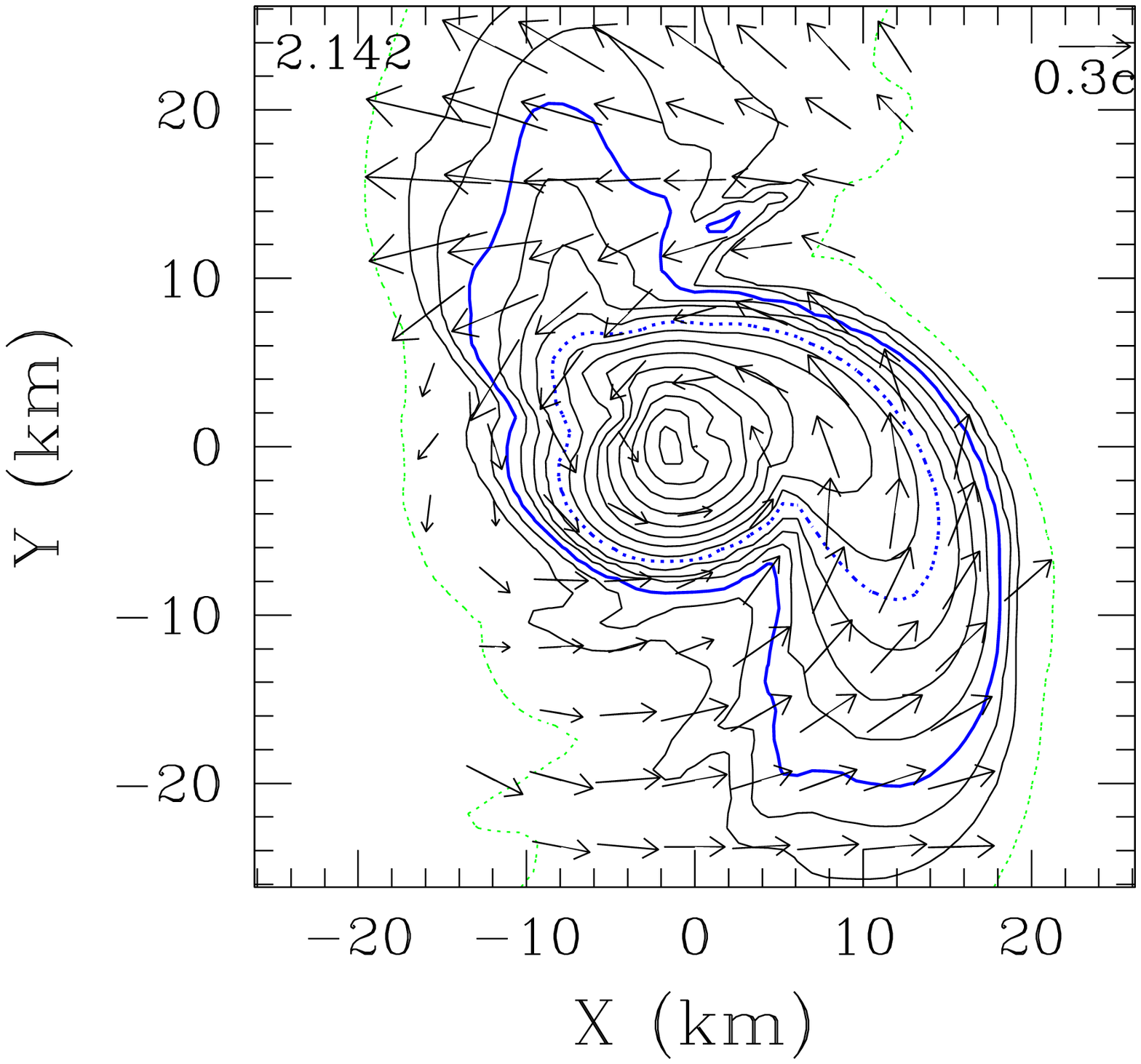}
\epsfxsize=2.2in
\leavevmode
\hspace{-1.73cm}\epsffile{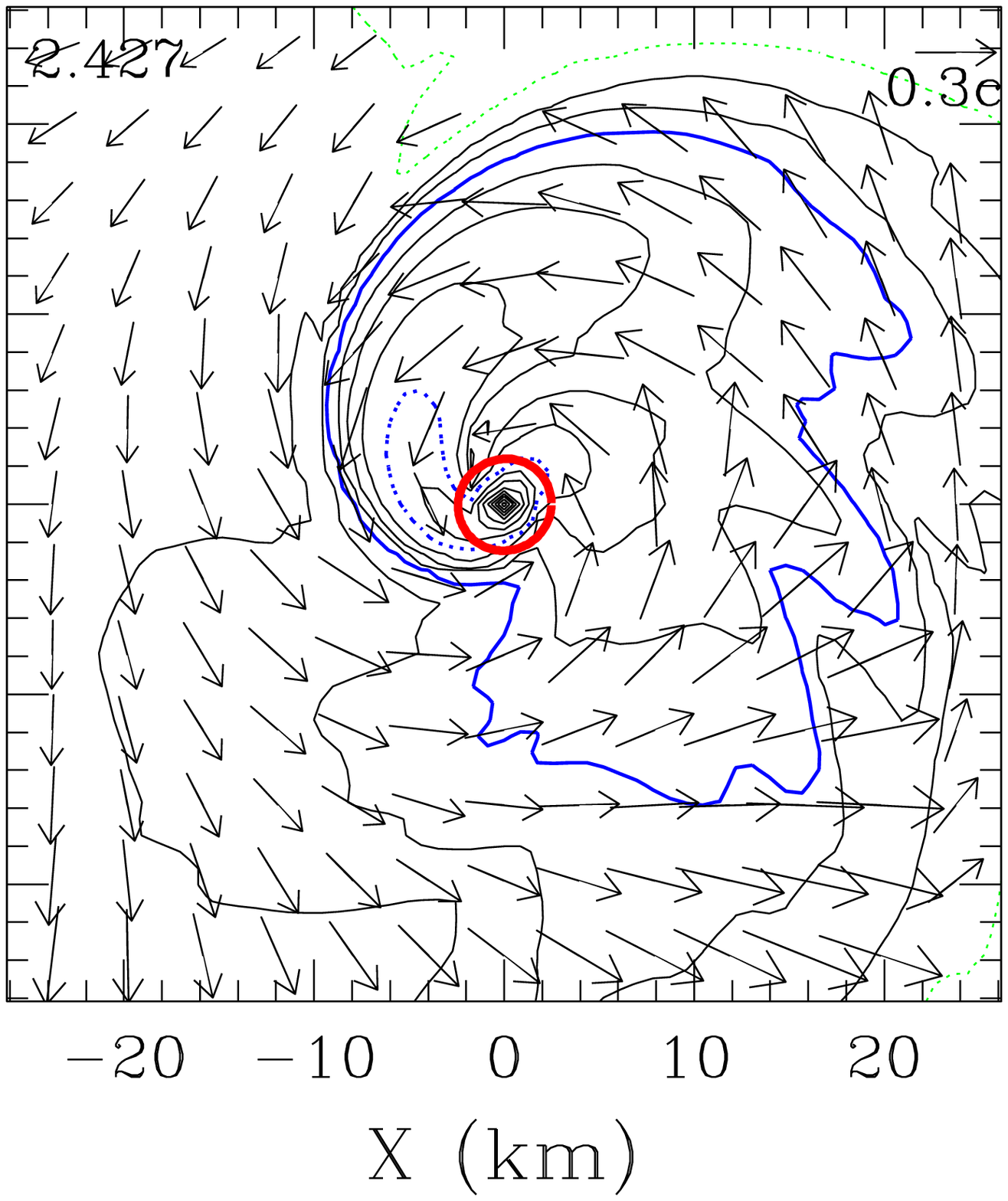} 
\epsfxsize=2.2in
\leavevmode
\hspace{-1.73cm}\epsffile{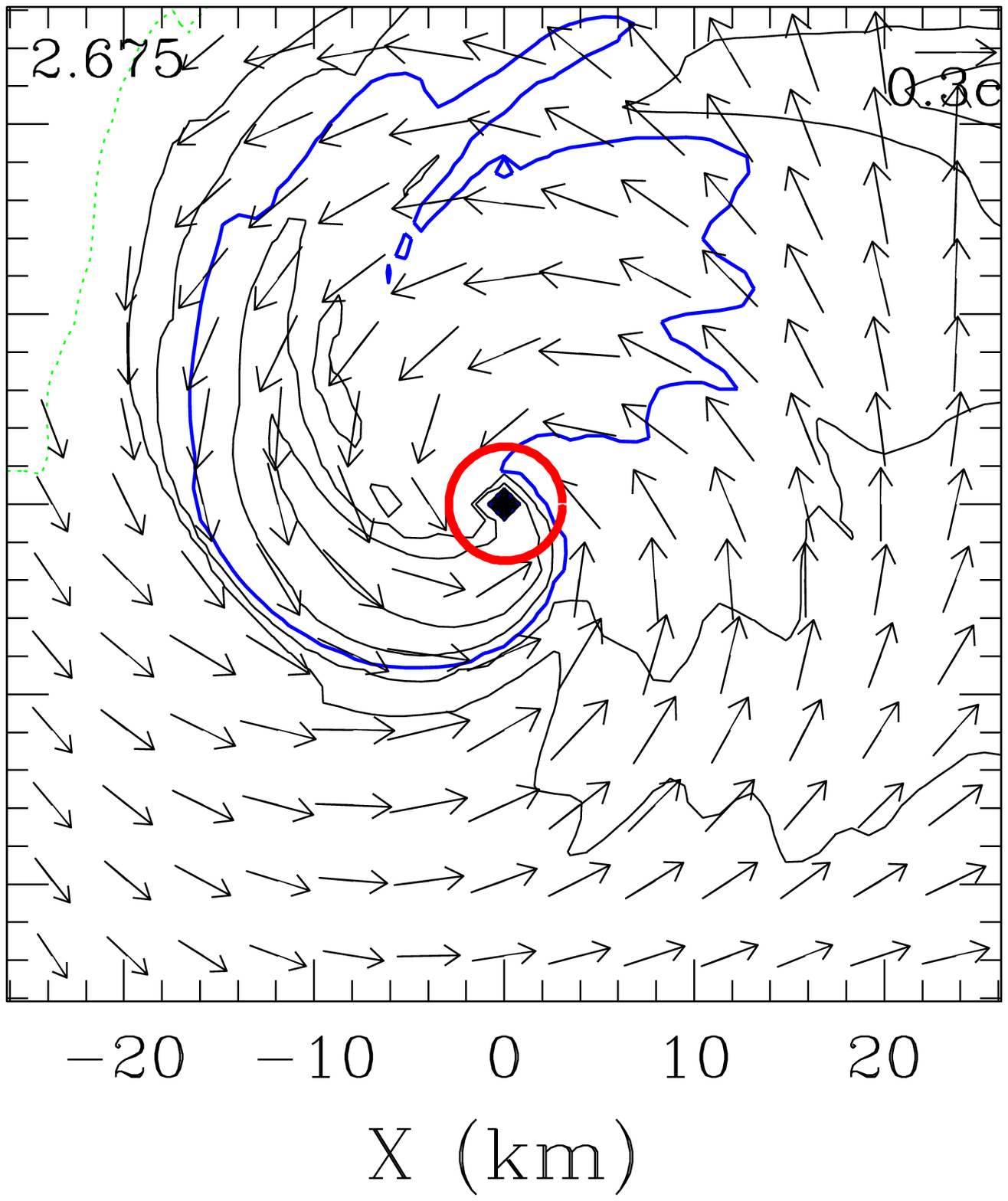}
\epsfxsize=2.2in
\leavevmode
\hspace{-1.73cm}\epsffile{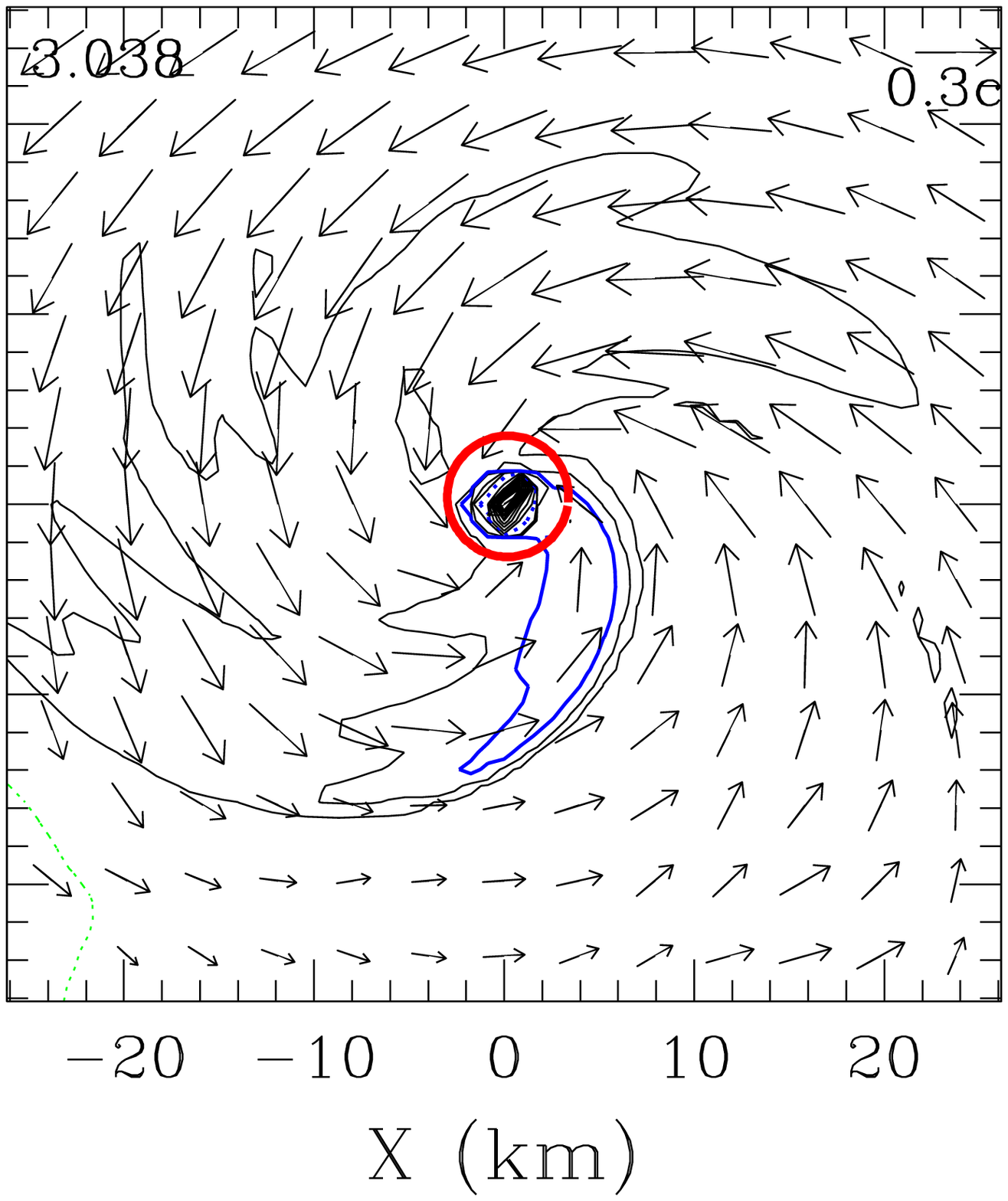} \\
\vspace{-6mm}
\caption{The same as Fig. \ref{FIG6} but for model APR1317. 
\label{FIG8}}
\end{center}
\end{figure*}

\begin{figure*}[t]
\begin{center}
\epsfxsize=2.2in
\leavevmode
(a)\hspace{-5mm}\epsffile{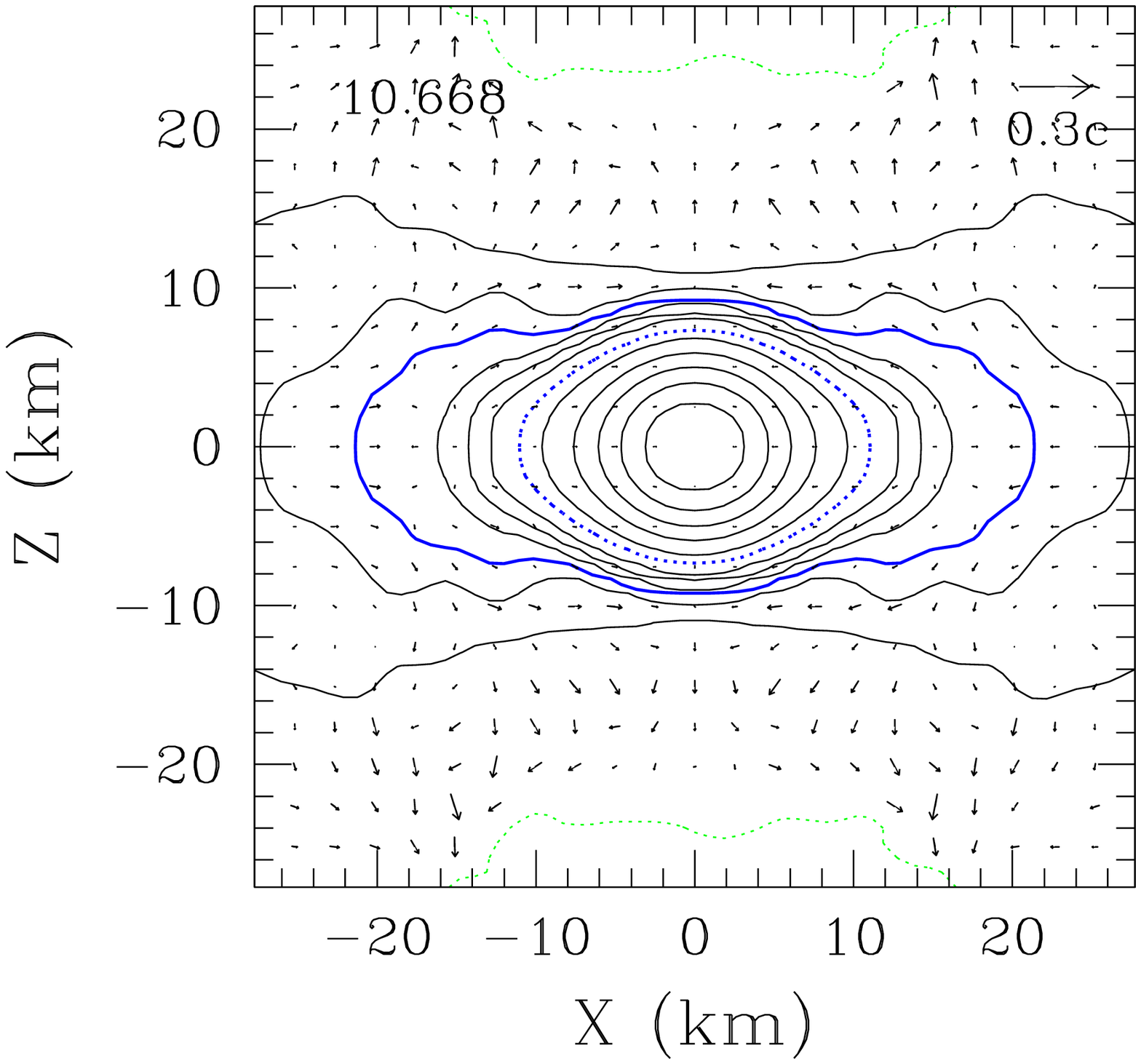}
\epsfxsize=2.2in
\leavevmode
(b)\hspace{-5mm}\epsffile{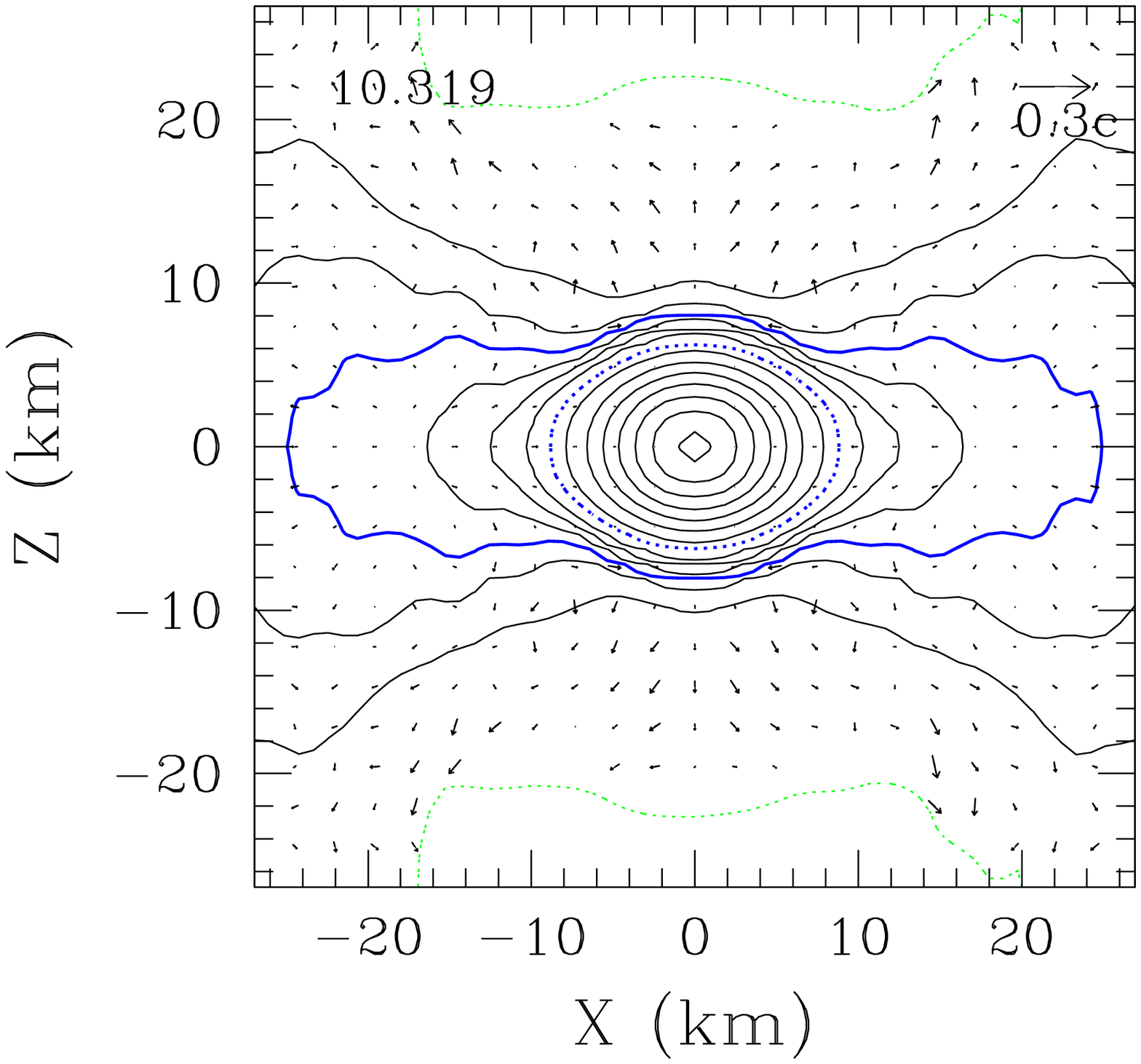}
\epsfxsize=2.2in
\leavevmode
(c)\hspace{-5mm}\epsffile{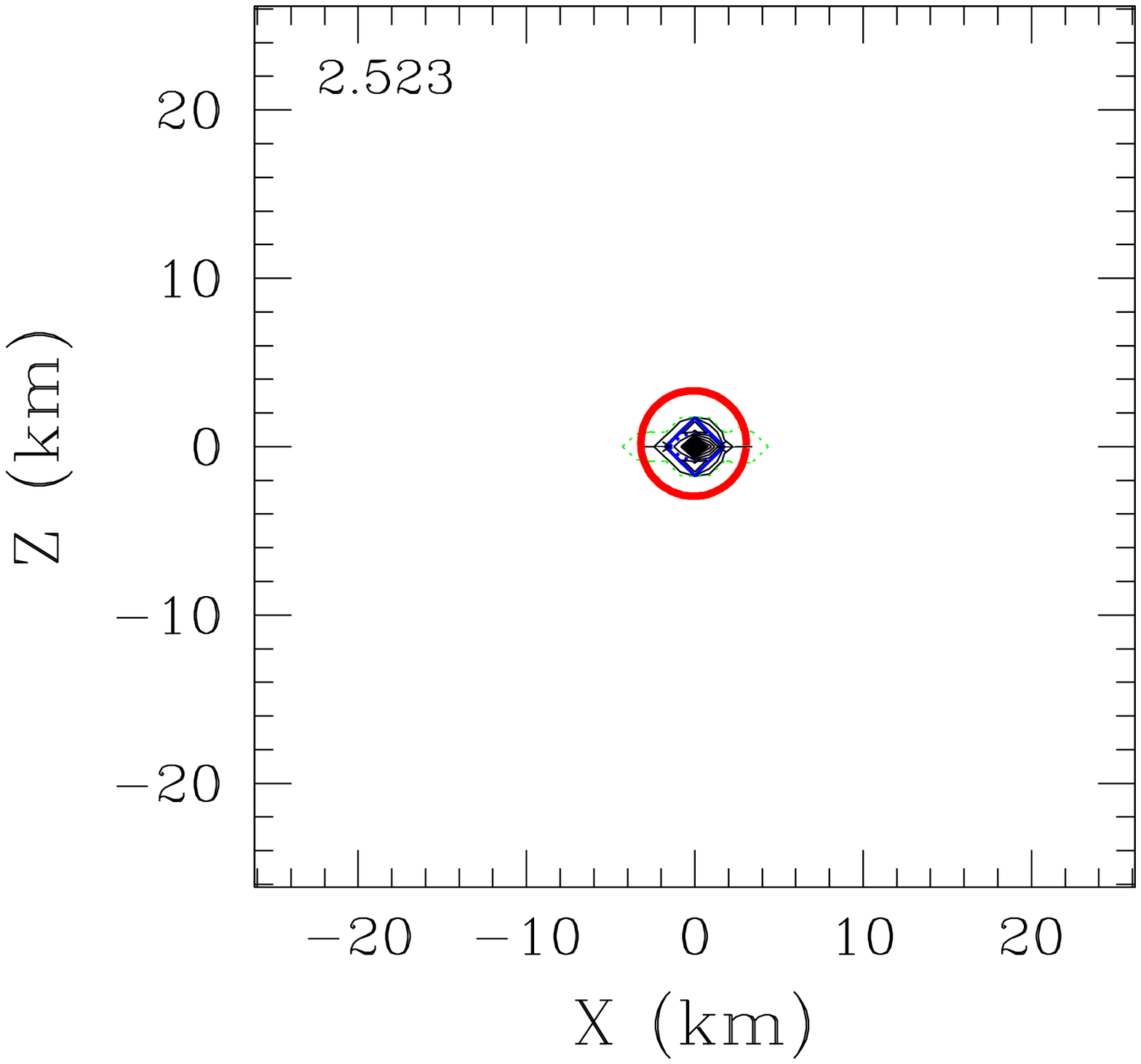} \\
\epsfxsize=2.2in
\leavevmode
(d)\hspace{-5mm}\epsffile{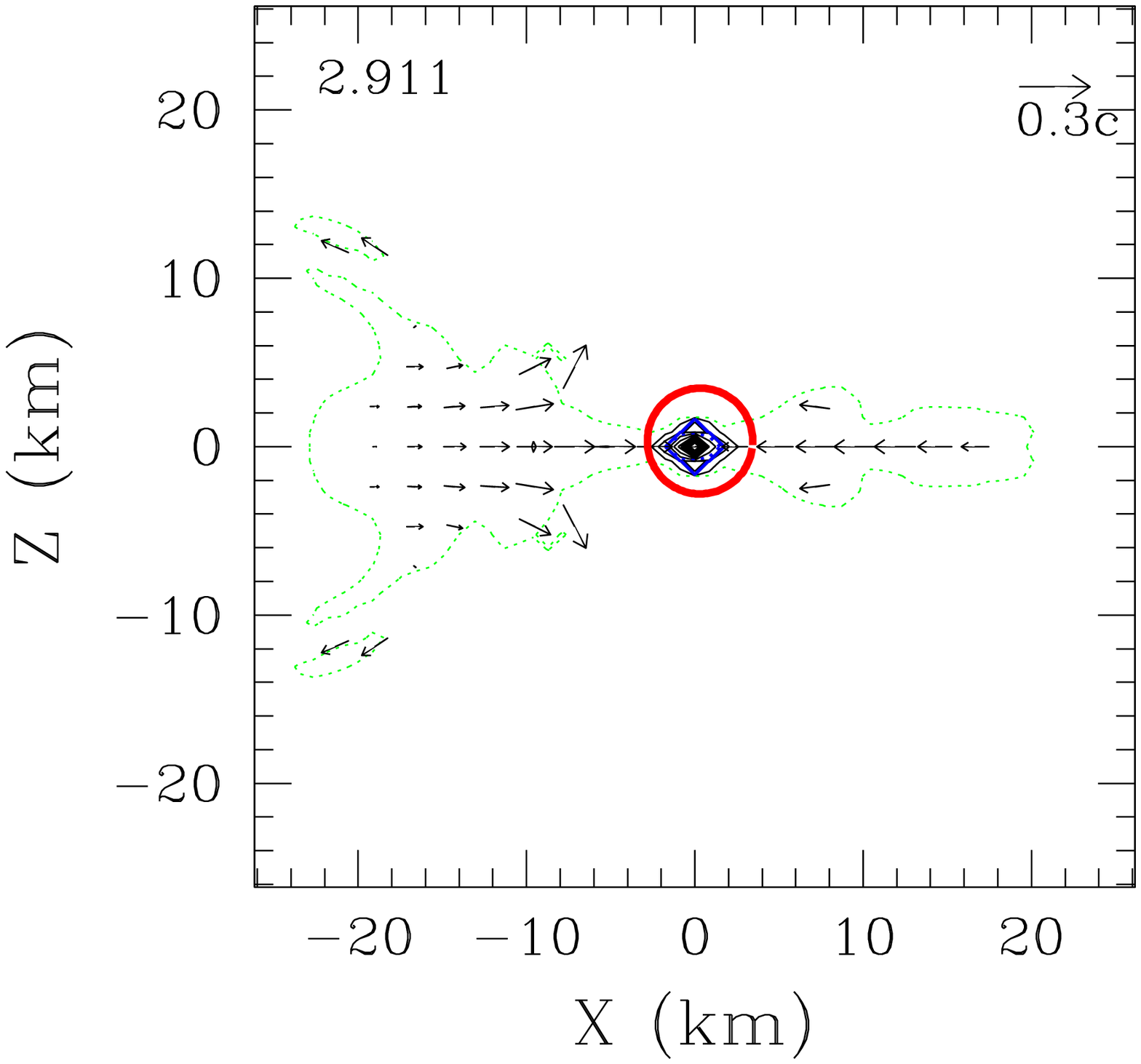}
\epsfxsize=2.2in
\leavevmode
(e)\hspace{-5mm}\epsffile{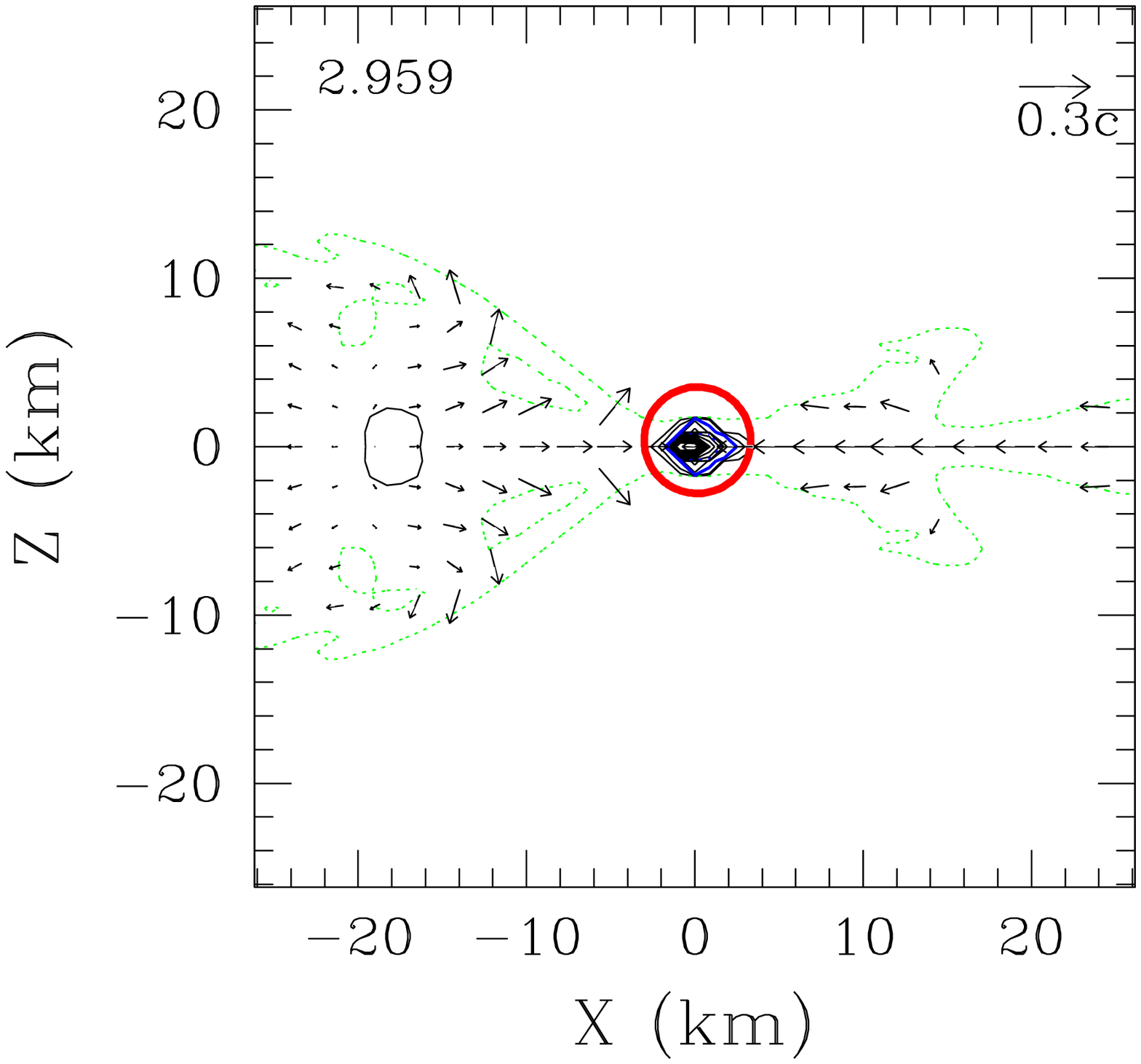}
\epsfxsize=2.2in
\leavevmode
(f)\hspace{-5mm}\epsffile{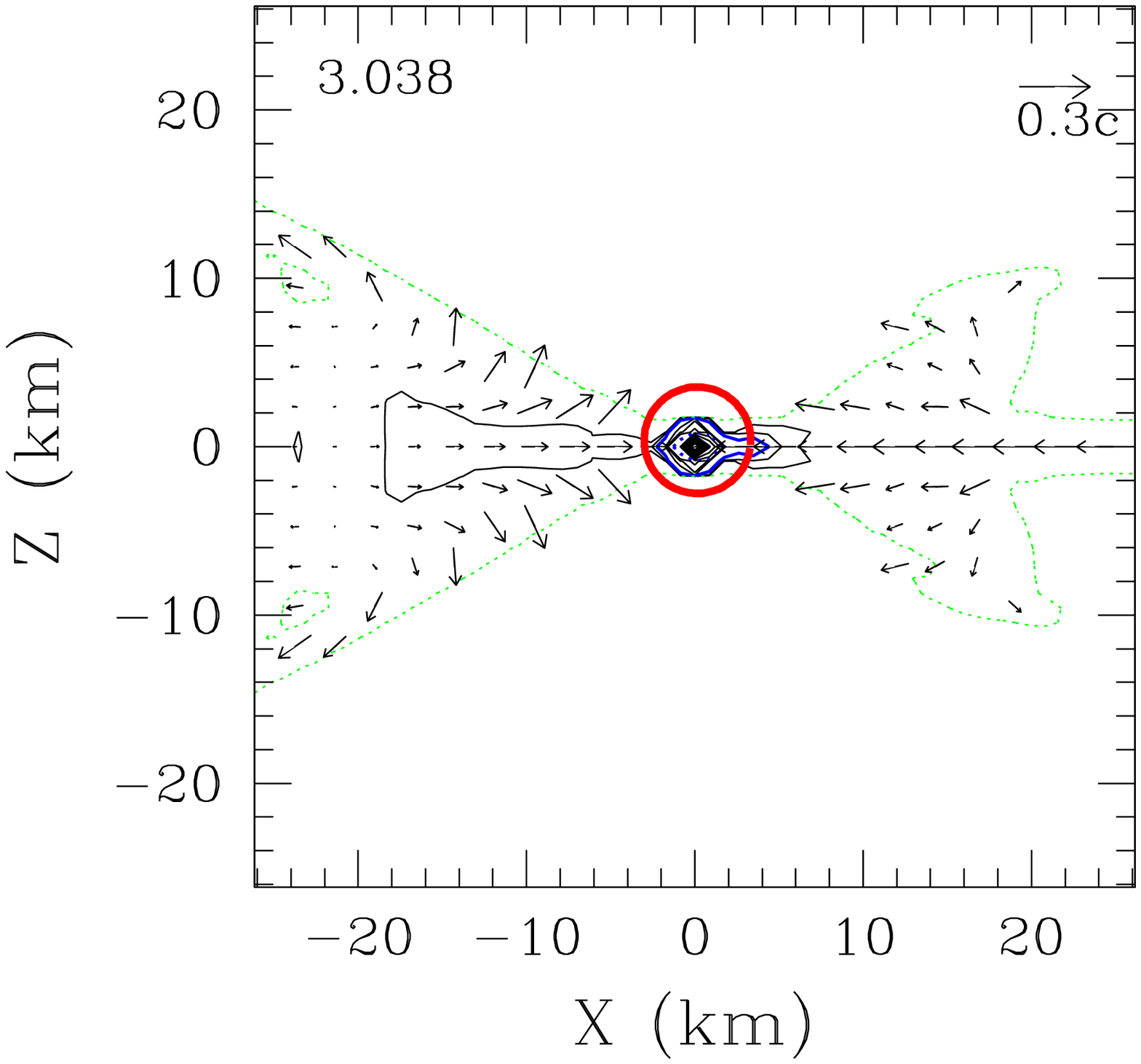}
\end{center}
\vspace{-6mm}
\caption{The density contour curves for $\rho$ and the
local velocity field $(v^x,v^z)$ in the $y=0$ plane
(a) at $t=10.668$ ms for model APR1313, (b) at $t=10.319$ ms for APR1414,
(c) at $t=2.523$ ms for APR1515,  (d) at $t=2.911$ ms for APR1416,
(e) at $t=2.959$ ms for APR135165, and (f) at $t=3.038$ ms for APR1317.
The contour curves and the velocity vectors are drawn in the
same way as in Fig. \ref{FIG6}.
\label{FIG9}}
\end{figure*}

\begin{figure*}[thb]
\vspace{-4mm}
\begin{center}
\epsfxsize=3.in
\leavevmode
\epsffile{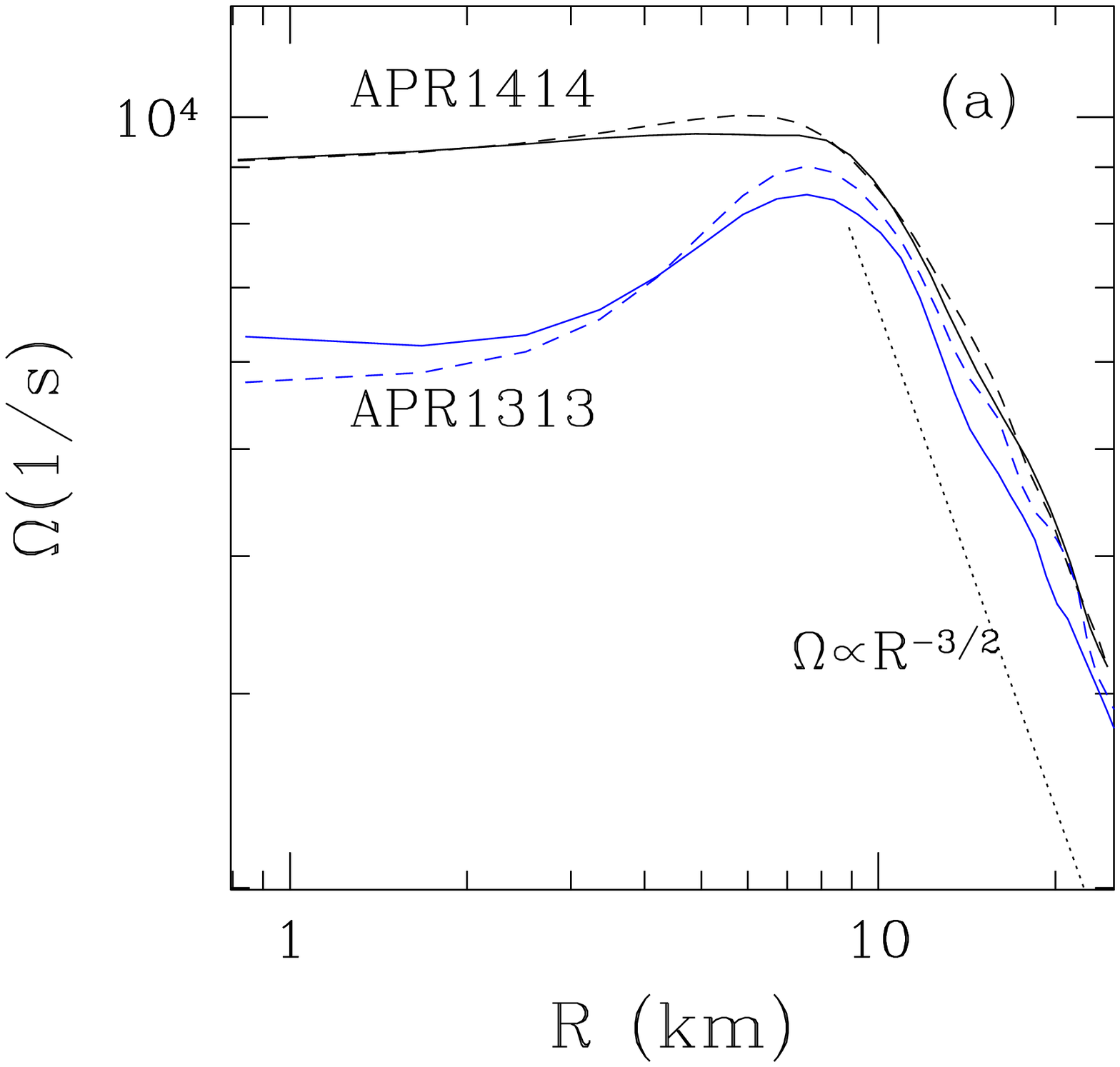}
\epsfxsize=3.in
\leavevmode
\hspace{1cm}\epsffile{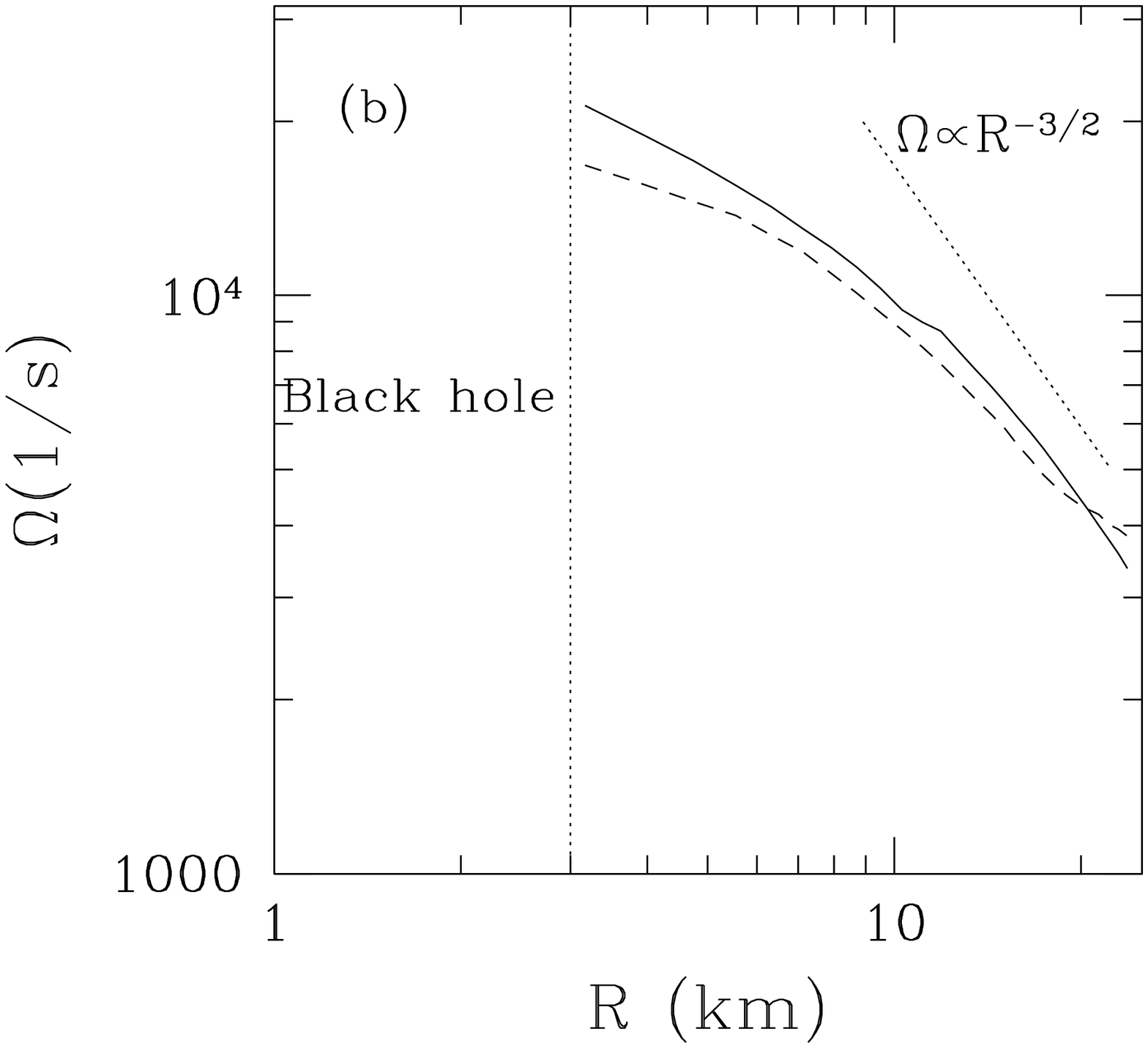}
\end{center}
\vspace{-12mm}
\caption{(a) The angular velocity $\Omega$ of hypermassive neutron stars
along $x$ (solid curves) and $y$ axes (dashed curves)
for model APR1313 (blue) at $t=10.668$ ms and for APR1414 (black)
at $t=10.319$ ms. 
(b) $\Omega$ of accretion disk around a central black hole
for model APR135165 along $x$ (solid curve) and $y$ (dashed curve) axes
at $t=2.959$ ms. The apparent horizon is located at $r \approx 3$ km.
For both figures, the horizontal axis denote the coordinate radius. 
\label{FIG10} }
\end{figure*}

In Figs. \ref{FIG4}--\ref{FIG8}, we display the snapshots of the
density contour curves and the velocity vectors in the equatorial
plane at selected time slices for models APR1313, APR1414, APR1515,
APR1416, and APR1317, respectively. In the first two cases, a
hypermassive neutron star is formed, while a black hole is a prompt
outcome in other three cases. For models APR1515, APR1416, and
APR1317, the ADM mass of the system is approximately identical while
the mass ratio is different. The structure and density of disk
surrounding the formed black hole depend significantly on the mass
ratio as found from Figs. \ref{FIG6}--\ref{FIG8}. Figure \ref{FIG9}
displays the density contour curves and the velocity vectors in the
$y=0$ plane for models APR1313, APR1414, APR1515, APR1416, APR135165,
and APR1317 at a late time when the mass accretion rate relaxes approximately
to a constant. This shows that (i) the hypermassive neutron
stars have a highly flattened structure and (ii) the disk surrounding
the black hole is geometrically thin for the mass ratio close to unity
but can be thick for the smaller mass ratios with $Q_M \alt 0.8$.

Figure \ref{FIG10} shows the angular velocity along the $x$ and $y$
axes (a) of relaxed hypermassive neutron stars for models APR1313 and
APR1414 and (b) of the disk surrounding the black hole for model
APR135165.  This illustrates that the hypermassive neutron stars, in particular
in their outer region, are differentially rotating and the central
part is rapidly rotating with the rotational period shorter than 1 ms. 
The disk surrounding the black hole has approximately a Kepler orbit
outside the ISCO (for which the estimated coordinate radius is $\sim 10$ km).

In the following two subsections, details about the formation process
of hypermassive neutron stars and black holes are discussed
separately. Implication of the results to formation of a central
engine of SGRBs is also discussed in the subsequent section \ref{sec:grb}.

\subsection{Formation of hypermassive neutron star}\label{sec:HMNS}

\subsubsection{Models APR1313 and APR135135}\label{sec:HMNS1}

In the formation of the hypermassive neutron stars from equal-mass
neutron stars with relatively low total mass $\sim
2.6$--2.7$M_{\odot}$ (models APR1313 and APR135135), a double core
structure is first formed (see the snapshot at $t=3.218$ ms of
Fig. \ref{FIG4}), and then, it relaxes to a highly nonaxisymmetric
ellipsoid [see the snapshots for $t > 4$ ms of Fig. \ref{FIG4} and
\ref{FIG9}(a)]. The contour plots, drawn for a high-density region
with the rest-mass density larger than the nuclear density $\sim 2
\times 10^{14}~{\rm g/cm^3}$, show that the axial ratio of the
ellipsoid measured in the equatorial plane is $\alt 0.8$ for $t \agt
4$ ms. Figure \ref{FIG9}(a) also shows that the axial length along the
$z$-axis for $\rho \geq 2 \times 10^{14}~{\rm g/cm^3}$ is about 7 km,
which is $\sim 70\%$ of the major axis. Namely, a rotating ellipsoid
of a large ellipticity is the outcome. This result is essentially the
same as that found in \cite{STU2} for models SLy1313 and
SLy125135. This large ellipticity is achieved due to the rapid
rotation and the high stiffness of the chosen EOS. 

The rapid rotation is found from Fig. \ref{FIG10}(a); it shows that
the rotational period of the central region is shorter than 1 ms at $t
\sim 10$ ms, at which the hypermassive neutron star relaxes
approximately to a stationary state. This indicates that the
rotational centrifugal force plays an important role for sustaining
the large self-gravity of the hypermassive neutron stars 
\cite{BSS,SBS}.

For the unequal-mass case with $Q_M \approx 0.85$ (model APR1214), the
smaller-mass neutron star is tidally elongated at the merger. As a
result, the double-core structure is not formed in contrast to model
APR1313. However, the formed hypermassive neutron star relaxes to an
ellipsoid in the similar way to model APR1313. Subsequent evolution
proceeds in essentially the same manner to that in the equal-mass case.

The merger for model APR135135 proceeds in qualitatively the same way
as that for model APR1313.  Quantitatively, the results such as the
ellipticity and the rotational period of the hypermassive neutron star
are slightly different reflecting the difference in the mass;
ellipticity is slightly smaller and the rotational period is slightly
shorter because of its larger compactness.

Because of the nonaxisymmetric structure, the hypermassive neutron
stars found for models APR1313, APR1214, and APR135135 emit
quasiperiodic gravitational waves and the angular momentum is
dissipated substantially.  However, the dissipation time scale is much
longer than 10 ms, implying that they remain the ellipsoidal star for
longer than 10 ms (cf. Sec. \ref{sec:gw}). These ellipsoids are also
differentially rotating (cf. Fig. \ref{FIG10}). Thus, they are subject
to angular momentum transport by  magnetic effects
\cite{BSS,MAG,DLSSS,DLSSS2}. However, note that the time scale of the
angular momentum transport is shorter than $\sim 100$ ms only when very
strong magnetic fields greater than $\sim 10^{15}$ G are present
\cite{BSS}.

Assume that the dissipation time scale by gravitational waves is
shortest among other processes.  Then, there are two possible fates
after a longterm emission of gravitational waves. One fate is that
after the angular momentum dissipation, the centrifugal force becomes
weak enough to induce gravitational collapse to a black hole. The
other is that the angular momentum (i.e., rotational kinetic energy)
is dissipated to be too small to maintain the nonaxisymmetric
structure and, consequently, a spheroidal star in a stationary state
is formed. (It is well known that an elliptical structure can be
achieved only in the case that the ratio of the rotational kinetic
energy to the gravitational binding energy is large enough
\cite{Chandra69}.) If the time scale for the decrease of the angular
momentum is shorter (longer) than that for the decrease of the
ellipticity, a black hole (a spheroidal hypermassive neutron star) is
the outcome. In the formation of a spheroid, subsequent evolution will be 
determined by other processes such as angular momentum transport by
the magnetic braking \cite{BSS} or the magnetorotational instability
(MRI) \cite{MAG}. In the previous paper \cite{STU2} in which the SLy
and FPS EOSs are adopted, we have found that a black hole is the
outcome (see the results for models SLy135135b and FPS125125b in
\cite{STU2}).  As the merger process depends weakly on the EOSs, the
black hole formation may be also the fate in the APR EOS. However,
this may not hold as illustrated in the next section for high-mass
model APR1414. To answer this question, an extremely longterm
simulation up to $t \sim 100$ ms is necessary. This problem is left as
the future issue. 

\subsubsection{Model APR1414}

In the formation of the hypermassive neutron stars with high mass
$\sim 2.8M_{\odot}$ (model APR1414), the evolution proceeds in a
different manner from that for models APR1313 and APR135135. Since the
mass is larger, a more compact merged object is formed soon after the
onset of the merger (second panel of Fig. \ref{FIG5}). Then, it
bounces back to a state with a large radius (or a small density;
cf. Fig. 3) and repeats quasiradial oscillations with a large
amplitude before relaxing to a hypermassive neutron star in a
quasistationary state.  At the state of the large radius, the angular
momentum is efficiently transported from the inner region to the outer
envelop (see discussion below). On the other hand, the compact state
achieved at maximum compressions (cf. second and fourth panels of Fig. 
\ref{FIG5}) results in a large amount of angular momentum dissipation
via gravitational radiation (cf. Sec \ref{sec:gw}). These effects
subsequently lead the hypermassive neutron star to a state with
relatively small angular momentum. 

The decrease of the angular momentum in a shorter time results in a
state of small rotational kinetic energy. On the other hand, the
gravitational binding energy is larger than in models APR1313 and
APR135135, because of its larger compactness.  As mentioned in the
last part of Sec. \ref{sec:HMNS1}, an elliptical structure can be
achieved only in the case that the ratio of the rotational kinetic
energy to the gravitational binding energy is large enough
\cite{Chandra69}. For the hypermassive neutron star formed in model
APR1414, this ratio should be smaller than those in models APR1313 and
APR135135. Consequently, a spheroid of small ellipticity is formed
[cf. the last three panels of Fig. \ref{FIG5} and \ref{FIG9}(b)].
This results in the facts that luminosity of gravitational waves is
much smaller than for model APR1313 for $t \agt 5$ ms
(cf. Sec. \ref{sec:gw}), and the dissipation time scale of the angular
momentum by gravitational radiation is much longer. In model APR1414,
probably, dissipation by gravitational radiation will not induce
gravitational collapse.

As in the hypermassive neutron star for model APR1313, the outer
region with the coordinate radius $R \agt 8$ km is differentially
rotating [see Fig. \ref{FIG10}(a)].  Such differentially rotating
region is subject to the magnetic braking and the MRI in the presence
of magnetic fields \cite{MAG,DLSSS,DLSSS2}. If the instabilities turn
on, the magnetic fields play an important role for transporting 
angular momentum outward, subtracting the angular momentum of the
central part. Hence, a plausible scenario for the hypermassive
neutron star of model APR1414 is the collapse to a black hole by the
magnetic effects. 

Finally, we note that in the previous paper \cite{STU2} in which the
SLy and FPS EOSs are used, we have not found the formation of
spheroids, but rather, gravitational radiation triggers the collapse
of hypermassive neutron stars to a black hole. The likely reason in
this difference is that the APR EOS is stiffer than others. Namely,
the hypermassive neutron stars can be more compact with this EOS
[cf. Fig. \ref{FIG2}(b)] escaping the collapse. In such extremely
compact state, the gravitational binding energy would be large enough
to reduce the ratio of the rotational kinetic energy to the binding
energy below the threshold value of the formation of an ellipsoid.

\subsubsection{Angular momentum transport and disk mass}
\label{sec:angmom_trans}

\begin{figure*}[thb]
\vspace{-4mm}
\begin{center}
\epsfxsize=3.in
\leavevmode
\epsffile{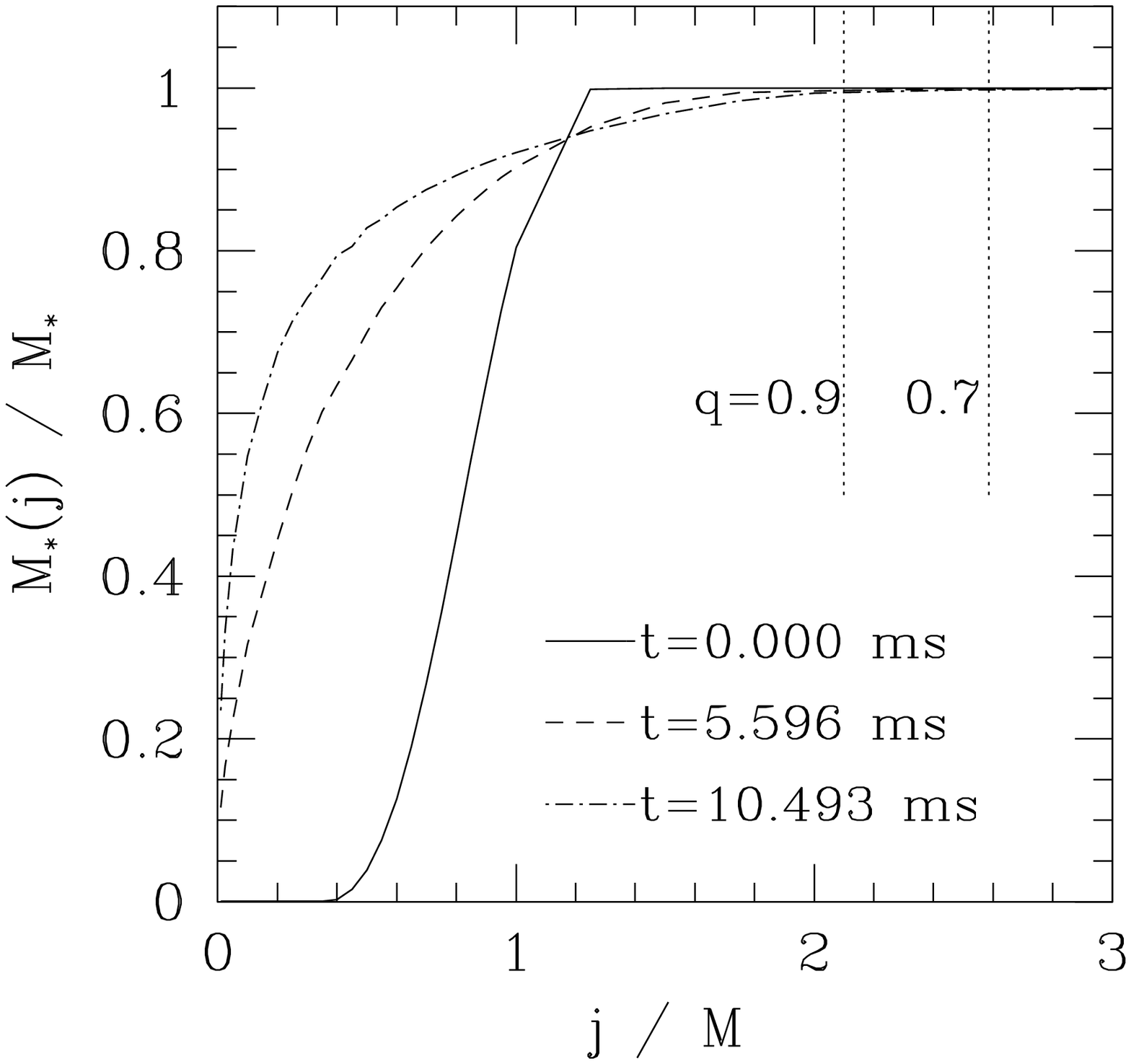}
\epsfxsize=3.in
\leavevmode
\hspace{1cm}\epsffile{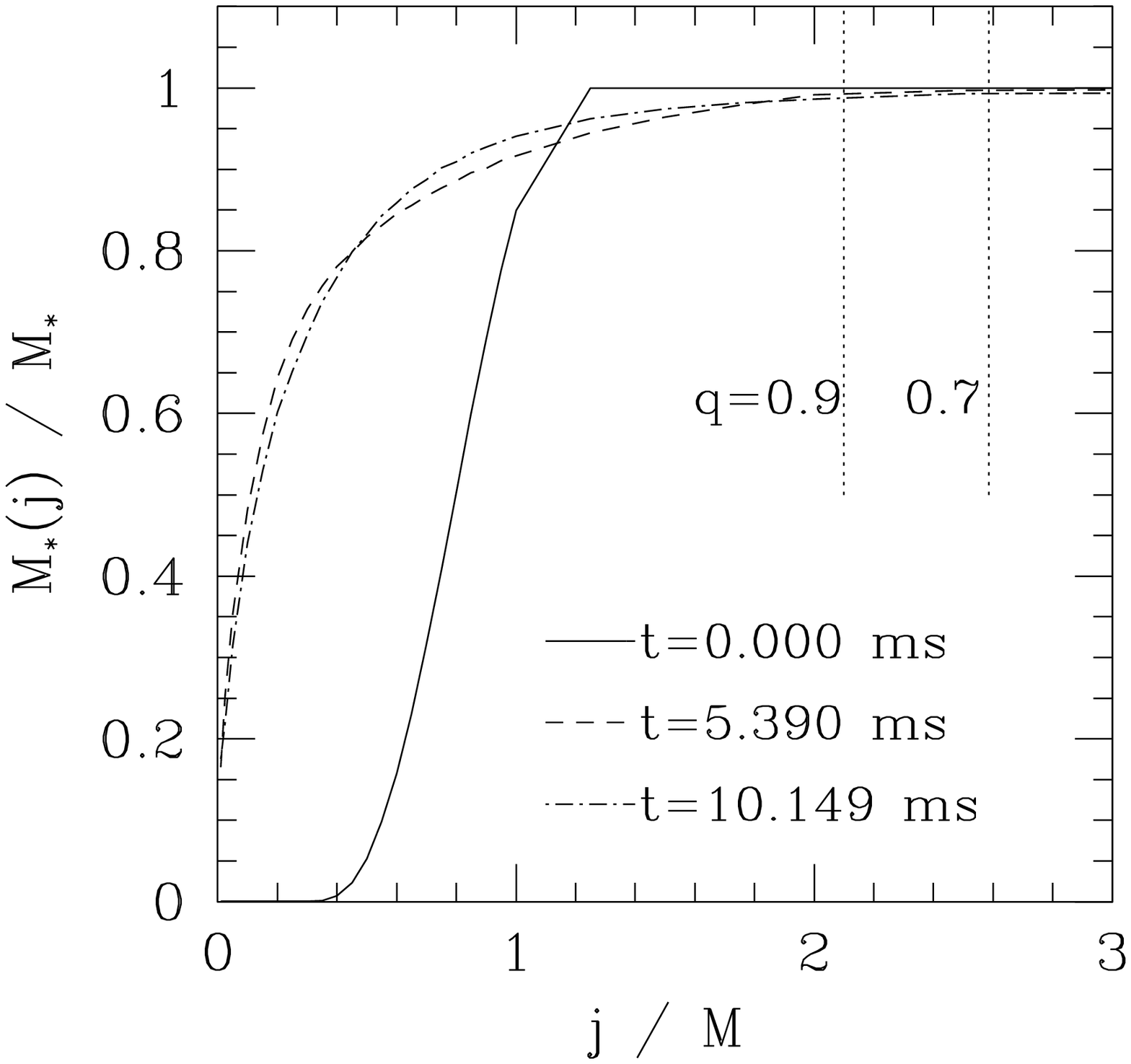}
\end{center}
\vspace{-12mm}
\caption{Evolution of $M_*(j)$ for models APR1313 (left) and APR1414
(right).  The dotted vertical lines labeled $q=0.9$ and 0.7 denote the
values of $j/M$ at ISCOs around a Kerr black hole of the spin
parameter $q$.
\label{FIG11} }
\end{figure*}

Due to a torque by the nonaxisymmetric structure of the merged object,
angular momentum is transported from the inner region to the outer
one. As a result, the rotational angular velocity $\Omega=v^{\varphi}$
near the center decreases and its profile is modified. Figure
\ref{FIG10} shows the angular velocity $\Omega$ of the hypermassive
neutron stars along $x$ and $y$ axes at $t \agt 10$ ms.  The
hypermassive neutron stars are differentially and rapidly rotating at
the birth \cite{STU2}; the rotational period around the central region
is $\approx 0.2$--0.3 ms at the birth. (The rotational period of the
double core found in Figs. \ref{FIG4} and \ref{FIG5}, which is formed
soon after the merger, is $\sim 0.2$--0.3 ms. This is found from the
frequency of gravitational waves; cf. Sec. \ref{sec:gw}).  The angular 
momentum is then redistributed by the torque, resulting in a fairly 
uniform profile of $\Omega$ near the rotational axis with $R \alt 8$ km.

The significance of the angular momentum transport is also found from
the evolution of the mass spectrum as a function of the specific
angular momentum, $M_*(j)$.  Here, the specific angular momentum $j$
is defined by $h u_{\varphi}$ and the mass spectrum $M_*(j_0)$ is
given by an integrated baryon rest-mass of fluid elements with $j \leq
j_0$;
\beq
M_*(j_0)=\int_{j<j_0} \rho_* d^3x.
\eeq
In Fig. \ref{FIG11}, we show this mass spectrum at selected time
slices for models APR1313 and APR1414. This illustrates that the
fraction of baryon rest-mass with $j \agt 1.4M_0$ is absent at $t=0$ but
increases with time. 
This is due to the angular momentum transport from the central
hypermassive neutron star of ellipsoidal shape to fluid elements in
the outer envelop by the torque associated with the nonaxisymmetric
structure. However, the fraction of the rest-mass with $j > 2M_0$ is
still $\sim 1\%$ of the total mass because such fluid elements are
absent initially, and also probably because of gravitational radiation
which carries away the angular momentum by $\sim 30\%$ in 10 ms
(cf. Sec. \ref{sec:gw}).

Assume a hypothetical case in which the central region collapses to
form a black hole at $t \approx 10$ ms. Then, the ADM mass and the
angular momentum of the black hole will be $\approx 0.97M_0$ and
$\approx 0.7J_0$, respectively, implying that the spin parameter is $q
\sim 0.7$ (cf. Sec. \ref{sec:gw}).  The specific angular momentum at
the ISCO, $j_{\rm isco}$, around a black hole with $q=0.7$ is $\approx
2.58M$ \cite{BPT} as plotted by the dotted vertical line in
Fig. \ref{FIG11}.  The mass of the fluid elements with $j \geq 2.5M$
is $\approx 0.01M_{\odot}$ at $t=10$ ms for model APR1313. Namely, a
disk of mass $\sim 0.01M_{\odot}$ will be formed in this
hypothesis. The value for the disk mass is approximately identical
with that presented in \cite{OJ} in which a similar nuclear EOS
\cite{shen} is adopted.

We note that the lifetime of the hypermassive neutron star is in
reality much longer than 10 ms (assuming no other strong dissipation
process than gravitational radiation), and hence, the angular momentum
will be dissipated much more by gravitational radiation before
formation of a black hole. Therefore, the value of $q$ will be smaller
and $j_{\rm isco}$ could be larger (e.g., for $q=0.5$, $j_{\rm isco}
\approx 2.90M$). This effect will slightly reduce the disk mass. On
the other hand, the angular momentum will be gradually transported in
the outer region with time due to the torque from the central region
of a nonaxisymmetric shape. This effect will slightly increase the
disk mass.

For model APR1414, the fraction of the rest-mass with $j \geq 2.5M$ is
slightly larger as $\approx 0.02M_{\odot}$ at $t \approx 10$ ms. The
reason is that in this case, the merged object quasiradially
oscillates with a larger amplitude soon after the onset of the merger
(cf. Fig. \ref{FIG3}).  During the high-amplitude oscillation, matter
in the outer region expands to a large radius and can gain a torque
from the ellipsoidal hypermassive neutron star for a longer
time. Actually, the fraction of the rest-mass with $j \geq 2.5M$
steeply increases for 4 ms $\alt t \alt$ 6 ms during which the
hypermassive neutron star repeats quasiradial oscillations of the high
amplitude. The high-amplitude oscillation results from the fact that
the ADM mass of the system is close to $M_{\rm thr}$. Hence, for the
system of $M \alt M_{\rm thr}$, the angular momentum transport
efficiently works.

The outcome for model APR1414 at $t \approx 10$ ms is a hypermassive
neutron star of nearly spheroidal shape. This implies that the angular
momentum will not be significantly dissipated any longer and also
outward angular momentum transport will not work efficiently for $t >
10$ ms. Therefore, in this case, the disk of mass at least $\approx
0.02M_{\odot}$ will be formed if the hypermassive neutron star
collapses to a black hole by other mechanisms.

For model APR1214, for which the ADM mass is approximately identical
with that for model APR1313, the less massive star is tidally
elongated by the primary at the merger and subsequently forms an
accretion disk around a formed hypermassive neutron star (evolution of
the contour curves is similar to those of model APR1416 for $t \alt 2$
ms; cf. Fig. \ref{FIG7}).  Because of the nonaxisymmetric density
profile, the angular momentum is transported outward more efficiently
than in model APR1313.  However, the baryon rest-mass with $j > 2.5M$
for model APR1214 at $t=10$ ms is $\approx 0.025M_{\odot}$ which is
only slightly larger than that for model APR1313, and hence, the
degree of the increase is not as drastic as that reported in \cite{OJ}
for $Q_M \sim 0.85$. The likely reasons for our small mass are (i) the
radius of neutron stars depends very weakly on each ADM mass in the
APR EOS (cf. Fig. \ref{FIG2}) and (ii) gravitational radiation carries
a large amount of angular momentum (cf. Sec. \ref{sec:gw}) which is
not taken into account in the work of \cite{OJ}. Because of the weak
dependence of the stellar radius on the mass, the tidal deformation
becomes important only for close orbits. At such small orbital radius,
the angular momentum is already dissipated by gravitational radiation
to be small, and hence, the angular momentum transport by a torque
associated with the nonaxisymmetric structure does not help
sufficiently increasing the mass with a large value of $j$.

Formation of hypermassive neutron stars from binaries of small values
of $Q_M \alt 0.8$ is unlikely since the low-mass binary neutron star
with such small value of $Q_M$ would be absent; i.e., the mass of
neutron stars should be larger than $\sim 1.2M_{\odot}$ while the
total mass has to be smaller than $M_{\rm thr}$, implying that $Q_M$
should be larger than $\sim 0.8$.  Therefore, the fraction of the mass
with $j \agt 2.5M$ will be at most $\sim 0.03M_{\odot}\approx 0.01M_*$
for the hypermassive neutron star formation case in the APR EOS. 

For model SLy1313 in which a hypermassive neutron star is also formed,
the merger process and subsequent evolution of the hypermassive
neutron star is quite similar to those for models APR1313 and
APR135135 \cite{STU2}. As a consequence, the rest-mass with $j > 2.5M$
at $t=10$ ms is the similar value $\sim 0.02M_{\odot}$. This suggests that
the value of the disk mass depends weakly on the EOS. However, we note
that the disk mass could be larger if we employ EOSs with which 
the neutron star radius is larger than that of the APR and SLy EOSs,
and hence, the gravitational radiation reaction at the merger is less
important. 

Before closing this section, we note that the discussion about the
disk mass here is based on the assumption that angular momentum is
transported only by a torque in the hypermassive neutron stars. In the
presence of magnetic fields with a sufficient strength, the angular
momentum could be transported efficiently in a short time scale
\cite{DLSSS,DLSSS2}, and as a result, the hypermassive neutron stars
such as formed for models APR1313, APR1414, APR1214, and SLy1313 may
collapse ejecting a large amount of the mass to form a disk. The
system eventually formed could be composed of a black hole and a
massive disk of $\agt 0.05M_{\odot}$ as illustrated in the recent
magnetohydrodynamic simulations \cite{DLSSS,DLSSS2}. 

\subsection{Formation of black hole}\label{sec:BH}

For binary neutron stars with $M \approx 2.96M_{\odot}$ in the APR EOS
and $M \approx 2.76M_{\odot}$ in the SLy EOS, a black hole is formed
soon after the onset of the merger irrespective of the mass ratio
$Q_M$. Until two stars come into contact, evolution proceeds in the
similar manner to that in the formation of hypermassive neutron stars.
However, after the contact, the merged objects quickly collapse to a
black hole. For the unequal-mass case, a less massive star is tidally
deformed just before the merger. The degree of the deformation as well
as lag angle, which is defined to be the angle in the equatorial plane
between the major axis of each star and the axis connecting the
centers of mass of two stars, are larger for smaller values of
$Q_M$. Because of the nonzero lag angle, angular momentum is
transported outward by a torque from the nonaxisymmetric merged
object, increasing the specific angular momentum of the matter located
in the outer region. Only in the case that the angular momentum transport works
efficiently, an accretion disk may be formed around the central
black hole since the specific angular momentum before the onset of the
merger is too small as illustrated in Sec. \ref{sec:angmom_trans} 
(see also a discussion below). 

\subsubsection{Disk mass around the central black hole}\label{sec:BHdisk}

\begin{figure}[thb]
\vspace{0mm}
\begin{center}
\epsfxsize=3.2in
\leavevmode
\epsffile{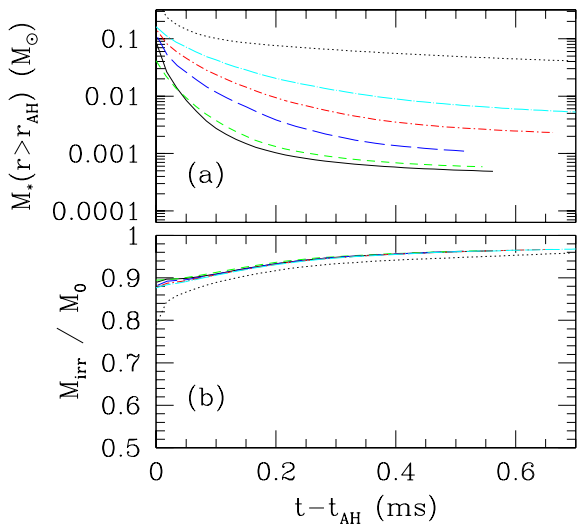}
\end{center}
\vspace{-12mm}
\caption{(a) Evolution of the baryon rest-mass of disks surrounding black
holes and (b) the evolution of irreducible mass of black holes for 
models APR1515 (solid curve), APR145155 (dashed curve), APR1416 
(long-dashed curve), APR135165 (dot-dashed curve), APR1317 
(dot-long-dashed curve), and SLy125155 (dotted curve).
$t_{\rm AH}$ denotes the time at the first formation of an apparent horizon. 
\label{FIG12} }
\end{figure}

\begin{figure}[thb]
\vspace{0mm}
\begin{center}
\epsfxsize=3.in
\leavevmode
\epsffile{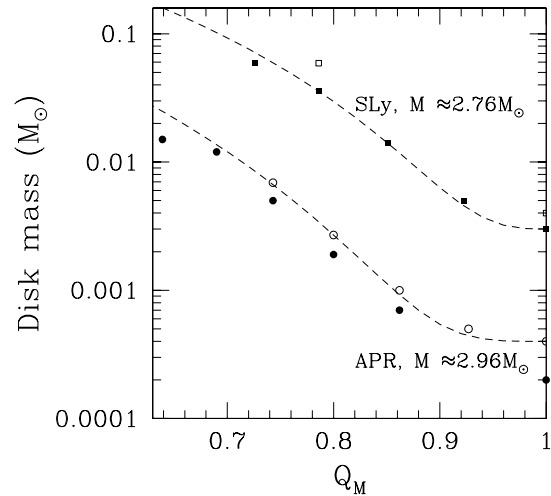}
\end{center}
\vspace{-10mm}
\caption{Baryon rest-mass of disks around a black hole as a function
of $Q_M=M_{*2}/M_{*1}$ for the black hole formation cases.  The disk mass is
evaluated at $t - t_{\rm AH}=0.5$ ms.  The open circles and squares
denote the results with (633, 633, 317) grid size
in the APR and SLy EOSs, respectively. The filled circles and squares 
denote the results with (377, 377, 189) grid size in the APR and SLy EOSs. 
The dashed curves denote the fitting formulae (\ref{diskmass}).
The results for the APR and SLy EOSs are derived for $M \approx 2.96M_{\odot}$
and $2.76M_{\odot}$, respectively. 
\label{FIG13}}
\end{figure}

In Fig. \ref{FIG12}(a), we plot the total baryon rest-mass located
outside the apparent horizon for models APR1515, APR145155, APR1416,
APR135165, APR1317, and SLy125155 as a function of $t-t_{\rm AH}$
where $t_{\rm AH}$ denotes the time at the first formation of an
apparent horizon.  In Fig. \ref{FIG13}, the baryon rest-mass at
$t-t_{\rm AH}=0.5$ ms as a function of $Q_M$ is also plotted for all
the models in which a black hole is formed. These figures show that
the final state is composed of a central black hole and a disk of a
small mass. A rapid accretion proceeds in the first $\sim 0.2$ ms
after formation of apparent horizon. In particular, for $Q_M \agt
0.9$, most of the fluid elements are swallowed into the black hole in
such a short time.  With decreasing the value of $Q_M$, the fraction of
the fluid elements which escape falling into the black hole 
steeply increases.

We note that for the smaller grid size, the disk mass is smaller even
for the same model (cf. Table III). As mentioned in Sec. \ref{sec:init},
the inspiral time is spuriously shorten with the smaller grid size 
since the radiation reaction is overestimated due to the situation
with $L \ll \lambda_0$. As a result, the time for transporting angular momentum
outward is shorten, resulting in the decrease of the disk mass. 

For $Q_M \agt 0.9$, the baryon rest-mass of the disk becomes smaller
than $0.01M_{\odot}$ within $\sim 0.3$ ms after formation of apparent
horizon ($\approx 5 \times 10^{-4}M_{\odot}$ for models APR1515 and
APR145155, and $\approx 5 \times 10^{-3}M_{\odot}$ for models SLy1414 and
SLy135145).  The fundamental reason is that the specific angular
momentum $j$ for all the fluid elements before the merger is too
small. In the present case, the maximum value of $j$ in the initial
conditions is $\sim 1.3M$, which is much smaller than the value of $j$
at the ISCO of a Kerr black hole of mass
$M$ and spin parameter 0.7--0.9.  Moreover, the specific angular
momentum decreases due to emission of gravitational waves. Hence, a
disk can be formed only in the case (i) an efficient mechanism for
angular momentum transport which works within $\sim 1$ ms after the
onset of the merger (before collapsing to a black hole) is present and
(ii) the transport mechanism works efficiently enough for overcoming
the dissipation by gravitational radiation in the outer region. For
$Q_M \sim 1$, angular momentum transport due to a torque by the
nonaxisymmetric central object does not work efficiently, resulting in
the small disk mass. 

We note that in the prompt formation of a black hole, the time for the
angular momentum transport is much shorter than in the formation of a
hypermassive neutron star. This is the main reason that the disk mass
is much smaller than that in Sec. \ref{sec:angmom_trans} for nearly
equal-mass binaries.

For $Q_M \alt 0.9$, the disk mass increases with decreasing the value
of $Q_M$ because the less massive neutron star is tidally elongated at the
merger, and hence, efficiency of the angular momentum transport is
enhanced.  However, the disk mass does not exceed $0.01 M_{\odot}$ even
for $Q_M =0.75$ for the APR EOS. The likely reason is that the stellar
radius is small irrespective of the mass of each star in this EOS. Due
to this property, the merged object is compact at the onset of
merger enough to promptly produce a black hole without 
transporting angular momentum efficiently. 

The disk mass is much larger in the SLy EOS for a given value of $Q_M$
than in the APR EOS. The reasons are (i) the mass ($M \approx
2.76M_{\odot}$) is smaller than for the APR EOS case ($M \approx
2.96M_{\odot}$) and (ii) with the SLy EOS, the stellar radius of
relatively small mass $M \alt 1.3M_{\odot}$ is larger 
(cf. Fig. \ref{FIG2}). With the smaller mass, the gravitational
radiation reaction time scale is longer. Consequently, the less
massive star in the binary is tidally elongated by a larger degree,
slightly postponing the collapse of the merged object to a black hole
and enhancing efficiency of the angular momentum transport.


It is interesting to note that the disk mass decreases during the
growth of a black hole in the following mechanism: In the merger of
unequal-mass binary, the less massive neutron star is always tidally
elongated. The inner part first collides the companion, but slips
through the companion's surface to form a small spiral arm around a
black hole which is subsequently born.  Outer part of larger angular
momentum also forms a large spiral arm. The phase difference between
these two spiral arms is about $180^{\circ}$ at their formation
(cf. Figs. \ref{FIG6}--\ref{FIG8}).  However, since the smaller spiral
arm rotates faster than the larger one, the two spiral arms collide each
other to generate shocks in one rotational period, which convert a large
fraction of the kinetic energy to the thermal energy. As a result, an
amount of matter is swallowed into the black hole quickly. The
collision of the spiral arms also leads to geometrical thickening of
the disk [cf. Fig. \ref{FIG9}(e) and (f)] as well as to the heat-up. 

For $M \approx 2.96M_{\odot}$ in the APR EOS and $M \approx 2.76M_{\odot}$
in the SLy EOS, the numerical results of the disk mass $M_d$ as a
function of $Q_M$ for $0.75 \alt Q_M \leq 1$ is approximately fitted by 
\beqn
M_d = M_{d0}+M_{d1}(1-Q_M)^p, \label{diskmass}
\eeqn
where $M_{d0}=0.0004M_{\odot}$, $M_{d1}\approx 1.44M_{\odot}$, and
$p=4$ for $M \approx 2.96M_{\odot}$ in the APR EOS with (633, 633,
317) grid size, and $M_{d0}=0.003M_{\odot}$, $M_{d1}\approx
3.33M_{\odot}$, and $p=3$ for $M \approx 2.76M_{\odot}$ in the SLy EOS
with (377, 377, 189) grid size. Since $M_{d0}$ is small, $M_d$
is approximately proportional to $(1-Q_M)^p$ for $Q_M \alt 0.8$, and hence,
the value of $M_{d1}$ essentially determines the relation. 
As mentioned previously, the value of $M_{d}$ varies by $\sim
30\%$ depending on the grid size by which the radiation reaction time
scale and resulting transport time of the angular momentum spuriously
change. 

As indicated by this fitting formulae (\ref{diskmass}), the disk mass
steeply increases with decreasing the value of $Q_M$
(cf. Fig. \ref{FIG13}) , although for $Q_M \agt 0.9$, the disk mass is
much smaller than $0.01M_{\odot}$. For $Q_M \alt 0.7$, the increase
rate of $M_d$ with decreasing value of $Q_M$ is not as steep as
Eq. (\ref{diskmass}), but it is still positive.

We note that the disk mass obtained here for the prompt black hole
formation case gives an approximate upper limit for both the APR and SLy
EOSs, since the ADM mass adopted is close to
$M_{\rm thr}$. For more massive case, the system at the merger 
becomes more compact and has a shorter dynamical time scale.
Therefore, outward transport of angular momentum works for a shorter
time scale, resulting in smaller disk mass.

Figure \ref{FIG10}(b) shows the angular velocity of the disk for model
APR135165 and illustrates that the disk approximately has the Kepler
angular velocity for the radius larger than $\sim 10$ km which is
approximately the radius of the ISCO around the black hole. The
angular velocity of the inner part with $R \alt 10$ km is not as large
as the Kepler one, indicating that matter in such region gradually
falls into the central black hole.  This is also seen from the
velocity fields in Fig. \ref{FIG9}(d)--(f).  In reality, the matter in
the accretion disk of the Kepler angular velocity will subsequently
evolve in a dissipation time scale, which is determined either by
viscosity or by magnetic fields.

For $Q_M \agt 0.9$, not only the disk mass is small but also the disk
is geometrically thin.  These properties are completely unfavored for
producing high-Lorentz factor jets of GRBs \cite{GRB}. Therefore,
prompt formation of a black hole in the merger of binary neutron stars
of the mass ratio $Q_M \agt 0.9$ is unlikely to be a scenario for
producing a central engine of SGRBs. On the other hand, for a
sufficiently small value of $Q_M$ ($Q_M \alt 0.75$ for the APR EOS and
$Q_M \alt 0.85$ for the SLy EOS), the disk mass will be larger than
$\sim 0.01M_{\odot}$. In addition, the disk is geometrically thick as
indicated in Fig. \ref{FIG9}(f) for model APR1317; i.e., a torus is
formed. Furthermore, the thermal energy, which is generated in the
shock heating between spiral arms in the disk and is evaluated from
$\varep-\varep_{\rm cold}$ \cite{STU2}, is high enough (typically
$\sim 1$--$2 \times 10^{11}$ K) for generating a large amount of
thermal neutrinos \cite{RJ0,RJ}. Viscous heating will also play a role
for subsequently keeping such high temperature. Implication of such
hot torus with mass $\agt 0.01M_{\odot}$ will be given in Sec. \ref{sec:grb}.

\subsubsection{Properties of black hole}

The area of the apparent horizon $A_{\rm AH}$ is determined in the black
hole formation cases. From the area $A_{\rm AH}$, the irreducible mass may 
be approximately defined by 
\beq
M_{\rm irr} \approx \sqrt{{A_{\rm AH} \over 16 \pi}}, 
\eeq
which varies from $0.9 M_0$ to $0.97 M_0$ for all the models in the
APR EOS as shown in Fig. \ref{FIG12}(b).  Since most of the fluid
elements are swallowed into the black hole and, also, the energy
carried out by gravitational radiation is $\approx 0.01M_0$ (see
Sec. \ref{sec:gw}), the mass of the formed black hole should be
approximately $\sim 0.99M_0$.  Assuming that the area of the apparent
horizon is equal to that of the event horizon, the non-dimensional spin
parameter of the black hole is defined by $q\equiv J_{\rm BH}/M_{\rm
BH}^2$ where $J_{\rm BH}$ and $M_{\rm BH}$ are the angular momentum and
the ADM mass of the black hole. 
Then the irreducible mass is defined in terms of $q$ and $M_{\rm
BH}$ by
\beqn M_{\rm irr}=\sqrt{{1+(1-q^2)^{1/2} \over 2}} M_{\rm BH}. 
\label{aaa}
\eeqn
Equation (\ref{aaa}) implies that for $M_{\rm irr}/M_0=0.9$--0.97,
$q=0.8$--0.4. This indicates that the small error in the estimation
for $M_{\rm irr}$ leads to a large error in $q$ estimated from the
area of the apparent horizon.
To reduce the error, the simulation should be performed with a better grid
resolution. For this purpose, an adaptive mesh refinement technique
will be helpful \cite{AMR}, but the simulation with such technique is
beyond the scope of this paper.

The value of $q$ is also approximately determined from the following
manner. As shown in Sec. \ref{sec:gw}, the angular momentum is
dissipated by $\sim 15$\% by gravitational radiation, while the ADM
mass decreases by $\sim 1\%$ throughout the simulation.  As listed in
Table II, the initial value of $q$ is $\approx 0.9$.  Therefore, the
value of $q$ in the final stage should be $\approx 0.75$ for which
$M_{\rm irr} \approx 0.91M_{\rm BH}$. Assuming that this is the
correct value, the error in the irreducible mass of the black hole
determined from the area of the apparent horizon is within $\sim 5\%$.

Given the spin parameter $q=0.7$--0.8 and the ADM mass $M_{\rm BH}
\approx 2.8M_{\odot}$, the frequency of the fundamental quasinormal
mode is about 6.5--7$(2.8M_{\odot}/M_{\rm BH})$ kHz \cite{Leaver}.  We
will show the ring-down gravitational waves with this frequency in
Sec. \ref{sec:gw}.

\subsection{Implication to SGRBs}\label{sec:grb}

As shown in sections \ref{sec:HMNS} and \ref{sec:BH} (also illustrated
in previous papers \cite{STU2,DLSSS2}), merger of binary neutron stars
can produce a system composed of a Kerr black hole and a hot torus of mass
$\agt 0.01M_{\odot}$ if certain condition is satisfied. To summarize,
there are following two scenarios to achieve the formation of a
massive disk with $M_d \geq 0.01M_{\odot}$: (i) In the first 
scenario, a hypermassive neutron star is required to be formed
first. Then, by some mechanisms, the angular momentum is transported 
from the inner region to the outer envelop which subsequently forms a
hot disk (torus), while the hypermassive neutron star eventually
collapses to a black hole either by angular momentum dissipation due to 
gravitational radiation or by angular momentum transport probably due to 
magnetic effects \cite{BSS}. In this scenario, the estimated torus mass will
be $\sim 0.01$--$0.03M_{\odot}$ in the absence of strong magnetic
effects, while in its presence, the mass can be $\agt 0.05M_{\odot}$
\cite{DLSSS,DLSSS2}.  (ii) In the second scenario, a black hole is
promptly formed after the onset of merger in an unequal-mass binary of
sufficiently small mass ratio $Q_M$. Then, a torus is formed from the
less massive neutron star which is tidally elongated at the merger and
subsequently constitutes a hot and geometrically thick torus. In this
case, the torus mass depends strongly on the value of $Q_M$ and
nuclear EOSs.

In \cite{DLSSS2}, a scenario through formation of a hypermassive
neutron star with strong magnetic fields is described. In this case, a
large accretion rate from the torus to the black hole with $\dot M
\agt 5M_{\odot}$/s is expected. In the rest of this section, we focus
mainly on other scenarios with no strong magnetic fields; in
particular, we focus on scenario (ii).

In the following, we assume that the disk mass is $M_d \sim
0.01$--$0.1M_{\odot}$ and mass accretion rate is $\dot M \sim
0.1$--$1M_{\odot}$/s. These are plausible values for a central engine
of SGRBs since the life time of torus becomes $t_{\rm dur} \sim
10$--1,000 ms with such choice. The transport time of the angular
momentum used here for $\dot M$ is assumed to be determined by an
appropriate value of the viscosity or moderately strong magnetic
fields as in the $\alpha$-viscosity model \cite{viscosity}. 

The gravitational binding energy $W$ of disk with mass
$\sim 0.01$--$0.1M_{\odot}$ in the black hole spacetime is
\beqn
W && \sim {M_{\rm BH} M_d \over r_d} \nonumber \\
&& \approx 2 \times 10^{51}~{\rm ergs} 
\biggl({10M_{\rm BH} \over r_d}\biggr)
\biggl({M_d \over 0.01M_{\odot}}\biggr), \label{eqW}
\eeqn
where $r_d$ denotes an averaged disk radius.  Due to viscous
dissipation in the accretion disk, the kinetic energy is converted to
the thermal energy. Equation (\ref{eqW}) indicates that the total
thermal energy produced during the accretion will be $\sim 10^{51}$--$10^{52}$ 
ergs, which is $10^2$--$10^5$ times larger than the total radiated energy in 
SGRBs (for the value after correction of a beaming factor) \cite{fox2005}.
Therefore, a relativistic fireball of sufficient energy may be formed 
if there is a mechanism by which a fraction of the thermal energy is 
converted to the fireball. In the past several years, scenarios for 
producing such fireball have been indeed proposed 
\cite{RJ0,RJ,PWF,NPK,KM,MPN,SRJ,LRP}. 
We describe a scenario following a semianalytic calculation \cite{MPN}.
More detailed numerical simulations have been performed 
recently \cite{SRJ,LRP}. According to these numerical works, the neutrino
luminosity can be by a factor of $\sim 10$ larger than that in
\cite{MPN} if the mass of disk is sufficiently large $\agt 0.05M_{\odot}$. 
Thus, the luminosity described below may be considered as a
conservative value. 

Because of its high temperature (typically $\sim 10^{11}$ K)
and density ($\sim 10^{11}~{\rm g/cm^3}$; cf.  Figs
\ref{FIG7}, \ref{FIG8}, and \ref{FIG9}(d)--(f)), the torus radiates
strongly in thermal neutrinos. The opacity inside the torus
(considering only neutrino absorption and scattering interactions with
nucleons) is $\kappa \sim 7\times 10^{-17} (T/10^{11}~{\rm K})^2~{\rm
cm}^2$~${\rm g}^{-1}$ \cite{PWF,MPN} where $T$ is the temperature. To
estimate the optical depth, we define the surface density of torus by
\beqn
\Sigma(x,y) = \int_{z \geq 0} \rho u^t \sqrt{-g} dz,
\eeqn
where the integral is carried out along lines of $x=y=$constant.  For $M_d
\agt 0.01M_{\odot}$ (e.g., for models APR1317, APR127175, APR1218,
SLy1315, SLy125155, SLy1216), $\Sigma \agt 10^{17}~{\rm g/cm^2}$, so
that the optical depth of the neutrinos $\sim \kappa \Sigma$ is $\agt
1$ for $r \alt 20$ km $\approx 5M_{\rm BH}$. This optical depth is in
approximate agreement with the model presented in \cite{MPN}.
(Note that in a high viscosity case, the significant increase of
the thermal pressure by the viscous dissipation may enforce the torus
to expand and to decrease the optical depth below unity \cite{SRJ}). 

In the optically thick case, approximate neutrino luminosity may be
estimated in the diffusion limit~\cite{ST} as $L_{\nu} \sim \pi r_d^2 F$
where $F$ is the neutrino flux from the torus surface,
approximately estimated by 
\beqn
F \sim {7N_{\nu} \over 3} {\sigma T^4 \over \kappa \Sigma}.
\eeqn
Here, $\sigma$ denotes the Stefan-Boltzmann constant and $N_{\nu}$ is
the number of neutrino species, taken as 3. Then, the neutrino
luminosity is expressed as 
\beqn
L_{\nu} &\sim & 2 \times 10^{52}~{\rm ergs/s}
\biggl({r_d \over 10~{\rm km}}\biggr)^2 \nonumber \\ 
&~&~~ \times \biggl( {T \over 10^{11}~{\rm K}} \biggr)^2
\biggl({\Sigma \over 10^{17}~{\rm g/cm^2}}\biggr)^{-1}, 
\eeqn
which is only slightly smaller than the neutrino Eddington 
luminosity~\cite{MPN}. Hence, our numerical results suggest formation
of a hot, hyperaccreting torus which is optically thick to neutrinos.
A model for the neutrino emission in a similar flow environment with
comparable $L_{\nu}$ is provided in \cite{MPN} as a neutrino-dominated
accretion flow (NDAF).

According to the model of \cite{MPN}, the luminosity due to
neutrino-antineutrino ($\nu\bar\nu$) annihilation is \cite{MPN}
\beq
L_{\nu\bar\nu} \sim 10^{49}{\hbox{--}}10^{50}~{\rm ergs/s}, \label{nunu}
\eeq
for $\dot M \sim 0.1$--$1M_{\odot}$/s. (The work in \cite{PWF,MPN} also 
indicates that for $\dot M < 0.1M_{\odot}$/s, the luminosity steeply
decreases far below $L_{\nu\bar\nu}=10^{49}$ ergs/s.) 
In our model, the black hole is rapidly rotating with $q \sim 0.75$, 
and hence, the luminosity may be enhanced by a factor of several due
to a GR effect \cite{PWF}. 
Also, in a more careful estimation for the neutrino annihilation and
for the neutrino opacity than those in \cite{MPN}, the value of
$L_{\nu\bar\nu}$ could be increased by a factor of $\sim 10$
\cite{SRJ,Takahashi}.

Because of the thick geometry of the torus, pair $\nu\bar\nu$
annihilation will be most efficient near the $z$-axis and just above
the surface of the inner region of the torus \cite{SRJ,Takahashi} for 
which the density is much smaller than that of the torus.
Consequently, the baryon-loading problem \cite{GRB} will be escaped,
and hence, a strongly relativistic fireball is likely to be produced.

Aloy et al.~\cite{ALOY} simulate the propagation of jets powered by
energy input along the rotation axis (as would be supplied by the
$\nu\bar\nu$ annihilation).  They find that if the half-opening angle
of the energy injection region is moderately small ($\alt 45^{\circ}$)
and the baryon density around the black hole is sufficiently low, jets
with the Lorentz factors in the hundreds can be produced given an
energy input $L_{\nu\bar\nu}\approx 10^{49}$--$10^{50}~{\rm ergs/s}$
lasting $\sim 100$ ms. In such case, the conversion rate from the
pair annihilation luminosity to the jet energy is a few $\times 10\%$ in
their results. They also indicate that the duration of SGRBs
may be $\sim 10$ times longer than the duration of the energy input
because of the differing propagation speeds of the jet head and tail;
namely, duration of the energy supply with $\sim 10$ ms may be
sufficient for explaining a SGRB with duration $\sim 100$ ms. 

Our numerical results, along with the accretion flow (NDAF) \cite{MPN}
and jet propagation models of~\cite{ALOY}, thus suggest that the
system composed of a black hole and hot torus with $M_d \agt
0.01M_{\odot}$ and accretion rate $\dot M \sim 0.1$--$1M_{\odot}$/s is
a possible candidate for the central engine of SGRBs. Since the
lifetime of the torus is $\agt 10$--100 ms in our model, the total
energy of the $\nu\bar\nu$ annihilation is $E_{\nu\bar\nu} \sim
10^{48}~{\rm ergs}(M_d/0.01M_{\odot})$. Typical total energy of SGRBs are
several $\times 10^{48}$ ergs \cite{fox2005}, and hence, a system of
the disk mass $\sim$ several $\times 0.01M_{\odot}$ is likely to be
appropriate for powering SGRBs as long as the emission is
beamed. Probably, beaming is encouraged by the thick geometrical
structure of the torus~\cite{ALOY}.

The above discussion suggests that if the mass ratio is sufficiently
small, prompt formation of a black hole may be a possible scenario for
producing the central engine of SGRBs. In particular, for a small
value of $Q_M \alt 0.8$ with the SLy EOS, a large SGRB energy is expected. 
Unfortunately, binary neutron stars of small mass ratio with $Q_M <
0.9$ have not been found so far \cite{Stairs}. Hence, it is not clear 
at present whether the formation rate of such unequal-mass binary is large
enough for explaining the event rate of SGRBs. This is a weak point
in this scenario. 

Through formation of a hypermassive neutron star with moderately
strong magnetic fields, a system composed of a black hole and a disk
of mass $\sim 0.01$--$0.03M_{\odot}$ may be formed as discussed in
Sec. \ref{sec:HMNS}.  The estimation described in this section is also
applied for such system, and indicates that a central engine of SGRBs
of relatively low total energy $\sim$ several $\times 10^{47}$--$10^{48}$
ergs could be formed. If the time scale of the angular momentum
transport by the magnetic effects (or viscosity) is longer than emission
time scale of gravitational waves, a black hole is formed (i.e., a 
SGRB should be generated) after a longterm emission of quasiperiodic
gravitational waves from the hypermassive neutron star. As we will
discuss in Sec. \ref{sec:gw}, such gravitational waves may be detected
by advanced laser-interferometric detectors. If quasiperiodic
gravitational waves and subsequent SGRB are detected coincidently in
the same direction with a small time lag, thus, this scenario may be
confirmed.

The results for the disk mass in this paper suggest that a merger
between a black hole and a neutron star could form a system composing
a black hole and surrounding massive disk by tidal disruption \cite{FBSTR}.
Indeed, if the mass of the black hole is small enough [$M_{\rm BH}
\alt 4M_{\odot}(R_0/10~{\rm km})^{3/2}$ for $q=0$ and $M_{\rm BH} \alt
20M_{\odot}(R_0/10~{\rm km})^{3/2}$ for $q=0.9$ where $q$ and $R_0$
are a spin parameter of the black hole and a neutron star radius
\cite{ISM}], the neutron star will be tidally disrupted before
reaching the ISCO. If the event rate (though it is not clear at
present because black hole-neutron star binary has not been observed
so far) is large enough, such a merger could be a promising source for
producing a central engine of SGRBs.  However, a detailed simulation
in full general relativity is necessary to confirm this scenario since
the disk mass depends crucially on the location of the ISCO.

\begin{figure*}[thb]
\vspace{-4mm}
\begin{center}
\epsfxsize=3.in
\leavevmode
\epsffile{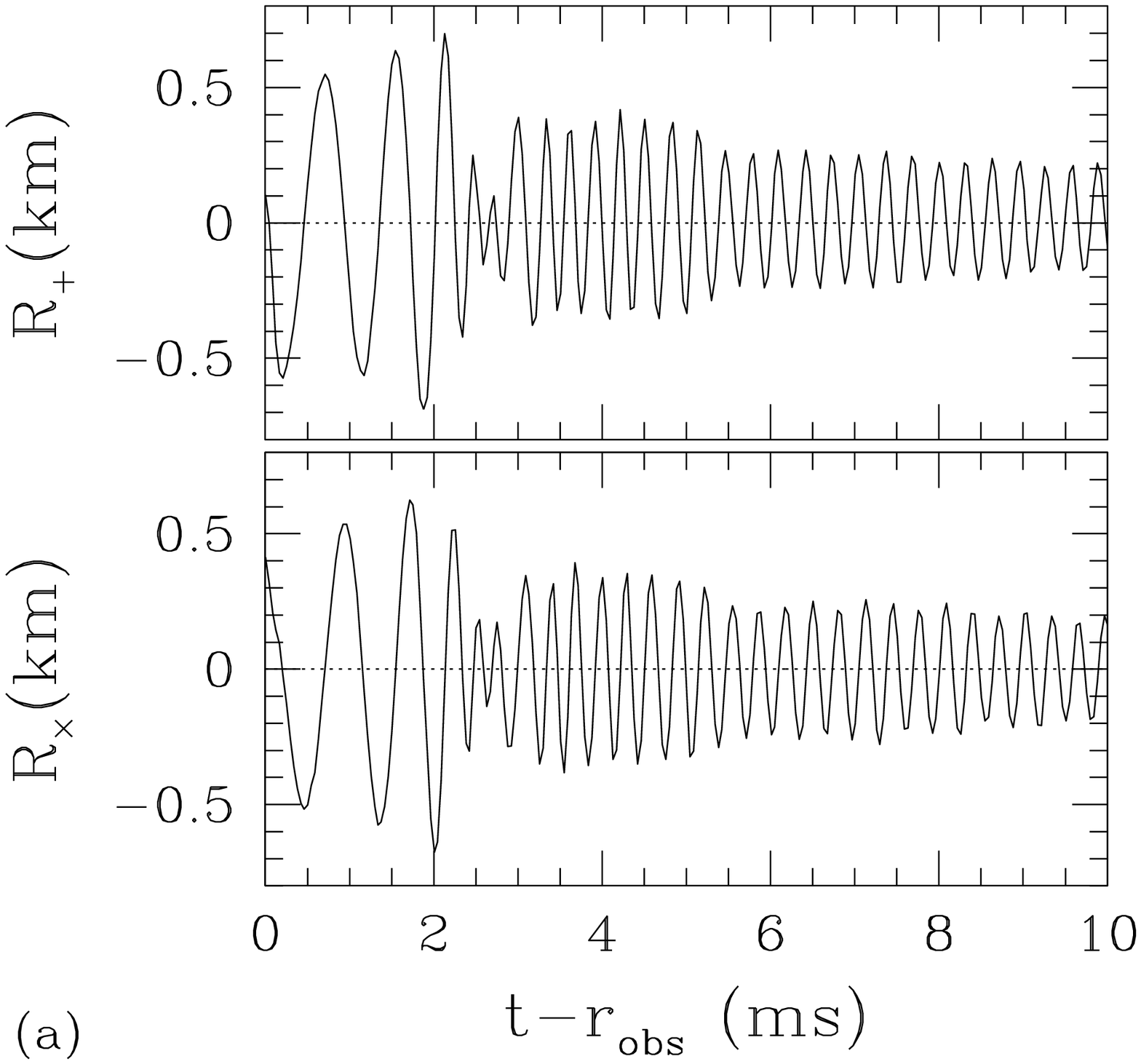}
\epsfxsize=3.in
\leavevmode
\hspace{1cm}\epsffile{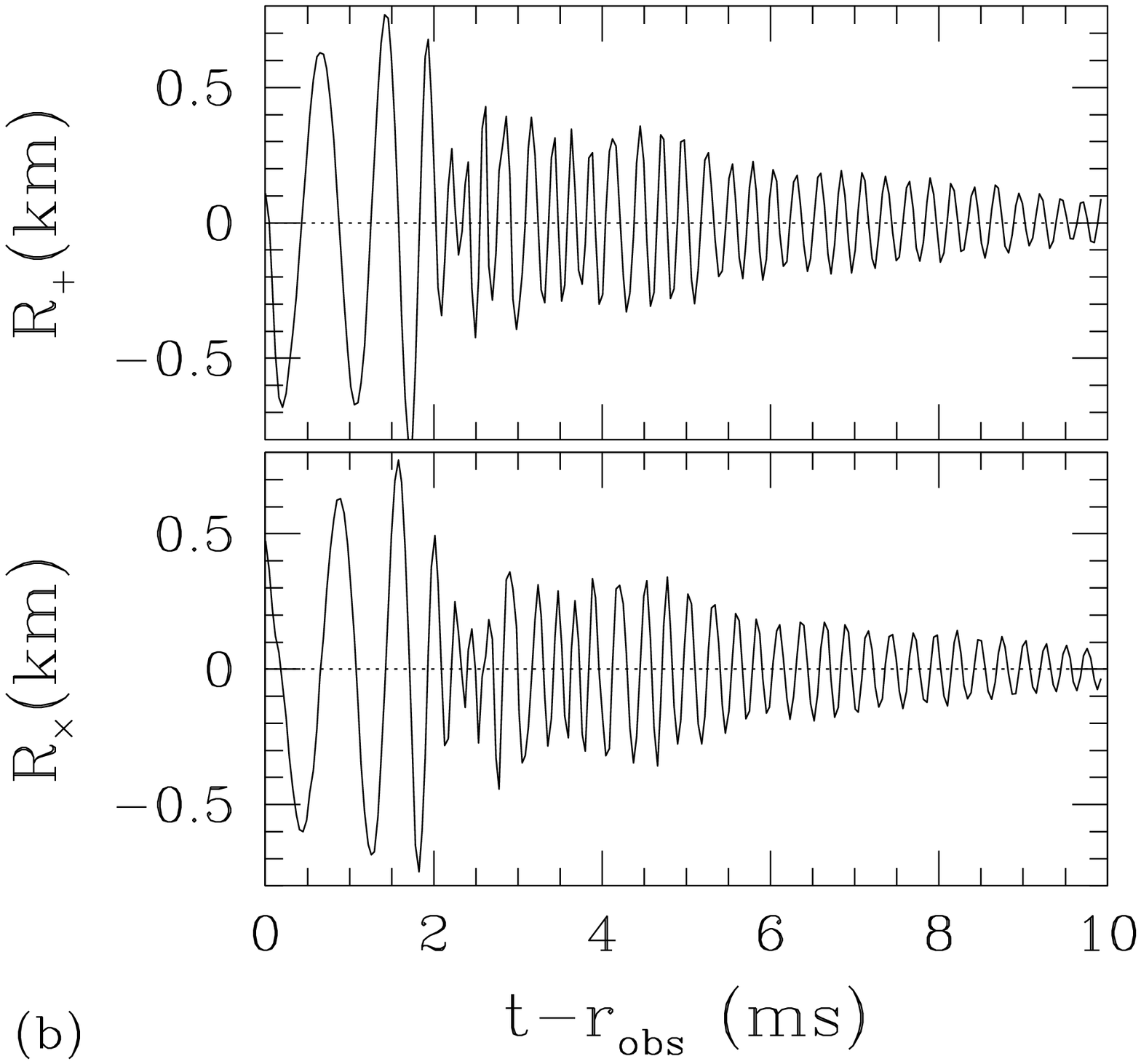}
\end{center}
\vspace{-8mm}
\caption{Gravitational waveforms,
$R_{+}$ and $R_{\times}$, (a) for model APR1313
at $r_{\rm obs}=36M_0$ and (b) for model APR1414 at $r_{\rm obs}=32M_0$. 
\label{FIG14}
}
\end{figure*}

\section{Gravitational waveforms}\label{sec:gw}

\subsection{Gravitational waves from hypermassive neutron stars}

\subsubsection{Waveform and luminosity} \label{sec:gwhmns}

In Fig. \ref{FIG14}, we present gravitational waveforms in the
formation of hypermassive neutron stars (models APR1313 and APR1414)
as a function of a retarded time. Throughout this paper, the retarded
time $t_{\rm ret}$ is defined by $t-r_{\rm obs}$ where $r_{\rm obs}$
is the coordinate radius of the wave extraction. In the early phase
($t_{\rm ret} \alt 2$ ms), gravitational waves associated with the
inspiral motion are emitted, while for $t_{\rm ret} \agt 2$ ms, those
by the rotating and oscillating hypermassive neutron star are emitted. 
In the following, we focus only on the waveforms for $t_{\rm ret} \agt 2$ ms.

For model APR1313, a hypermassive neutron star of ellipsoidal shape is
formed after the merger sets in. As a result, quasiperiodic
gravitational waves with an approximately constant frequency $\approx
3.2$ kHz are emitted. Also, the amplitude remains approximately
constant for $t \agt 5$ ms. These properties are essentially the same
as those found in \cite{STU2} for models SLy1313 and SLy125135. 

For model APR1414, on the other hand, the amplitude decreases with
time in particular for $t > 6$ ms. This reflects the fact that the
ellipticity of the formed hypermassive neutron star steeply decreases
for $t \agt 6$ ms. This indicates that only less massive binaries can
produce a longterm emitter of quasiperiodic gravitational waves. For
model APR1414, the frequency is not constant either, but modulates
with time as in model SLy135135 \cite{STU2}. The reason is that the
formed hypermassive neutron star quasiradially oscillates with a large
amplitude and the characteristic frequency varies with the change of
the characteristic radius. Due to this, the shape of the Fourier
spectra for models APR1313 and APR1414 are different even
qualitatively (cf. Fig. \ref{FIG17}). Namely, for $M \alt M_{\rm
thr}$, a slight difference in the total ADM mass results in a
significant difference in gravitational waveforms.

\begin{figure}[thb]
\vspace{-2mm}
\begin{center}
\epsfxsize=3.in
\leavevmode
\epsffile{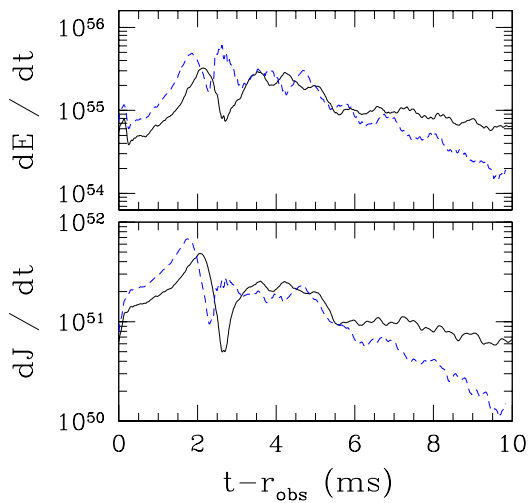}
\end{center}
\vspace{-10mm}
\caption{Energy and angular momentum emission rates,
$dE/dt$ and $dJ/dt$, of gravitational waves 
for models APR1313 (solid curves) and APR1414 (dashed curves). 
The units are erg/s and ${\rm g~cm^2/s^2}$, respectively. 
\label{FIG15}
}
\end{figure}

In Fig. \ref{FIG15}, the emission rates of the energy and the angular
momentum by gravitational radiation are shown for models APR1313
(solid curves) and APR1414 (dashed curves).  In the inspiral phase for
$t_{\rm ret} \alt 2$ ms, the emission rates increase with time
(besides initial unphysical bump associated with the conformal flat
initial condition in which gravitational waves are neglected), since
the amplitude and the frequency of the chirp signal increase. Then the
peak is reached at $t_{\rm ret} \sim 2$ ms. The height of the peak is
larger for model APR1414 since the compactness of each star is larger. 
After the peak is reached, the emission rates once quickly decrease
since the merged object becomes a spheroidal transient object.
However, because of its large angular momentum and stiff EOS, the
formed hypermassive neutron star soon changes to a highly ellipsoidal
object which emits gravitational waves of a large amplitude. The
luminosity from the ellipsoidal neutron star is as high as the first
peak at $t_{\rm ret} \sim 2.5$ ms for model APR1414 and at $t_{\rm
ret} \sim 3.5$ ms for model APR1313.

After the second peak, the emission rates of the energy and the
angular momentum via gravitational waves gradually decrease with time,
since the degree of the non-axial symmetry decreases. However, for
model APR1313, the decrease rates are small and the emission rates at
$t_{\rm ret} \sim 10$ ms remain to be as high as that in the late
inspiral phase as $dE/dt \sim 6 \times 10^{54}$ erg/s and $dJ/dt \sim
6 \times 10^{50}~{\rm g~cm^2/s^2}$ reflecting the high ellipticity of
the hypermassive neutron star.  The angular momentum at $t \sim 10$ ms
is $J \sim 0.7J_0 \sim 4 \times 10^{49}~{\rm g~cm^2/s}$.  Assuming
that the emission rate of the angular momentum remains $\sim 5 \times
10^{50}~{\rm g~cm^2/s}$ and that emission stops when about half of $J$
is dissipated, the emission time scale is approximately evaluated as
$J/(2dJ/dt) \sim 40$ ms. If the nonaxisymmetric structure is
maintained and $dJ/dt$ does not vary much, thus, the hypermassive
neutron star will collapse to a black hole within $\sim 50$ ms even in
the absence of other dissipation processes (cf. the discussion in the
last paragraph of \ref{sec:HMNS1}).

For model APR1414, on the other hand, the emission rates decrease quickly
since the hypermassive neutron star relaxes to an approximately
axisymmetric spheroid for $t \agt 6$ ms. For this model, the dissipation
time scale of the angular momentum is much longer than 50 ms at $t=10$ ms. 
Therefore, other dissipation processes such as magnetically induced
angular momentum transport will trigger the collapse to a black hole. 

\begin{figure}[thb]
\vspace{-4mm}
\begin{center}
\epsfxsize=3.in
\leavevmode
\epsffile{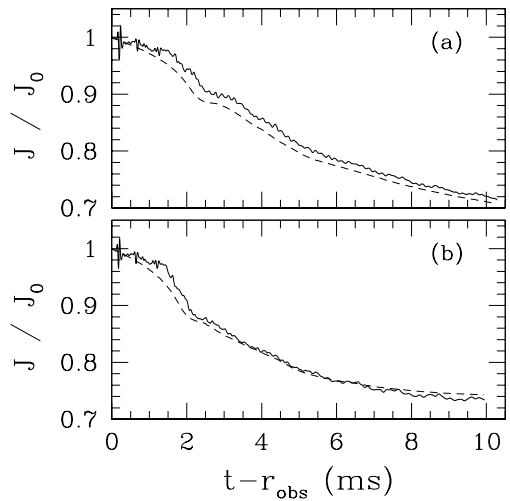}
\end{center}
\vspace{-10mm}
\caption{The solid curve denote evolution of angular momentum for 
models (a) APR1313 and (b) APR1414.  The dashed curves denote 
$J_0-\Delta J$ where $J_0$ denotes the angular momentum at $t=0$. 
\label{FIG16}
}
\end{figure}

By the time integral of $dE/dt$ and $dJ/dt$, the total energy and
angular momentum radiated are computed and found to be about $0.03M_0$
and $0.30J_0$ for model APR1313 and $0.03M_0$ and $0.26J_0$ for model
APR1414, respectively. This indicates that the angular momentum is
significantly dissipated, illustrating that the angular momentum
dissipation plays an important role in the evolution of the system. To
confirm that the radiation reaction is followed in the simulation, we
display $J(t)$ and $J_0-\Delta (t)$ as a function of time for models
APR1313 and APR1414 in Fig. \ref{FIG16}. This shows that the angular
momentum computed from Eq. (\ref{eqj00}) agrees approximately with
$J_0-\Delta J$ (within $\sim 2\%$ error), proving that radiation
reaction is computed with a good accuracy.

\subsubsection{Fourier spectrum}

In the real data analysis of gravitational waves, a matched filtering
technique \cite{KIP} is employed.  In this method, the signal of the
identical frequency can be accumulated using appropriate templates. As
a result, the effective amplitude increases by a factor of $N^{1/2}$
where $N$ denotes an approximate number of the cycle of gravitational
waves for a given frequency.

\begin{figure}[thb]
\vspace{-4mm}
\begin{center}
\epsfxsize=3.in
\leavevmode
\epsffile{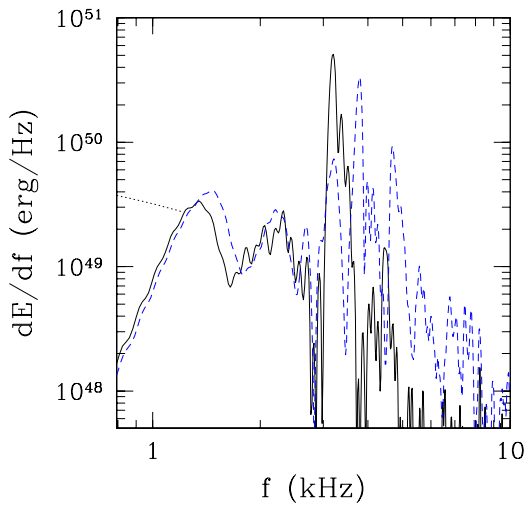}
\end{center}
\vspace{-10mm}
\caption{Fourier power spectrum of gravitational waves $dE/df$ for
models APR1313 (solid curve) and APR1414 (dashed curve). Since the
simulations are started when the characteristic frequency of
gravitational waves is $\sim 1$ kHz, the spectrum for $f < 1$ kHz
cannot be presented.  The dotted curve in the panel denotes the
analytical result of $dE/df$ in the second post Newtonian and
point-particle approximation by which the real spectrum for $f \alt 1$
kHz is approximated.
\label{FIG17}
}
\end{figure}

To determine the characteristic frequency of gravitational waves, we
carry out a Fourier analysis. In Fig. \ref{FIG17}, the power
spectrum $dE/df$ is presented for models APR1313 and APR1414. Since
the simulations were started with the initial condition of the orbital
period $\sim 2$ ms (i.e., frequency of gravitational waves is $\sim 1$
kHz), the spectrum of inspiraling binary neutron stars for $f < 1$ kHz
cannot be correctly computed. Thus, only the spectrum for $f \agt 1$
kHz should be paid attention. As a plausible spectrum for $f \alt 1$
kHz, we plot the Fourier power spectrum of two point particles in
circular orbits in the second post Newtonian approximation (the dotted curve)
\cite{blanchet} (the third post Newtonian terms does not significantly
modify the spectrum since their magnitude is $\sim 0.01$ of the
leading-order term). 

Figure \ref{FIG17} shows that a sharp characteristic peak is present
at $f \approx 3.2$ and 3.8 kHz for models APR1313 and APR1414,
respectively.  This is associated with quasiperiodic gravitational
waves emitted by the formed hypermassive neutron stars. Two side-band
peaks are present at $f \approx 3.2$ and 4.7 kHz for model
APR1414. Thus, the spectral shape is qualitatively different from that
for model APR1313. The reason is that the amplitude of the quasiradial
oscillation of the hypermassive neutron star is outstanding and the
characteristic radius varies for a wide range for model APR1414,
inducing the modulation of the wave frequency. 

\begin{figure}[thb]
\vspace{-2mm}
\begin{center}
\epsfxsize=3.in
\leavevmode
\epsffile{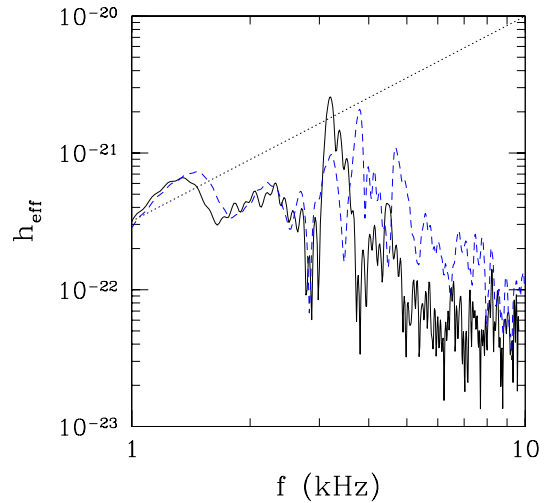}
\end{center}
\vspace{-10mm}
\caption{Non-dimensional effective amplitude of gravitational waves from
hypermassive neutron stars for models APR1313 (solid curve) and
APR1414 (dashed curve). The assumed distance is 50 Mpc. 
The dotted line denotes the planned noise level of the advanced LIGO. 
\label{FIG18}
}
\end{figure}

An effective amplitude of gravitational waves observed 
from the most optimistic direction (which is parallel to the 
axis of the angular momentum) is proportional to $\sqrt{dE/df}$ 
in the manner 
\beqn
h_{\rm eff} & \equiv & \sqrt{|\bar R_+|^2 + |\bar R_{\times}|^2}f \nonumber \\
&=&1.8 \times 10^{-21}
\biggl({dE/df \over 10^{51}~{\rm erg/Hz}}\biggr)^{1/2}
\biggl({100~{\rm Mpc} \over r}\biggr), \label{heff}
\eeqn
where $r$ denotes the distance from the source, and $\bar
R_{+,\times}$ are the Fourier spectrum of $R_{+,\times}$.  In
Fig. \ref{FIG18}, we show $h_{\rm eff}$ as a function of $f$ for a
hypothetical distance of 50 Mpc. This shows that the effective
amplitude of the peak is $\sim 3$ times larger than that at $\sim
1.3$--1.5 kHz which corresponds to the frequency of the last inspiral
motion.

For model APR1313, furthermore, the amplitude of the peak in reality
should be larger than that presented here, since we stopped
simulations at $t \sim 10$ ms to save the computational time, and
hence, the integration time ($\sim 10$ ms) is much shorter than the
realistic value. Extrapolating the decrease rate of the angular
momentum, the hypermassive neutron star will dissipate sufficient
angular momentum by gravitational radiation until a black hole or a
spheroidal star is formed.  As indicated in Sec. \ref{sec:gwhmns}, the
duration of the angular momentum dissipation would be $\sim 50$
ms. Thus, we may expect that the emission will continue for such time
scale and the effective amplitude of the peak would be in reality
amplified by a factor of $\sim 5^{1/2} \approx 2$ to be $\sim 6 \times
10^{-21}$ at a distance of 50 Mpc. Although the sensitivity of
laser-interferometric gravitational wave detectors for $f > 1$ kHz is
limited by the shot noise of the laser, this value is by a factor of
3--4 larger than the planned noise level of the advanced LIGO $\approx
10^{-21.5}(f/1~{\rm kHz})^{3/2}$ \cite{KIP}. It will be interesting to
search for such quasiperiodic signal of high frequency if the chirp
signal of gravitational waves from inspiraling binary neutron stars of
distance $r \alt 50$ Mpc are detected in the near future. As discussed
in \cite{S2005}, the detection of such quasiperiodic signal leads to
constraining nuclear EOSs for neutron star matter. Furthermore, as
mentioned in Sec. \ref{sec:grb}, the detection may also lead to
confirming a scenario for producing the central engine of SGRBs.

\begin{figure*}[tbh]
\vspace{-4mm}
\begin{center}
\epsfxsize=3.3in
\leavevmode
\epsffile{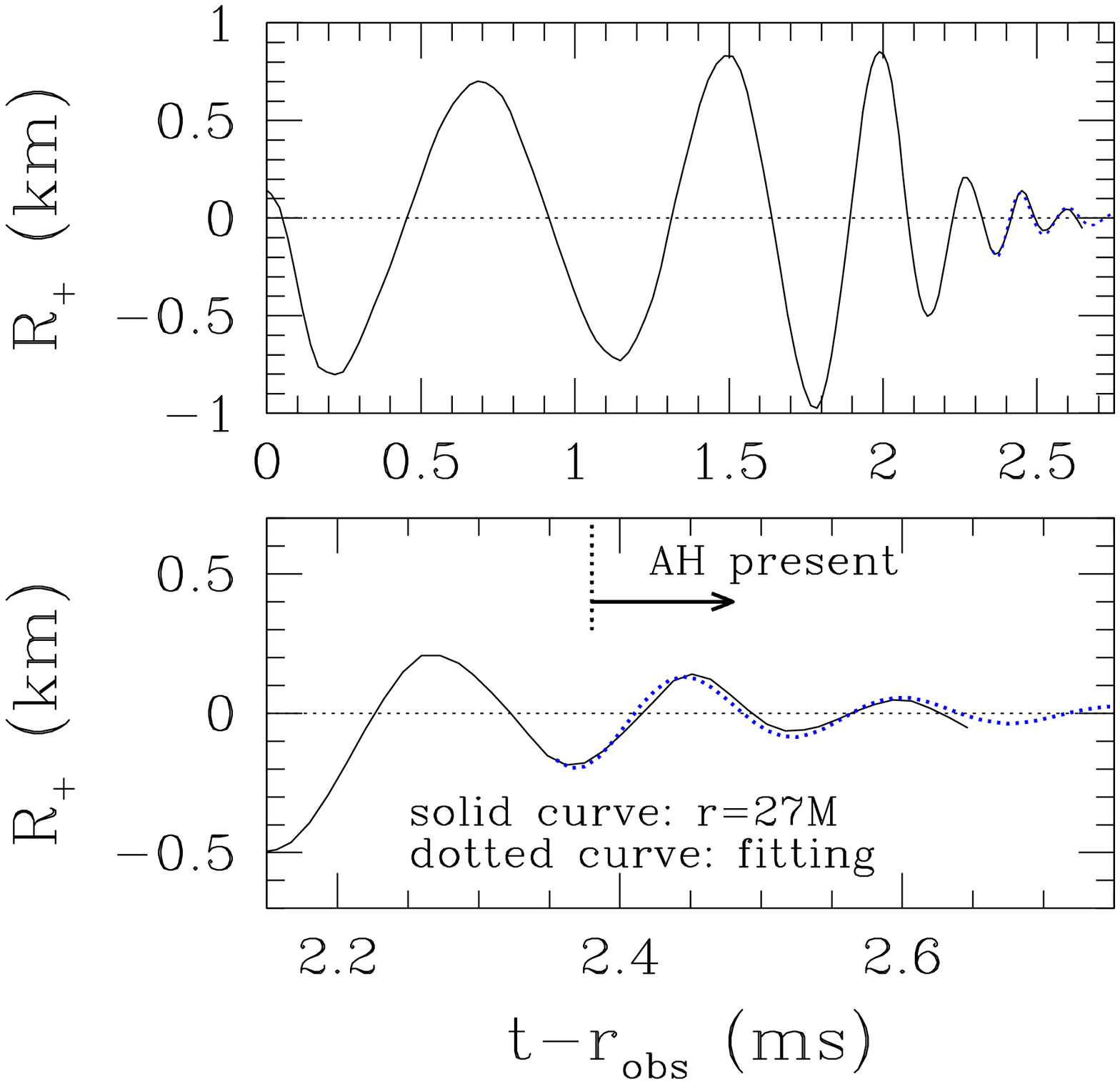}
\epsfxsize=3.3in
\leavevmode
\hspace{1cm}\epsffile{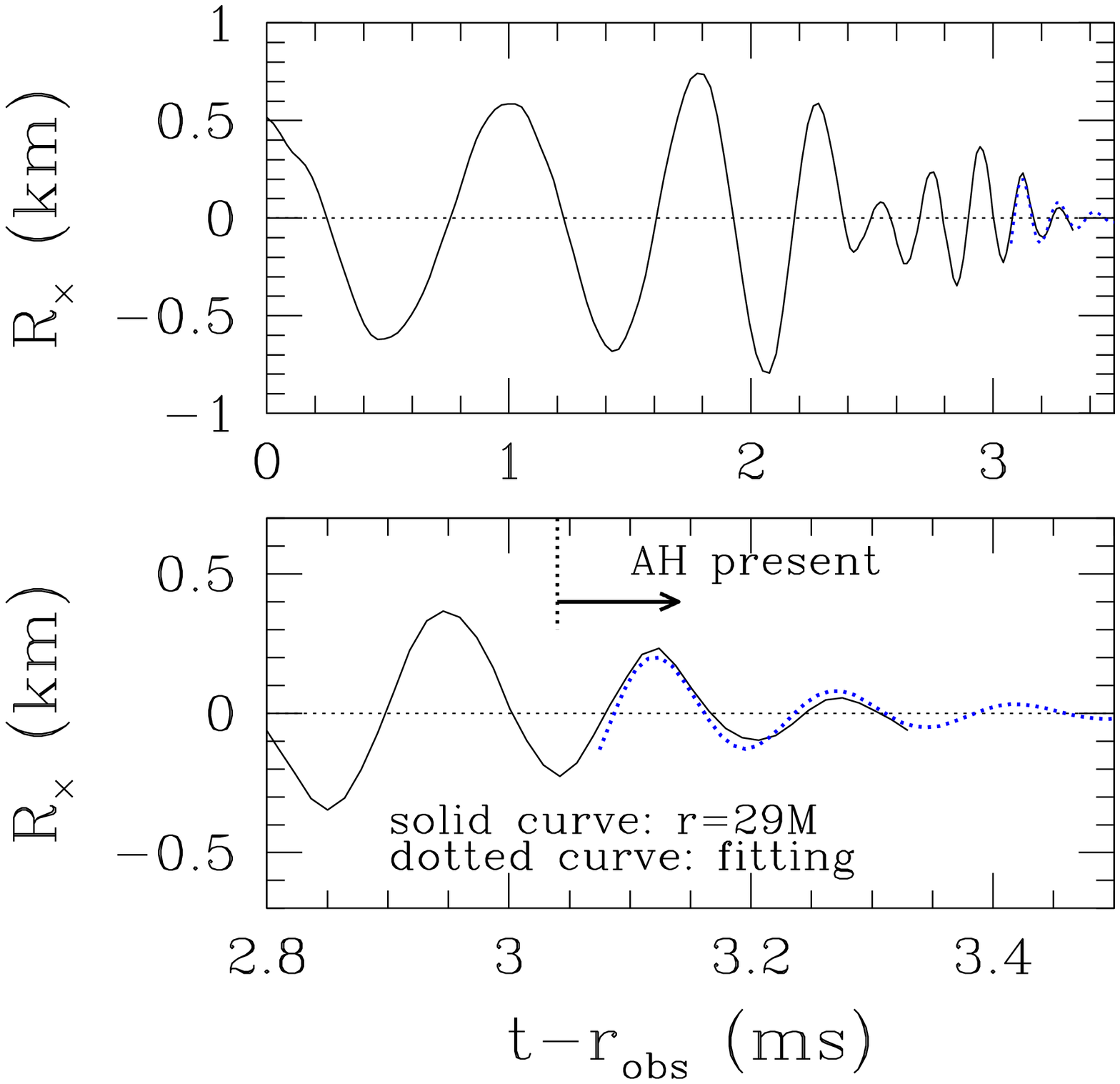}
\end{center}
\vspace{-10mm}
\caption{The same as Fig. \ref{FIG14}, but for models APR135165 (left)
at $r_{\rm obs}=27M_0$ and model SLy125155 at $r_{\rm obs}=29M_0$
(right).  The lower panel is the enlargement of the upper panel for
ring-down gravitational waves associated with a quasinormal mode of
the formed black hole. The formation time of the apparent horizon
$t_{\rm AH}$ is 2.38 ms for model APR135165 and 3.04 ms for model
SLy125155, and hence, the quasinormal mode is excited for $t-r_{\rm
obs} \agt t_{\rm AH}$. The thick dotted (blue) curves denote the
fitting formulae described by Eq. (\ref{fitting}).
\label{FIG19}
}
\end{figure*}

On the other hand, for model APR1414, the quasiperiodic gravitational
waves damp at $t \sim 6$ ms, and hence, the expected amplitude is
$h_{\rm eff} \sim 2 \times 10^{-21}$ at a distance of 50 Mpc with $f
\sim 4$ kHz. If the required signal to noise ratio of gravitational
waves for confirming the detection is $\sim 5$ in the advanced LIGO,  
such signal will be detected only for an event within the distance $\sim 
10$ Mpc. Since the predicted event rate of the merger for such distance
is $\alt 0.01$ per years \cite{BNST}, frequent detection is not 
expected. However, the fortunate detection may lead to confirming
a scenario for producing the central engine of SGRBs.

\subsection{Gravitational waves in the black hole formation}

In the prompt formation of a black hole, ring-down gravitational
waveforms associated with a quasinormal mode is emitted.  Figure
\ref{FIG19} shows $R_+$ for model APR135165 and $R_{\times}$ for model
SLy125155 as a function of retarded time. In the early time ($t_{\rm
ret} \alt 2$ ms for model APR135165 and $t_{\rm ret} \alt 2.3$ ms for
model SLy125155), gravitational waveforms are determined by the
inspiraling motion. After this, a merger waveform, which is emitted
due to hydrodynamic interaction between two stars, is seen for a
short time scale $\sim 0.5$ ms. The merger waveform is emitted for a
longer time for model SLy125155 simply because the time from the onset
of the merger to formation of a black hole is longer. Finally, a black
hole is formed and the waveforms are determined by the fundamental
quasinormal mode. The formation time of the apparent horizon $t_{\rm
AH}$ is 2.38 ms for model APR135165 and 3.04 ms for model
SLy125155. Assuming that the formation time of the event horizon is
approximately the same, ring-down waveforms associated with the
quasinormal mode should be induced only for $t_{\rm ret} \agt t_{\rm
AH}$. In our numerical results, this condition indeed holds, and
therefore, we have confirmed that the quasinormal mode is extracted
\cite{footnote9}.

\begin{figure}[thb]
\vspace{-4mm}
\begin{center}
\epsfxsize=3.in
\leavevmode
\epsffile{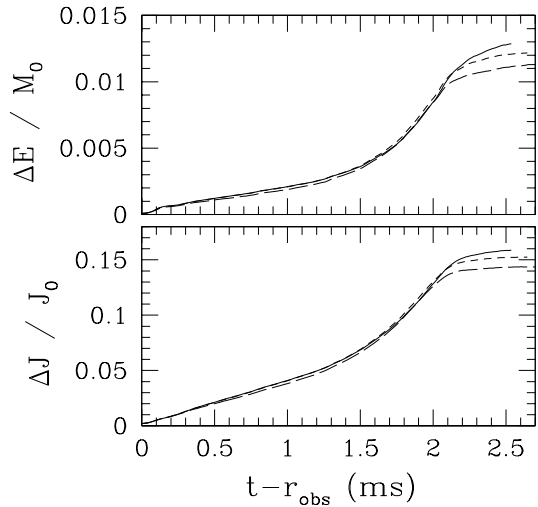}
\end{center}
\vspace{-10mm}
\caption{Evolution of the total radiated energy $\Delta E$ and 
angular momentum $\Delta J$ for models APR1416 (solid curves),
APR135165 (dashed curves), and APR1317 (long dashed curves) 
\label{FIG20}
}
\end{figure}

For models APR1416, APR135165, APR1317, and SLy125155, the initial
values of the non-dimensional angular momentum parameter $q_0$ is
$\approx 0.9$.  The energy and the angular momentum are dissipated by
gravitational waves by $\sim 0.01M_0$ and $0.15J_0$ until formation of
the black holes (cf. Fig. \ref{FIG20}). Taking into account the fact
that a tiny fraction of mass and angular momentum is distributed to
the surrounding disk, the spin parameter of the formed black hole
should be $\sim 0.75$. The dotted curve in Fig. \ref{FIG19} denotes a
model of ring-down gravitational waveforms
\beqn
A e^{-(t-t_{\rm AH})/t_d}\cos (2\pi f_{\rm qnm} t +\delta), \label{fitting}
\eeqn
where $A$ denotes the maximum amplitude and $\delta$
a phase constant. $f_{\rm qnm}$ and $t_d$ are the frequency and
damping rate of ring-down gravitational waves of $l=m=2$ mode,
which are approximately given by \cite{Leaver}
\beqn
&& f_{\rm qnm} \approx 10.8
\biggl({M \over 3M_{\odot}}\biggr)^{-1} [ 1-0.63(1-q)^{0.3}]~{\rm
kHz},~~ \label{eqbh} \\
&& t_d \approx {2(1-q)^{-0.45} \over \pi f_{\rm qnm}}. 
\label{QNMf}
\eeqn
For $M=2.9M_{\odot}$ and $q=0.75$, $f_{\rm qnm}=6.5$ kHz and
$t_d=0.183$ ms which are used for the fitting to the waveform of model
APR135165.  For model SLy125155, we plot the curve with
$M=2.7M_{\odot}$ and $q=0.7$, and hence, $f_{\rm qnm}=6.7$ kHz and
$t_d=0.163$ ms. This illustrates that with $M
=2.7$--$2.9M_{\odot}$, the frequency will be in a small range
between 6.5 and 7 kHz. 

In Fig. \ref{FIG19}, the fitting formulae are plotted together. It
shows that the computed waveforms agree approximately with the fitting
curves.  This implies that gravitational waves numerically extracted
carry the information of the formed black hole correctly. The maximum
value of $R_{+,\times}$ in the ring-down phase is $\sim 0.15$ km for
model APR135165 and $\sim 0.25$ km for model SLy125155, implying that
the maximum amplitude is $\sim 1$--$2 \times 10^{-22}$ at a distance
of 50 Mpc [cf. Eq. (\ref{hamp})].  These values depend weakly on the
mass ratio. The amplitude in model SLy125155 is slightly larger than
that in the APR EOS.  This is probably because the accretion rate of
the matter which can excite the quasinormal mode is larger in the SLy
EOS.

Since the frequency is too high and the amplitude is too small,
it will be difficult to detect ring-down gravitational waves even by
advanced laser-interferometric gravitational wave detector. Only in the
case that a merger happens in our local group, it may be detected.
However, the expected event rate is less than $0.001$ /yrs \cite{BNST}. 

Figure \ref{FIG20} shows the evolution of radiated energy and angular
momentum by gravitational waves for models APR1416, APR135165, and
APR1317. It is found that $\sim 1\%$ of the total mass energy and
$\sim 15\%$ of the angular momentum are dissipated by gravitational
waves.  These gravitational waves are mostly emitted in the last one
inspiral orbits and the contribution from the ring-down phase is
negligible. (For model SLy125155, the results are quantitatively the
same as those of APR models.)  The significance of the dissipation of
the angular momentum in the last orbit should be emphasized since it
affects the mass of disks surrounding the formed black hole. Another
important property found from Fig. \ref{FIG20} is that for smaller
mass ratios, the radiated energy and angular momentum are smaller.
This results from the fact that with the smaller mass ratios, the
tidal elongation sets in at a larger orbital separation, and hence,
the inspiral waveform shuts off earlier \cite{RS,FR,STU}. 

\section{Summary}\label{sec:summary}

\begin{figure*}
\vspace{1mm}
\begin{center}
\epsfxsize=5in
\leavevmode
\epsffile{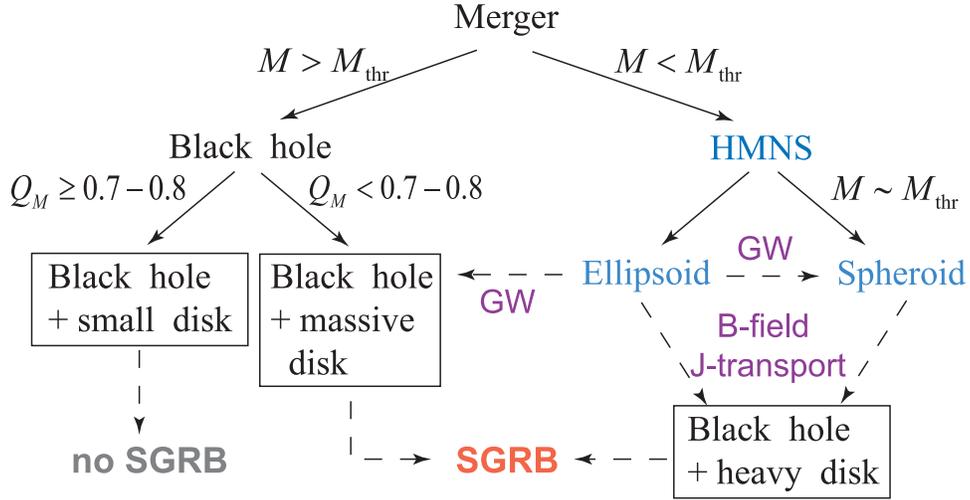}
\end{center}
\vspace{-4mm}
\caption{Summary about the outcome after the merger. ``HMNS'', ``GW'',
``B-field'', and ``J-transport'' are hypermassive neutron star,
gravitational wave emission, magnetic field, and angular momentum
transport, respectively. ``small disk'', ``massive disk'', and ``heavy
disk'' imply that the disk mass is $M_d \ll 0.01M_{\odot}$,
$0.01M_{\odot} \alt M_d \alt 0.03M_{\odot}$, and $M_d \agt
0.05M_{\odot}$, respectively. The solid arrow denotes the route found
in this paper and the dashed arrow is the possible route based on the
results of \cite{DLSSS2} and the speculation. There are three possible
fates of the ellipsoidal hypermassive neutron star depending on the
EOS and the time scale of dissipation and angular momentum transport
processes.  We note that a spheroid is formed probably only for the
APR EOS. See the summary 1--4 for details.
\label{FIG21}
}
\end{figure*}

As an extension of our previous work \cite{STU2}, we performed fully 
general relativistic simulations for the merger of binary neutron 
stars adopting stiff nuclear EOSs. We focus particularly on the 
formation of a black hole and surrounding disk in this paper. 
The following is the summary of the new results described in this paper.
The summary about the outcome after the merger is also described
in Fig. \ref{FIG21}. 

\noindent
1. \underline{Threshold mass for black hole formation} 
\vskip 1mm

In the APR EOS which is mainly adopted in this paper, the threshold
mass for prompt formation of a black hole is $M_{\rm thr} \approx
2.8$--$2.9M_{\odot}$ which is about 30--35 \% larger than the maximum
mass for spherical neutron stars with identical EOS, $M_{\rm
sph}$. Collecting the results obtained in \cite{STU2}, we conclude that
the following relation holds for stiff nuclear EOSs such as FPS,
SLy, and APR EOSs: $M_{\rm thr}=1.3$--1.35$M_{\rm sph}$.

\vskip 1mm
\noindent
2. \underline{Formation of hypermassive neutron star ($M < M_{\rm thr}$)}
\vskip 1mm

If $M < M_{\rm thr}$, a hypermassive neutron star is formed as an
outcome of the merger.  If the mass is not close to $M_{\rm thr}$, the 
outcome has an ellipsoidal shape, and hence, is a strong emitter of 
quasiperiodic gravitational waves. Because of angular momentum
dissipation by gravitational waves, the ellipsoidal hypermassive
neutron star may collapse to a black hole within $\sim 50$ ms, as
found in \cite{STU2}, even in the absence of any other transport
mechanism. On the other hand, if the ellipticity reduces to zero within
$\sim 50$ ms, the hypermassive neutron star will settle to a spheroid
in a stationary state. Then, the collapse to a black hole will be
triggered by other mechanisms such as the magnetic braking and the MRI 
\cite{MAG,DLSSS,DLSSS2}.

Since a torque works from the ellipsoidal hypermassive neutron star,
matter in the outer envelop receives angular momentum from the central part. 
This process helps producing disk, which may be an accretion disk
around a black hole eventually formed. If gravitational radiation
triggers the collapse, the expected disk mass is $\sim
0.01$--$0.03M_{\odot}$ for merger of equal-mass as well as
unequal-mass binaries for $Q_M \agt 0.85$ with the black hole mass
$\sim 2.6$--$2.8M_{\odot}$ and spin $q < 0.7$. Such system may be a
central engine of SGRBs of relatively small burst energy \cite{PWF,MPN}.

Alternatively, collapse to a black hole may be triggered by other
mechanisms such as the magnetic braking and the MRI
\cite{MAG,DLSSS,DLSSS2}. In the collapse by these magnetic effects,
the mass of a torus surrounding the black hole will be larger than
$0.05M_{\odot}$ because an efficient angular momentum transport works,
as illustrated in \cite{DLSSS,DLSSS2}. Furthermore, the MRI generates 
shocks in the torus, heating up the material to $\sim 10^{11}$ K
\cite{DLSSS2}. Therefore, this scenario is more favorable for
producing a central engine of SGRBs.

If the time scale of the angular momentum transport in the
hypermassive neutron stars is longer than $\sim 50$ ms, a black hole
or a spheroidal hypermassive neutron star is formed after a longterm
emission of quasiperiodic gravitational waves.  Such gravitational
waves may be detected by advance laser-interferometric detectors (see
summary 6). If a black hole is subsequently formed, it may also
produce a SGRB. Thus, if quasiperiodic gravitational waves and
subsequent SGRB are detected coincidently in the same direction with
small time lag, the merger scenario through formation of a
hypermassive neutron star for the central engine of SGRBs may be
confirmed. (However, for detecting gravitational waves, the event
should be within $\sim 50$ Mpc; see summary 6).

In the presence of strong magnetic fields with $\agt 10^{16}$ G, a 
black hole may be formed before a longterm emission of quasiperiodic
gravitational waves from the hypermassive neutron star
\cite{DLSSS,DLSSS2}. In this case, a smaller value of $h_{\rm eff}$
for the quasiperiodic gravitational waves is expected, and hence, a
smaller-distance event is required for confirmation of this scenario
by gravitational wave detection.

\vskip 1mm
\noindent
3. \underline{Formation of hypermassive neutron star ($M \alt M_{\rm thr}$)} 
\vskip 1mm

For $M$ only slightly smaller than $M_{\rm thr}$, the hypermassive
neutron star relaxes to a spheroidal state in $\sim 10$ ms since its
small angular momentum and compact structure lead to a too small ratio
of the rotational kinetic energy to the gravitational binding energy
to form an ellipsoid \cite{Chandra69}.  Here, the small angular
momentum is a result from the effects that the angular momentum is
significantly dissipated by gravitational waves soon after the onset
of the merger and that the angular momentum transport works
efficiently during an early quasiradial oscillation with a large
amplitude in the hypermassive neutron star. Since the emissivity of
gravitational waves by the hypermassive neutron stars of small
ellipticity is not high, it may survive for more than 50 ms in
contrast to the smaller-mass case.  Collapse to a black hole will be
triggered by other mechanism such as the magnetic braking and the MRI
\cite{MAG,DLSSS,DLSSS2}. After the collapse, disk of mass $\agt
0.05M_{\odot}$ could be formed \cite{DLSSS,DLSSS2}, and hence, such
system is also a candidate for the central engine of SGRBs.  In this
scenario, the amplitude of quasiperiodic gravitational waves is not as
high as that for the smaller-mass case, and hence, frequent detection
of this signal is not expected (see summary 6).

The formation of a spheroid is found only for the APR EOS (in
\cite{STU2} in which the SLy and FPS EOSs are used, we have not found 
the formation of spheroids). The likely reason in this difference is
that the APR EOS is stiffer than others. Namely, the hypermassive neutron
stars can be more compact with this EOS [cf. Fig. \ref{FIG2}(b)]
escaping the collapse to a black hole. In such compact state, 
the gravitational binding energy could be large enough to reduce the
ratio of the rotational kinetic energy to the binding energy below the
threshold value of the formation of an ellipsoid. 

\vskip 1mm
\noindent
4. \underline{Prompt formation of black hole ($M > M_{\rm thr}$)}
\vskip 1mm

In the black hole formation from the nearly equal-mass merger, most of
mass elements are swallowed into the horizon, and hence, the disk mass
around the black hole is much smaller than $0.01M_{\odot}$. However,
the disk mass steeply increases with decreasing the value of $Q_M$; we
find an empirical relation in which the disk mass is written 
approximately by $M_{d0}+M_{d1}(1-Q_M)^p$ with $p=3$--4 for given EOS
and mass for a range $0.7 \alt Q_M \leq 1$.

For $Q_M \sim 1$, the disk not only has small mass but also is
geometrically thin. Therefore, the merger will not produce a central
engine of SGRBs for nearly equal-mass case. On the other hand, for a
sufficiently small value of $Q_M$ ($Q_M \alt 0.75$ for the APR EOS and
$Q_M \alt 0.85$ for the SLy EOS), the disk mass is larger than
$0.01M_{\odot}$.  Furthermore, the disk is geometrically thick and the
thermal energy is large enough (typically $\sim 1$--$2 \times 10^{11}$
K) for producing the large amount of thermal neutrinos.  In addition,
the black hole spin is large $q \sim 0.75$. Therefore, prompt
formation of a black hole in the merger of unequal-mass binary neutron
stars with a small value of $Q_M$ is a scenario for
formation of a central engine of SGRBs.  Unfortunately, binary neutron
stars of small mass ratio with $Q_M < 0.9$ have not been found so far
\cite{Stairs}. Hence, it is not clear at present whether the merger rate of
such unequal-mass binary is large enough for explaining the event rate
of SGRBs. The present results also indicate that the merger between a
neutron star and a black hole with small mass will be a possible
candidate for producing central engine of SGRBs.

\vskip 1mm
\noindent
5. \underline{Gravitational waves from black hole}
\vskip 1mm

The non-dimensional angular momentum parameter ($J/M^2$) of the formed
Kerr black hole is $\sim 0.75$ in the prompt formation case.  Then,
for the system of mass 2.7--$2.9M_{\odot}$, the frequency of ring-down
gravitational waves associated with the fundamental quasinormal mode
of $l=m=2$ is $\sim 6.5$--7 kHz.  We extract the ring-down
gravitational waveform and confirm this frequency.  The amplitude of
gravitational waves is $\sim 1$--$2 \times 10^{-22}$ at a distance of
50 Mpc which is too small to be detected even by advanced
laser-interferometric detectors unless the merger event happens within
the local group of galaxies.

\vskip 1mm
\noindent
6. \underline{Gravitational waves from hypermassive neutron star}
\vskip 1mm

The effective amplitude of quasiperiodic gravitational waves from
hypermassive neutron stars of ellipsoidal shape can be $\agt 5 \times
10^{-21}$ at a distance of 50 Mpc with the frequency 3--3.5 kHz. This
property agrees with that found in \cite{STU2,S2005}, and hence, we
conclude that this fact holds irrespective of stiff nuclear EOSs with
$M_{\rm sph} \agt 2 M_{\odot}$.  Quasiperiodic gravitational waves may
be detected by advanced laser-interferometric detectors. Detection will 
lead to constraining the nuclear EOS as discussed in \cite{S2005}. On the
other hand, for $M$ only slightly smaller than $M_{\rm thr}$, the
expected effective amplitude is $h_{\rm eff} \sim 2 \times 10^{-21}$
at a distance of 50 Mpc with frequency $f \sim 4$ kHz; only for an
event with a distance within $\sim 10$ Mpc, detection of such signal
will be possible by the advanced laser-interferometric detectors.

\acknowledgments

MS is grateful to Stu Shapiro for discussion about the fate of
hypermassive massive neutron stars during collaboration in the
magnetohydrodynamic simulation in general relativity. Numerical
computations were performed on the FACOM VPP5000 machine at the data
analysis center of NAOJ and on the SX6 machine at the data analysis
center of ISAS in JAXA. This work was in part supported by
Monbukagakusho Grant (Nos. 17030004 and 17540232).


\begin{thebibliography}{99}

\bibitem{NEW} M. Burgey et al., Nature {\bf 426}, 531 (2003). 

\bibitem{BNST} V. Kalogera et al., Astrophys. J. {\bf 601},
L179 (2004); {\em ibid}, {\bf 614}, L137 (2004). 

\bibitem{KIP} C. Cutler and K. S. Thorne, in {\em Proceedings of the 16th
International Conference on General Relativity and Gravitation},
eds. N. T. Bishop and S. D. Maharaj (World Scientific, 2002), p.72,
and references therein.

\bibitem{Ando} M. Ando et al. (the TAMA collaboration), 
Phys. Rev. Lett. {\bf 86}, 3950 (2001). 

\bibitem{NPT} R. Narayan, B. Paczynski, and T. Piran, Astrophys. J. Lett.
{\bf 395}, L83 (1992).

\bibitem{GRB} 
B. Zhang and P. M\'esz\'aros, Int.~J.~Mod.~Phys.~A {\bf 19}, 2385 (2004);
T. Piran, Rev. Mod. Phys. {\bf 76}, 1143 (2005). 

\bibitem{short} J. S. Bloom, et al., Astrophys. J. {\bf 638}, 354 (2006); 
N. Gehrels, et al., Nature, {\bf 437}, 851 (2005); 
E. Berger et al., Nature, {\bf 438}, 988 (2005); 
S. D. Barthelmy et al., Nature, {\bf 438}, 994 (2005). 

\bibitem{fox2005} D.\ B.\ Fox, et al., Nature, {\bf 437}, 845 (2005).

\bibitem{sgrb724} J. S. Villasenor, et al., Nature, {\bf 437}, 855 (2005); 
J. Hjorth, et al., Nature, {\bf 437}, 859 (2005). 

\bibitem{RJ0} E.g., 
M. Ruffert and H.-Th. Janka, Astron. Astrophys. {\bf 380}, 544 (2001). 

\bibitem{gr3d} M. Shibata, Phys. Rev. D {\bf 60}, 104052 (1999).

\bibitem{bina} M. Shibata and K. Ury\=u, Phys. Rev. D {\bf 61}, 064001
(2000). 

\bibitem{bina2} M. Shibata and K. Ury\=u, Prog. Theor. Phys. {\bf 107},
265 (2002). 

\bibitem{other} J. A. Font et al., Phys. Rev. D {\bf 65}, 084024 (2002).

\bibitem{Font} J. A. Font, Living Rev. Relativ. {\bf 6}, 4 (2003). 

\bibitem{shiba2d} M. Shibata, Phys. Rev. D {\bf 67}, 024033 (2003).

\bibitem{STU} M. Shibata, K. Taniguchi, and K. Ury\=u,
Phys. Rev. D {\bf 68}, 084020 (2003). 

\bibitem{marks} M. Miller, P. Gressman, and W.-M. Suen, Phys. Rev. D
{\bf 69}, 064026 (2004).

\bibitem{illinois} M. D. Duez, P. Marronetti, T. W. Baumgarte, and 
S. L. Shapiro, Phys. Rev. D {\bf 67}, 024004 (2003). 

\bibitem{Baiotti} L. Baiotti, I. Hawke, P. J. Montero, F. L\"offler, 
L. Rezzolla, N. Stergioulas, J. A. Font, and E. Seidel, 
Phys. Rev. D {\bf 71}, 024035 (2005). 

\bibitem{STU2} M. Shibata, K. Taniguchi, and K. Ury\=u,
Phys. Rev. D {\bf 71}, 084021 (2005).

\bibitem{footnote1}
The hypermassive neutron star is defined as a differentially
rotating neutron star for which the total baryon rest-mass is larger
than the maximum allowed value of rigidly rotating neutron stars for a
given EOS: See \cite{BSS} for definition.

\bibitem{BSS} T. W. Baumgarte, S. L. Shapiro, and M. Shibata, 
Astrophys. J. Lett. {\bf 528}, L29 (2000). 

\bibitem{S2005} M. Shibata, Phys. Rev. Lett. {\bf 94}, 201101 (2005). 

\bibitem{nice} D. J. Nice et al., Astrophys. J., {\bf 634}, 1242 (2005). 

\bibitem{EOS1} A. Akmal, V. R. Pandharipande, and D. G. Ravenhall,
Phys. Rev. C {\bf 58}, 1804 (1998). 

\bibitem{EOS2} F. Douchin and P. Haensel, Astron. 
Astrophys.  {\bf 380}, 151 (2001).

\bibitem{Unruh} W. Unruh, unpublished (1984); 
E. Seidel and W.-M. Suen, Phys. Rev. Lett. {\bf 69}, 1845 (1992). 

\bibitem{AB}
M. Alcubierre and B. Br\"ugmann, Phys. Rev. D {\bf 63}, 104006 (2001).

\bibitem{DLSS} 
M. D. Duez, S. L. Shapiro, and H.-J. Yo, Phys. Rev. D. {\bf 69}, 104016 
(2004). 

\bibitem{OJ} R. Oechslin and H.-Th. Janka, Mon. Not. R. astr. Soc.,
submitted (astro-ph/0507099). 

\bibitem{shen} H. Shen, H. Toki, K. Oyamatsu, and K. Sumiyoshi,
Nuclear physics A {\bf 637}, 435 (1998). 

\bibitem{SN} M. Shibata and T. Nakamura, Phys. Rev. D {\bf 52} (1995),
5428; see also, T. W. Baumgarte and S. L. Shapiro,
Phys. Rev. D {\bf 59} (1999), 024007; 
M. Alcubierre et al., Phys. Rev. D {\bf 61} (2000), 041501. 

\bibitem{kurganov-tadmor} 
A. Kurganov and E. Tadmor, J.~Comput.\ Phys. {\bf 160}, 214 (2000).

\bibitem{SFont} M. Shibata and J. A. Font, Phys. Rev. D {\bf 72}, 047501
(2005). 

\bibitem{gw3p2} M. Shibata, Prog. Theor. Phys. {\bf 101}, 1199 (1999). 

\bibitem{S03} M. Shibata, Astrophys. J. {\bf 595}, 992 (2003). 

\bibitem{AH} M. Shibata, Phys. Rev. D {\bf 55}, 2002 (1997): 
M. Shibata and K. Ury\=u, Phys. Rev. D {\bf 62}, 087501 (2000). 

\bibitem{WM} J. R. Wilson and G. J. Mathews, Phys. Rev. Lett. {\bf 75},
4161 (1995).

\bibitem{CBS} 
C. S. Kochanek, Astrophys. J. {\bf 398}, 234 (1992): \\
L. Bildsten and C. Cutler, Astrophys. J. {\bf 400}, 175 (1992).

\bibitem{irre} M. Shibata, Phys. Rev. D {\bf 58}, 024012 (1998).\\
S. A. Teukolsky, Astrophys. J. {\bf 504}, 442 (1998). 

\bibitem{GBM} E. Gourgoulhon et al., Phys. Rev. D {\bf 63}, 064029 (2001).

\bibitem{TG}
K. Taniguchi and E. Gourgoulhon, Phys. Rev. D {\bf 66}, 104019 (2002):
{\em ibid} {\bf 68}, 124025 (2003). 

\bibitem{moncrief} V. Moncrief, Ann. of Phys. {\bf 88}, 323 (1974): 
The Moncrief formalism was originally derived for the 
Schwarzschild spacetime. We here apply his formalism 
in a flat spacetime. 

\bibitem{SS3} M. Shibata and Y.I. Sekiguchi, Phys. Rev. D {\bf 71}, 024014 
(2005).


\bibitem{HP} P. Haensel and A. Y. Potekhin, 
Astron. Astrophys. {\bf 428}, 191 (2004). 

\bibitem{footnote2}
The tables for the SLy and APR EOSs, which were involved in
the LORENE library in Meudon group (http://www.lorene.obspm.fr),
were implemented by Haensel and Zdunik.
(See \cite{BGGHTZ} for quasiequilibrium sequences of binary neutron stars
using these tables.)

\bibitem{Chandra} S. Chandrasekhar, {\em Stellar Structure} 
(Dover, 1967), Chap. 10. 

\bibitem{Stairs} I. H. Stairs, Science {\bf 304}, 547 (2004). 

\bibitem{Tsuruta} S. Tsuruta, Phys. Rep. {\bf 292}, 1 (1998). 

\bibitem{USE} K. Ury\=u, M. Shibata, and Y. Eriguchi, Phys. Rev. D {\bf 62}, 
104015 (2000). 



\bibitem{SU01} M. Shibata and K. Ury\=u, Phys. Rev. D {\bf 64}, 
104017 (2001). 

\bibitem{LBS} N. D. Lyford, T. W. Baumgarte, and S. L. Shapiro,
Astrophysical J. {\bf 583}, 410 (2003): 
I. A. Morrison, T. W. Baumgarte, and S. L. Shapiro,
Astrophys. J. {\bf 610}, 941 (2004). 

\bibitem{SBS}
M. Shibata, T. W. Baumgarte, and S. L. Shapiro, Astrophys. J. {\bf 542}, 453 
(2000). 

\bibitem{MAG} S. A. Balbus and J. F. Hawley, Astrophys. J. {\bf 376},
214 (1991); Rev. Mod. Phys. {\bf 70}, 1 (1998). 

\bibitem{DLSSS} M. D. Duez, Y. T. Liu, S. L. Shapiro, M. Shibata, and
B. C. Stephens, Phys. Rev. Lett. {\bf 96}, 031101 (2006). 

\bibitem{DLSSS2} M. Shibata, M. D. Duez, Y. T. Liu, S. L. Shapiro, and
B. C. Stephens, Phys. Rev. Lett. {\bf 96}, 031102 (2006). 

\bibitem{Chandra69} S. Chandrasekhar, {\em Ellipsoidal Figures of Equilibrium} 
(Yale University Press, New Haven, 1969); 
J.-L. Tassoul, {\em Theory of Rotating Stars}
(Princeton Univ. Press, Princeton, 1978).

\bibitem{BPT} J. M. Bardeen, W. H. Press, and S. A. Teukolsky, 
Astrophys. J. {\bf 178}, 347 (1972).

\bibitem{RJ} M. Ruffert, H.-Th. Janka, and G. Sch\"afer, Astron. Astrophys. 
{\bf 311}, 532 (1996): M. Ruffert et al., {\em ibid} {\bf 319}, 122 (1997); 
H.-Th. Janka and M. Ruffert, Astron. Astrophys. {\bf 307}, L33 (1996). 

\bibitem{AMR} M. Berger and J. Oliger, J. Comp. Phys. {\bf 53}, 484 (1984): 
E. Evans, S. Iyer, E. Schnetter, W.-M. Suen, J. Tao, R. Wolfmeyer, 
and H.-M. Zhang, Phys. Rev. D {\bf 71}, 081301 (2005):
B. Zink, N. Stergioulas, I. Hawke, C. D.
Ott, E. Schnetter, and E. M\"uller, gr-qc/0501080. 

\bibitem{Leaver} E. W. Leaver, Proc. R. Soc. London {\bf A402}, 285 (1985).\\
T. Nakamura, K. Oohara, and Y. Kojima, Prog. Theor. Phys. Suppl. No.~
{\bf 90}, 1 (1987).

\bibitem{viscosity} I. D. Novikov and K. S. Thorne, in {\em Black Holes},
editted by B. DeWitt and C. DeWitt (Gordon Breach, New York, 1973). 

\bibitem{PWF} R. Popham, S. E. Woosley, and C. Fryer, Astrophys. J.
{\bf 518}, 356 (1999).

\bibitem{NPK} R. Narayan, T. Piran, and P. Kumar, Astrophys. J. {\bf 557},
949 (2001). 

\bibitem{KM} K. Kohri and S. Mineshige, Astrophys. J. {\bf 577}, 311 (2002). 

\bibitem{MPN} Di Matteo, R. Perna, and R. Narayan, Astrophys. J. {\bf 579}, 
706 (2002).

\bibitem{SRJ} S. Setiawan, M. Ruffert, and H.-Th. Janka, Mon. Not. R.
astr. Soc. {\bf 352}, 753 (2004).

\bibitem{LRP} W. H. Lee, E. R. Ruiz, and D. Page, Astrophys. J.
{\bf 608}, L5 (2004); Astrophys. J. {\bf 632}, 421 (2005). 

\bibitem{ST} See, e.g., Appendix I of S. L. Shapiro and S. A. Teukolsky, 
{\em Black Holes, White Dwarfs, and Neutron Stars}, Wiley Interscience
(New York, 1983).

\bibitem{Takahashi} R. Takahashi, private communication. 

\bibitem{ALOY} M. A. Aloy, H.-T. Janka, and E. M\"uller, Astron. Astrophys. 
{\bf 436}, 273 (2005). 

\bibitem{FBSTR} J. A. Faber, T. W. Baumgarte, S. L. Shapiro, K. Taniguchi,
and F. A. Rasio, Phys. Rev. D {\bf 73}, 024012 (2006). 

\bibitem{ISM} M. Ishii, M. Shibata, and Y. Mino, Phys. Rev. D {\bf 71}, 
044017 (2005). See also, \cite{taniguchi}.

\bibitem{blanchet} L. Blanchet, G. Faye, B. R. Iyer, and B. Joguet,
Phys. Rev. D {\bf 65}, 061501(R) (2002): 
L. Blanchet, T. Damour, G. Esposito-Farese, and B. R. Iyer, Phys. Rev.
Lett. {\bf 93}, 091101 (2004).

\bibitem{footnote9} It is important to check that the quasinormal mode
of a black hole is seen for $t_{\rm ret} \agt t_{\rm AH}$ for
confirming that it is different from the w-mode \cite{AK} which has a
similar damping waveform but is emitted before the formation of a
black hole. For example, see \cite{Luca} in which the authors seem to
misrelate the gravitational waveform extracted to a quasinormal mode
ringing; namely, the gravitational waveform is extracted only for
$t_{\rm ret} < t_{\rm AH}$ in their simulation, and hence, it cannot
correspond to a quasinormal mode of a black hole.

\bibitem{AK} N. Andersson and K. D. Kokkotas, Phys. Rev. Lett. {\bf 77},
4134 (1996); K. D. Kokkotas and B. Schmidt, 
Living Rev. Relativ. {\bf 2}, 2 (1999). 

\bibitem{Luca} L. Baiotti, I. Hawke, L. Rezzolla,
and E. Schnetter, Phys. Rev. Lett. {\bf 94}, 131101 (2005). 

\bibitem{RS} 
F. A. Rasio and S. L. Shapiro, Astrophys. J. {\bf 432}, 242 (1994); 
X. Zhuge, J. M. Centrella, and S. L. W. McMillan, 
Phys. Rev. D {\bf 54}, 7261 (1996). 

\bibitem{FR}  J. A. Faber and F. A. Rasio, Phys. Rev. 
D {\bf 62}, 064012 (2000); {\it ibid} {\bf 63}, 044012 (2001); 
{\em ibid} {\bf 65}, 084042 (2002). 

\bibitem{BGGHTZ} 
M. Bejger, D. Gondek-Rosi\'nska, E. Gourgoulhon, P. Haensel, 
K. Taniguchi, and J.L. Zdunik, Astron. Astrophys. {\bf 431}, 297 (2005). 

\bibitem{taniguchi} K. Taniguchi, T. W. Baumgarte, J. A. Faber, S. L.
Shapiro, Phys. Rev. D {\bf 72}, 044008 (2005). 

\end{thebibliography}
\end{document}